RESEARCH ARTICLE

# Internet of Things-Based Smart Precision Farming in Soilless Agriculture: Opportunities and Challenges for Global Food Security

MONICA DUTTA[1], DEEPALI GUPTA[1], SUMEGH THAREWAL[2,3], DEEPAM GOYAL[1], JASMINDER KAUR SANDHU[4], MANJIT KAUR[5], AHMAD ALI ALZUBI[6], AND JAZEM MUTARED ALANAZI[6]

[1]Chitkara University Institute of Engineering and Technology, Chitkara University, Rajpura, Punjab 140401, India
[2]Faculty of Engineering, Free University of Bozen-Bolzano (Unibz), 39100 Bolzano, Italy
[3]School of Advance Computing, DBS Global University, Dehradun, Uttarakhand 248011, India
[4]School of Computer Science and Engineering, IILM University, Greater Noida 201306, India
[5]School of Computer Science and Artificial Intelligence, SR University, Warangal, Telangana 506371, India
[6]Department of Computer Science, Community College, King Saud University, Riyadh 11451, Saudi Arabia

Corresponding authors: Deepali Gupta (deepali.gupta@chitkara.edu.in) and Manjit Kaur (manjit@ieee.org)

This work was supported by King Saud University, Riyadh, Saudi Arabia, through the Researchers Supporting Project, under Grant RSP2025R395.

**ABSTRACT** The rapid growth of the global population and the continuous decline in cultivable land pose significant threats to food security. This challenge worsens as climate change further reduces the availability of farmland. Soilless agriculture, such as hydroponics, aeroponics, and aquaponics, offers a sustainable solution by enabling efficient crop cultivation in controlled environments. The integration of the Internet of Things (IoT) with smart precision farming improves resource efficiency, automates environmental control, and ensures stable and high-yield crop production. IoT-enabled smart farming systems utilize real-time monitoring, data-driven decision-making, and automation to optimize water and nutrient usage while minimizing human intervention. This paper aims to explore the opportunities and challenges of IoT-based soilless farming. It also highlights its role in sustainable agriculture, urban farming, and global food security. These advanced farming methods ensure greater productivity, resource conservation, and year-round cultivation. However, these methods also face challenges such as high initial investment, technological dependency, and energy consumption. Through a comprehensive study, bibliometric analysis, and comparative analysis, this research highlights current trends and research gaps. It also outlines future directions for researchers, policymakers, and industry stakeholders to drive innovation and scalability in IoT-driven soilless agriculture. By highlighting the benefits of vertical farming and Controlled Environment Agriculture (CEA)-enabled soilless techniques, this paper supports informed decision-making to address food security challenges and promote sustainable agricultural innovations.

## I. INTRODUCTION

The recent trend of industrialization and urbanization has led to increased employment opportunities and innovation. As a result, a large portion of the rural population is migrating to urban areas. Urban expansion into rural regions has caused deforestation and loss of farmland, depleting groundwater and increasing soil, air, and water pollution. It has also contributed to greenhouse gas emissions, soil fertility degradation, worsening climatic conditions, and negative impacts on public health [1], [2], [3]. According to estimates by the United Nations Population Division, the global population is expected to reach 9.7 billion by 2050 [4], posing severe challenges for the agricultural sector. These include imbalanced food demand and supply for humans and livestock, compromised agricultural quality, and food security concerns. Therefore, increasing food production on available land while minimizing water and resource usage—to ensure sustainability and reduce environmental hazards—remains a critical issue [5], [6].

The complexity of this challenge is further intensified by agriculture's unavoidable dependence on water, particularly in regions facing desertification [7], [8]. Sustainable Development Goals (SDGs) such as SDG-2 (Zero Hunger), SDG-6 (Clean Water and Sanitation), SDG-12 (Responsible Consumption and Production), SDG-13 (Climate Action), SDG-14 (Life Below Water), and SDG-15 (Life on Land) emphasize sustainable agriculture and natural resource management, underscoring the vital role of agriculture [9], [10]. In [9], the status of SDGs in EU Member States has been investigated by focusing on the latest progress in SDG-2 (Zero Hunger) and the projected attainment of its targets by 2030.

The primary motivation for transitioning from traditional cultivation methods to soilless cultivation was to make farming more feasible for urban communities while introducing an affordable, sustainable, and efficient approach to meet the growing demand for agricultural produce [11]. Food security has become a pressing issue in modern society. Vertical farming was reviewed as a sustainable cultivation method that also helps address global food security challenges [12]. Similarly, integration of smart sustainable vertical farming with Internet of Things (IoT) was also used to enhance efficiency [13]. Another major factor driving the shift to soilless cultivation is the reduction in the use of fertilizers, growth regulators, pesticides, herbicides, and other harmful chemicals. This method not only benefits national economies but also ensures adulteration-free food [14].

Rapid urbanization and industrialization have led to climate inconsistencies and depleting natural resources. Smart precision farming, a sensor-based system, addresses this issue by eliminating dependence on natural resources and climatic conditions. Both small-scale and commercial farming setups can achieve year-round, healthier agricultural yields by automating soilless precision farming with IoT [15], [16], [17]. The role of Information and Communication Technology (ICT), IoT, Machine Learning (ML), Artificial Intelligence (AI), and robotics has been explored in modern farming techniques [18]. The dedicated sensors are programmed with the permissible ranges for agronomic variables, and real-time data is collected and analyzed via the cloud. The system automatically adjusts temperature, moisture, light intensity, and humidity to maintain optimal growing conditions [19].

The key contributions of this paper are as follows. This paper explores IoT-based soilless farming as a sustainable solution to global food security challenges. It specifically addresses land scarcity and the impact of climate change on agriculture. It also analyzes the integration of IoT with smart precision farming to improve agricultural practices. It highlights IoT's role in optimizing resource efficiency, automating environmental control, and enhancing crop yields. Additionally, this paper evaluates the benefits and challenges of advanced farming techniques such as Controlled Environment Agriculture (CEA) and vertical farming. Finally, this paper conducts a comprehensive bibliometric and comparative analysis of IoT-driven soilless agriculture. It identifies research trends, gaps, and future directions in this emerging field.

## II. SOILLESS CULTIVATION

The concept of cultivating plants without soil is termed as soilless cultivation. In this method, an alternative growing medium is used for cultivation. It may be either air or water. Researchers have identified many alternative agricultural methods [20]. Soilless cultivation methods have emerged as a viable alternative to reduce dependence on soil, water, and other natural resources. It also responds positively to an eco-friendly agricultural practice. It ensures better qualitative and quantitative yield with a shorter vegetation period, thus addressing the emerging issue of food security [14], [21], [22].

However, while soil cultivation requires minimal energy and supports natural pollination, it is entirely dependent on environmental conditions and vulnerable to natural calamities [14], [20], [21], [22], [23], [24]. The advantages and disadvantages of both cultivation methods are presented in Table 1.

A key advantage of soilless cultivation methods is their efficient water use. Experiments on various plants have proved that the traditional cultivation consumes multiple times more water than soilless cultivation methods [25]. In [26], tomatoes were grown in three different setups. Water and nutrient usage were monitored and controlled in all setups. It was found that hydroponic cultivation is more water efficient and those plants transpired less water than those cultivated in soil. The results showed that hydroponic systems are more productive qualitatively. In [27], sodium hypochlorite was applied to the reused nutrient mix. It acted as a chemical disinfectant to analyze the effect on plant growth, composition of minerals in tissue, accumulation of residues in tomatoes, and tomato yield. The results showed that the application of 2.5 mg/L of chlorine caused a remarkable enhancement in the total fruit yield.

Soilless cultivation supports only herbaceous types of plants due to its constraint of supporting big woody perennial trees. Maximum types of herbaceous plants are supported in soilless cultivation methods. In [28], a method was described for extracting nutrients from food waste for hydroponics. The study evaluated the feasibility of using this fertilizer for cultivating hydroponic lettuce and cucumber. It was found that lettuce showed no change in the growth parameters when grown using this fertilizer and commercial fertilizer.

**TABLE 1.** Merits and demerits of soilless cultivation methods over conventional soil-based cultivation methods.

| Method | Merits | Demerits |
|---|---|---|
| Soilless Cultivation | • Requirement of technical expertise to assemble sensors and integrate with cloud<br>• Sustainable<br>• Water conservation<br>• Optimal use of fertilizers and pesticides<br>• High yield per unit area<br>• Reduced cultivation time<br>• Better qualitative and quantitative produce<br>• Automatic farm monitoring<br>• Year-round and reliable yield | • Expensive setup<br>• Complex irrigation and nutrient supply to plants<br>• Highly dependent on energy and backup system |
| Soil Cultivation | • Natural pollination<br>• More damage caused<br>• Minimal energy consumption | • Dependency in weather conditions and natural resources<br>• Water consuming method<br>• High transportation time and cost overhead<br>• Manual, labor-dependent farm monitoring |

However, cucumber showed significant depletion in growth using the food waste fertilizer as compared to commercial fertilizer. In [29], the zucchini crop was subjected to salt and water stresses, and the effect of these stresses on cultivation and gas exchange was evaluated. In [30], aquaponic okra cultivation was conducted to evaluate the effects of foliar application of micro and macronutrients. Its effects on the vegetative growth of okra caused a higher plant yield including leaf count, length, and perimeter.

In [31], an experiment was conducted on hydroponic eggplant cultivation using different growing media and varying ratios of nitrogen (N) and potassium (K) in a modified Hoagland solution. The results showed that plants grown in cocopeat substrate with 298 ppm N and 330 ppm K achieved the greatest height, denser branching, and the highest leaf count. In [32], hydroponic potato cultivation was introduced as an alternative farming system, where potatoes were grown in wood fiber medium using a drip irrigation system in plastic containers. This method resulted in higher tuber production compared to conventional farming. In [33], crops such as onion, sweet basil, black cabbage, and lettuce were cultivated in a soilless medium, and their toxic element content was evaluated. In [34], onion shoot count, perimeter, and length were monitored to assess yield. The peat-perlite mixture produced the highest yield, while the peat medium resulted in the lowest yield.

There are five different types of hydroponic setup: Nutrient Film Technique (NFT), Deep Water Culture (DWC), ebb and flow, wick system, and drip irrigation. In [35], cauliflower cultivation was carried out in two experiments in an NFT setup in the winter-spring and spring-summer seasons, respectively. The plants were cultivated subject to six EC values with and without NaCl. The qualitative yield was found to be better in the winter-spring season. In [36], the growth responses of cauliflowers and broccoli were compared in NFT hydroponics. The fresh weight of broccoli was comparatively higher in spring than that of cauliflower in summer. In [37], a hydrogel substrate was developed for soilless cultivation of mung beans, resulting in a 33.69% increase in total biomass.

In [38], NaCl was added to the Hogland solution with EC mS/cm. The dependent parameters, such as nutrient content, evapotranspiration, and leaf color were observed. The maximum growth responses were obtained with EC 25 mS/cm. In [39], a comparative growth response analysis was presented of lettuce in soil and NFT hydroponics. The NFT setup depicted the best root and shoot growth response. In [40], the growth responses of pak choy and spinach were analyzed in aeroponics and hydroponic systems. In [41], the growth evaluation of coriander and rockets was presented in conditions of saline stress and thermal stress. Out of the six treatments, the coriander plant was more stress-tolerant than the rocket. In [42], the growth of parsley, cress, and lettuce in aquaponics was evaluated under three different lighting sources: Light-Emitting Diode (LED) (200W), High-Pressure Sodium (HPS) lamp (200W), and fluorescent (FLO) lamp (200W). A total of 43 koi fish with an initial biomass of 3,628g were cultured, achieving 36% growth and reaching an average weight of 132.7g. After 45 days, parsley grew to 8.76g under HPS, 7.45g under LED, and 2.04g under FLO. After 54 days, lettuce reached 54.09g under HPS, 25.60g under LED, and 60.83g under FLO. After 42 days, cress grew to 1.03g under HPS, 1.15g under LED, and 1.31g under FLO.

In [43], the growth performance of mint was compared in horizontal and vertical A-type NFT setups, using three NPK fertilizer combinations: 40:65:40, 50:75:50, and 60:85:60 per hectare. The results indicated that the vertical A-type setup with an NPK ratio of 40:65:40 per hectare produced the best growth, showing superior leaf count, height, branch count, root length, and both dry and fresh weights of root and shoot. In [44], hydroponic green basil, red basil, rocket, and mint were subjected to high pH and low nutrient concentrations. The findings revealed that green basil, followed by mint, showed the highest tolerance to these conditions, while red basil and rocket exhibited negative growth responses. In [45], the study primarily focused on cultivation conditions and medicinal crops grown in various hydroponic setups. In [46], the research elaborated on artificial farming methods under CEA and their multiple advantages.

Table 2 provides a concise overview of herbs that thrive in soilless cultivation systems.

TABLE 2. Crops compatible with soilless cultivation.

| References | Category | Crop Names |
|---|---|---|
| [27-31, 47-52] | Fruit vegetables | Cucumber, Tomato, Zucchini, Okra, Eggplant |
| [32-34, 53-56] | Root vegetables | Potato, Radish, Turnip, Carrot, Onion |
| [55, 56, 57-59] | Flower vegetables | Cauliflower, Broccoli |
| [37, 60] | Pulses | Beans |
| [58] | Stem vegetables | Asparagus |
| [28, 39-42, 61-63] | Leafy vegetables | Lettuce, Spinach, Coriander, Parsley, Celery, Cabbage, Bok choy |
| [43-46, 64] | Herbs | Mint, Aloe Vera, Basil |

The soilless cultivation methods produce enhanced crop yield in a shorter vegetation period using minimum water for irrigation. Table 3 shows the comparison of water consumed during the cultivation of various crops along with their corresponding vegetation periods when cultivated in soil and soilless mediums respectively.

TABLE 3. Water consumption and vegetation period of crops in soilless cultivation methods.

| References | Crops | Water Requirement (liters/kg) | | Vegetation Period (days) | |
|---|---|---|---|---|---|
| | | Soil Medium | Soilless Medium | Soil Medium | Soilless Medium |
| [25, 40] | Spinach | 100 | 6.[illegible] | 45-60 | 40-50 |
| [25, 39] | Lettuce | 7[illegible] | 1.[illegible] | 60-90 | 30-40 |
| [25, 56, 58] | Brassica | 129 | 5 | 90-120 | 65-90 |
| [25, 48] | Tomatoes | 7[illegible] | 3.[illegible] | 70-100 | 55-65 |
| [25, 29, 50] | Zucchini | 9[illegible] | 3[illegible] | 50-60 | 45-55 |

Various factors influence the success of soilless cultivation. Selecting appropriate crops and sensors, managing installation costs, assessing environmental impact, and optimizing energy consumption are crucial for minimizing risks, preventing agricultural losses, and reducing environmental impact. The key factors are as follows:

a) Selecting compatible crops: Crop selection is the pivot matter of consideration for cultivation in any soilless cultivation method is crop selection. Short herbaceous plants are generally found to thrive in a soilless cultivation setup, due to their small size and shorter growing cycle, which helps in the cultivation of numerous plants in vertical stacks in multiple batches [10].
b) Component selection: The sensor module and the actuator module components of any soilless cultivation setup have to be selected precisely and vigilantly. All the components have to be selected to obtain optimized use and maximum efficiency. The sensors and other hardware components must be efficiently able to monitor the parameters to be monitored according to the selected plant [10], [11].
c) Expenditure for initial installation: There is an initial expenditure during the initial installation of the setup which depends upon the type of plant selected and the compatible soilless cultivation setup required to cultivate it. This is one of the major constraints in the process of soilless cultivation but the profit obtained from soilless cultivation of plants surpasses the setup expenses [12].
d) Environmental footprint: The elimination of the need for tractors for plowing, vehicles for transportation, etc. in the case of soilless cultivation methods reduces the environmental footprint to a considerable extent [13].
e) Energy consumption: The soilless cultivation setups implement a controlled environment throughout the setup, which causes a huge consumption of energy consumption. Every agronomic and environmental parameter is monitored and maintained using sensors, actuators, and other hardware devices with the help of energy. This causes a huge energy consumption that surpasses that of traditional substrate cultivation methods [12], [14].

The economic analysis of smart soilless farming in comparison to traditional soil-based cultivation methods plays a vital role in calculating profitability, sustainability, and resource use efficiency [25], [39]. This can be greatly be beneficial for decision makers before beginning with cultivation [25], [39], [40], [41], [48], [49], [50]. Table 4 represents a concise representation of the comparative economic analysis of soilless vs traditional cultivation methods.

## A. CLASSIFICATION OF SOILLESS CULTIVATION METHODS

Soilless cultivation techniques can be further classified into three categories [65], [66]. These are aeroponics (cultivation in air medium) [67], [68], [69], hydroponics (cultivation in water medium) [70], [71], [72], [73], [74], and aquaponics (hydroponic cultivation with a balanced ecosystem of aquaculture) [75], [76] (see Figure 1). A detailed analysis of various types of soilless cultivation methods aeroponics, hydroponics, and aquaponics is mentioned in [20] and [65].

### 1) AEROPONICS

Aeroponics is a soilless cultivation method where plants grow suspended in the air, with their roots exposed and nourished by a fine mist of nutrient-rich water. It offers several advantages, such as faster plant growth, improved oxygenation, reduced water consumption, and minimal pest damage. However, aeroponics also comes with challenges such as technical complexity and reliance on a continuous power supply, which can limit its widespread adoption [67], [68], [69].

TABLE 4. Economic analysis of soilless vs soil based cultivation methods.

| Factor | Soilless farming | Traditional farming |
|---|---|---|
| Initial investment | Moderate as compared to traditional farming. | Low as compared to soilless farming. |
| Water usage | 90-95% less than traditional farming. | More as compared to soilless farming. |
| Yield per square foot | 2-5 times higher than traditional farming. | Standard yield. |
| Labor costs | Negligible as compared to traditional farming. | High as compared to soilless farming. |
| Resource efficiency | High as compared to traditional farming. | Low as compared to soilless farming. |
| Yield | Year-round. | Seasonal. |
| Vegetation period | Very short as compared to traditional farming. | Prolonged as compared to soilless farming. |
| Damage in yield due to infections | Negligible as compared to traditional farming. | High as compared to soilless farming. |
| Profitability | High as compared to traditional farming. | Moderate to low as compared to soilless farming. |

*All the factors for economic analysis vary with the crop type and the type of soilless farming.

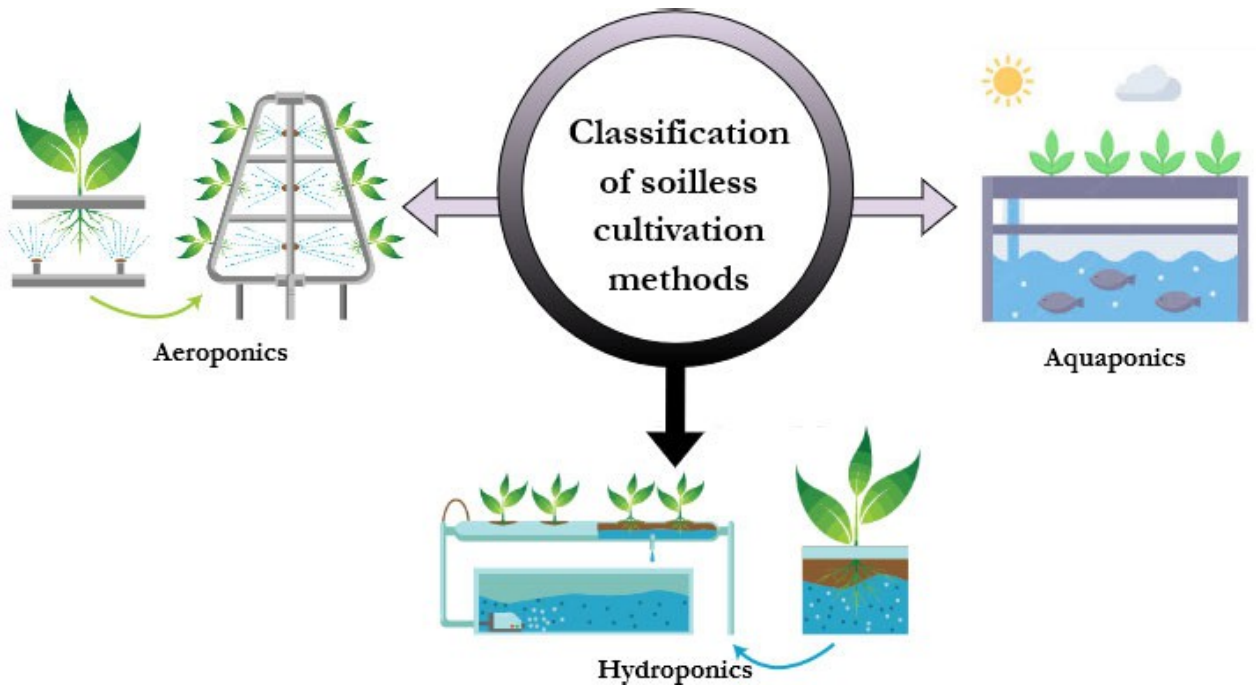

FIGURE 1. Classification of soilless cultivation methods.

Unlike traditional or hydroponic cultivation, aeroponics does not use soil or any solid medium. Instead, nutrients are delivered directly to the plant roots via mist sprays. Many aeroponic systems incorporate sensor modules to track key agronomic variables, ensuring optimal growth conditions. In [77], research on potato mini tuber production in an aeroponic setup demonstrated that the vegetation period was significantly shorter compared to soil-based farming. However, results varied depending on the plant genotype: Cleopatra and Sinora had shorter growth cycles but produced fewer mini tubers, while Kennebec and Agria had longer cycles but yielded larger and more abundant tubers. Among them, Sinora produced the heaviest mini tubers.

One of the major challenges in aeroponics is ensuring that plants receive all essential nutrients for growth and high yields. In [78], researchers evaluated two types of nutrient solutions to determine the optimal nitrate levels for potato plants. Another study [79] investigated the impact of LED light exposure on medicinal plant roots. The results revealed that different light wavelengths influence the accumulation of phytochemicals. For example, artemisinin was found to accumulate in the shoots of sweet wormwood when exposed to white, red, and blue lights, while coumaroylquinic acid was concentrated in St. John's wort leaves under similar conditions.

To enhance aeroponic farming efficiency, researchers have developed smart aeroponic systems equipped with IoT-based sensor modules. In [68], a system was described that continuously monitors and controls environmental parameters. According to [80], each sensor operates within a predefined range, and if values exceed the permissible limits, alerts are sent to users via a connected device.

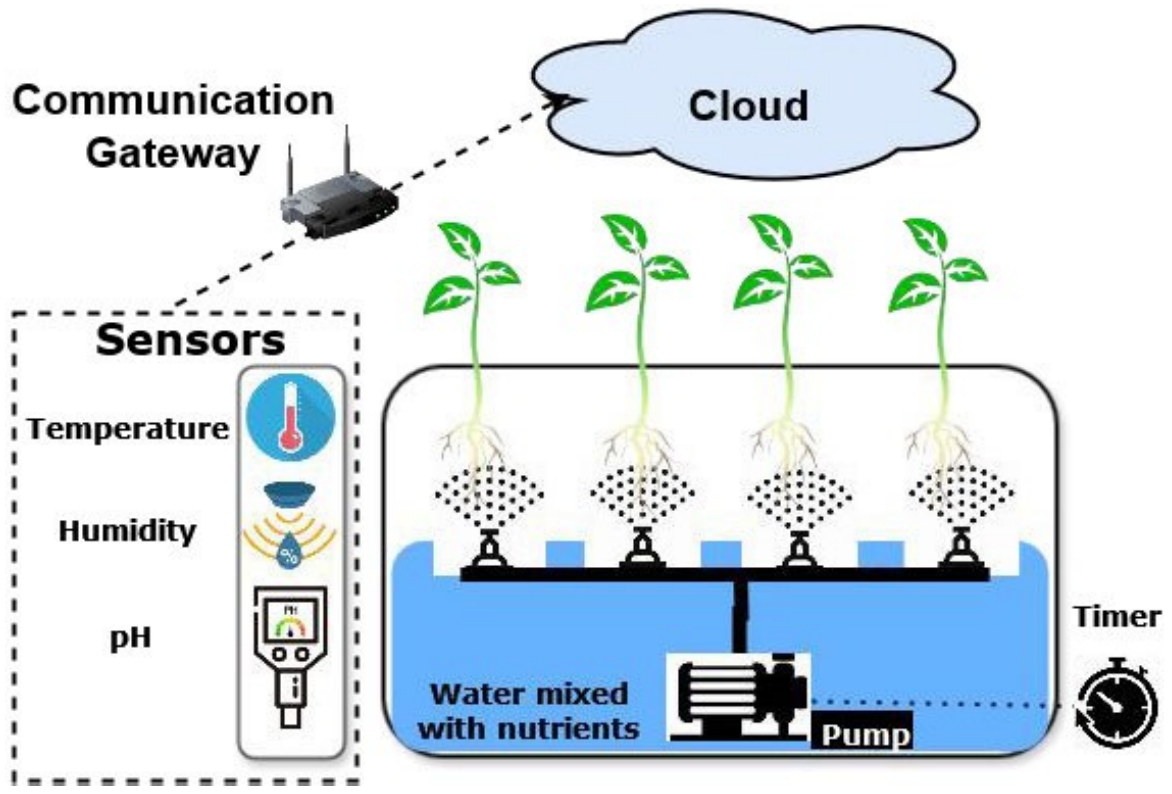

FIGURE 2. A smart aeroponic setup.

A smart aeroponic setup consists of three primary components is shown in Figure 2. The sensor module continuously monitors key environmental factors such as temperature, humidity, and nutrient levels to ensure optimal growing conditions. The cultivation setup with actuators regulates the mist spraying system and nutrient distribution, providing essential nourishment directly to the plant roots. Additionally, the cloud-based data processing system analyzes real-time sensor data and makes automatic adjustments as needed, enhancing efficiency and maintaining ideal growth parameters.

A notable smart greenhouse system integrating hydroponics and aeroponics was developed in Egypt [81]. It was tested on Batavia lettuce and demonstrated impressive results, achieving 80% savings in water and energy. Additionally, it doubled productivity per unit area and reduced the vegetation period from 75 to 45 days. The greenhouse is designed to automatically adjust environmental conditions based on the plant type and season. Pest control is managed through automated pesticide spraying to ensure healthy plant growth.

Recent research has focused on AI-driven aeroponic solutions. In [82], an IoT and AI-powered system was designed to optimize resource usage and crop production. Another study [83] proposed a four-layer IoT-based monitoring system, using Thingspeak and Firebase servers to enable real-time tracking of climate conditions and automated irrigation management via the Aeroponics Monitor mobile app.

In [84], Sugeno fuzzy logic was implemented to automate fertigation, allowing temperature and humidity levels to be analyzed for precise irrigation control. Furthermore, a ML model was developed to predict lettuce yields with an accuracy of 95.86% [85].

### 2) HYDROPONICS

Hydroponics is a soilless cultivation technique where plants grow in a nutrient-rich water solution instead of soil. This method is classified into five main types, such as NFT, drip system, Ebb and flow, wick system, and Deep-Water Culture (DWC). In most hydroponic systems, substrates are used to provide structural support for plants [70], [71]. Hydroponic farming has been shown to increase agricultural productivity and contribute to economic growth [72]. Although it requires high energy input and may have environmental impacts, its year-round cultivation, higher productivity, and shorter growth cycles make it a promising alternative to traditional soil-based farming [73], [74].

A smart hydroponic setup, as shown in Figure 3, automates and optimizes plant growth. This setup often integrates sensors that track agronomic variables such as water levels, electrical conductivity (EC), pH, and nutrient availability. These sensors communicate with cloud-based data processing systems, allowing real-time monitoring and automatic adjustments to ensure optimal growing conditions.

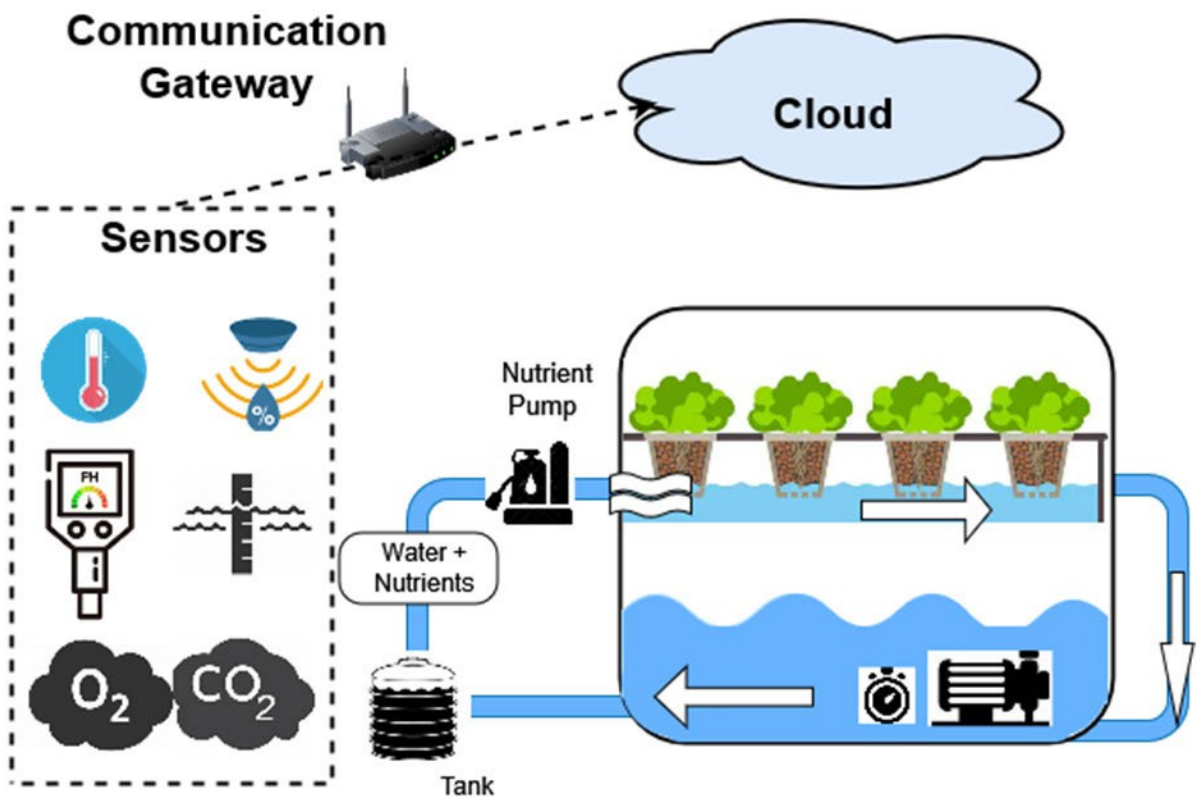

FIGURE 3. A smart hydroponic setup.

An example of an IoT-based smart hydroponic system was implemented at the National Water Research Centre in Egypt [81]. This setup, tested on Batavia lettuce, resulted in 80% reduction in water usage and 50% reduction in cultivation time.

To further improve efficiency, weather-independent hydroponic setups have been developed. In [86], an IoT-enabled hydroponic system was designed to minimize fertilizer and water consumption. This system uses relays and alarms to maintain appropriate nutrient levels, while sensors continuously measure water levels, pH, and EC. Several advancements in hydroponics leverage IoT, AI, and ML to enhance precision and automation. A spectroscopic sensor system was developed to detect nitrogen variations in hydroponic solutions [87]. A hybrid IoT-based hydroponic monitoring system was also proposed for indoor farms, integrating wireless and wired components for better efficiency [88]. Another prototype continuously monitors real-time sensor data and alerts farmers about any unfavorable conditions [89]. A smart hydroponic setup proposed in [90] was designed to control environmental factors such as humidity, temperature, pH, and moisture levels. This system allows remote farm access through IoT, sending real-time crop data and yield predictions to farmers using the Random Forest (RF) algorithm. Another study [91] tested lettuce cultivation in an NFT hydroponic setup, where smart sensors continuously monitored and optimized growth conditions, significantly improving lettuce productivity.

In [92], a large-scale hydroponic greenhouse was developed for growing various crops under controlled climatic conditions. The K-Nearest Neighbors (KNN) algorithm was used to predict the growth rate of leafy vegetables, achieving 93% accuracy. It was also found that the NFT system with coconut coir as a growing medium provided the best results. An indoor vertical hydroponic setup was designed to function independently of external environmental conditions [93]. This setup, controlled through IoT, helps meet domestic food demands within limited space. Another smart hydroponic setup [94] was successfully implemented to maximize crop yield, proving to be a more efficient method for cultivating leafy crops compared to soil-based farming.

ML-powered hydroponic system was introduced in [95] to monitor and control key environmental factors, including water level, nutrient concentration, temperature, and light. The system is trained using an ML model, while a Raspberry Pi is used to set suitable growth conditions. A web portal was also developed to allow users to monitor and manage their hydroponic farms remotely. In [96], a secure IoT-based framework for smart hydroponic farming was presented. This system features a four-layered IoT architecture, utilizing supervised data analytics to manage the fog layer efficiently for cloud-based processing. Additionally, blockchain technology was integrated to enhance security and protect data integrity, ensuring precise monitoring and control over hydroponic farming operations.

### 3) AQUAPONICS

Aquaponics is an advanced form of hydroponics that integrates hydroponic plant cultivation with aquaculture, creating a self-sustaining ecosystem [75]. This system allows for the simultaneous production of fish and plants, often referred to as the ‘‘coupling’’ of hydroponics and aquaculture. It enhances sustainability while improving both crop yield and fish farming efficiency [76]. Aquaponics provides a two-fold benefit by combining plant cultivation with fish farming. However, large-scale implementation is more effective due to high initial costs, technological dependence, and limited compatibility between certain fish and plant species. Despite these challenges, a properly managed aquaponic system ensures higher productivity and profitability by efficiently utilizing resources [97], [98].

TABLE 5. Comparison of soilless cultivation techniques.

| References | Soilless Cultivation Techniques | Merits | Demerits |
|---|---|---|---|
| [20, 66-68, 77-85] | Aeroponics | • Minimum water consumption.<br>• No substrate dependency.<br>• Easy transplantation process.<br>• No damage due to pests.<br>• Easy to observe the plants remotely.<br>• Reduced cultivation period.<br>• Increased aeration delivers more oxygen to plant roots.<br>• Lesser energy consumption as compared to the other soilless cultivation methods, due to gravitational force draining the excess liquid automatically. | • Complete dependency on power.<br>• Expensive and complicated setup.<br>• Expensive power backups and spray equipment.<br>• Requires constant pH, nutrient, etc. monitoring.<br>• Suitable for herbaceous leafy plants only.<br>• Susceptible to power outages.<br>• Requires technical expertise. |
| [66, 67, 86-97] | Hydroponics | • Independent of soil-related issues.<br>• Enhanced feasibility.<br>• Consumes multiple times less water and optimal quantity of fertilizer as compared to conventional methods.<br>• No damage due to pests and pesticides.<br>• No soil-related, labor-intensive activities like plowing, are needed.<br>• Shortens vegetation cycle.<br>• Cost-efficient soilless cultivation method. | • Vulnerable to power outages, generator backup is needed.<br>• Requires technical expertise.<br>• Susceptible to waterborne diseases. |
| [20, 98-101] | Aquaponics | • A two-way business benefit of hydroponics and aquaculture is ensured from a single unit.<br>• Ensures reduced vegetation cycle.<br>• The fish provide organic liquid fertilizer for the plants as a replacement for the chemical fertilizers.<br>• Sustainable and more efficient soilless cultivation method, as it recycles waste products of two ecosystems.<br>• Water is conserved by recycling. | • Installation and maintenance of the setup is expensive.<br>• Demands technical expertise.<br>• Limited varieties of crops and fish are compatible.<br>• Heavy dependency on electricity.<br>• Low resilience to the risk of unexpected power failure.<br>• Requires more attention and care than other VF methods. |

A sensor-equipped aquaponic system is shown in Figure 4, where various agronomic parameters are continuously monitored. The system consists of two interconnected setups: one for fish culture and the other for hydroponic plant growth. Wastewater from the aquaculture section, rich in organic waste and ammonia, is processed by nitrification bacteria to convert ammonia into nitrates. These nitrates serve as essential nutrients for the hydroponic plants. After nutrient absorption, the filtered water is recirculated back to the fish tank, maintaining a closed-loop system that minimizes water wastage.

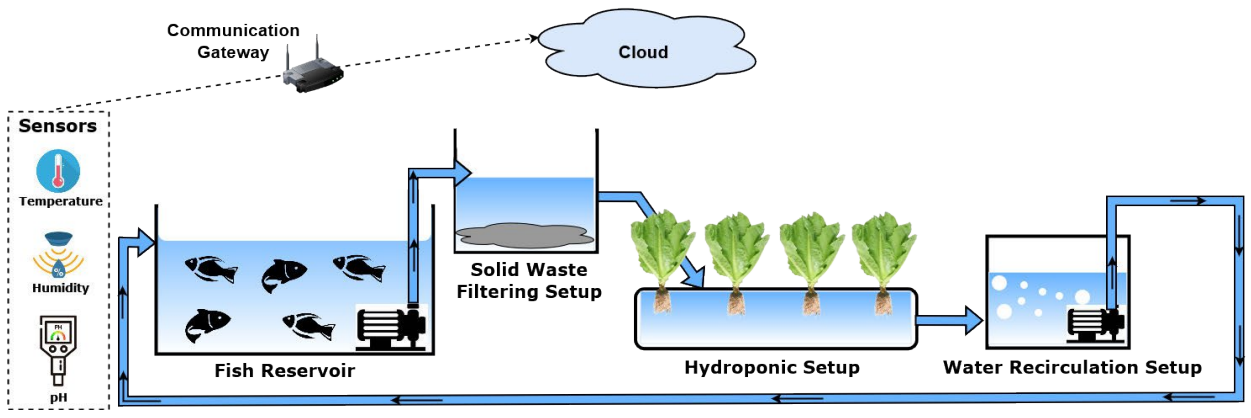

FIGURE 4. A smart aquaponic setup.

In [99], an IoT-based aquaponic system was proposed to optimize farming conditions by predicting water temperature, plant health, fish growth, and bacterial activity. Sensor data was analyzed using XGBoost and Random Forest (RF) algorithms, achieving 99.13% accuracy in predicting system parameters. Another ML-enhanced aquaponic system was developed in [100], training datasets with multiple ML models to achieve 99% accuracy in system monitoring.

Several studies have integrated IoT and AI into aquaponic farming. In [101], IoT was used to monitor critical water quality parameters, including pH, ammonia, dissolved oxygen, and temperature. The system employed a Blynk cloud server for real-time data analysis and remote access via the internet. In [102], a neuromorphic computing system utilizing Spiking Neural Networks (SNNs) was designed to estimate fish length and weight with high accuracy. This system achieved real-time processing of 84 million fish samples per second while consuming minimal resources, making it an efficient alternative to traditional monitoring methods. An IoT-integrated aquaponic system was developed in [103] to monitor water quality in real time using neural networks. The system also detected nutrient deficiencies in plants at an early stage using Convolutional Neural Networks (CNNs). Additionally, an automated fish-feeding mechanism was implemented to improve efficiency and profitability by eliminating the drawbacks of manual feeding.

**TABLE 6.** SWOT analysis of soilless cultivation methods.

| Strengths | WEAKNESSES | Opportunities | THREATS |
|---|---|---|---|
| • Increased yield in minimal cultivation area [1, 40, 61].<br>• Independent of soil or natural resources [2, 3, 23].<br>• Better qualitative and quantitative produce [54, 55, 57, 58].<br>• Less water consumption [60, 92-95].<br>• Efficient nutrient utilization [81-84].<br>• Year-round and shorter vegetation periods [24, 25].<br>• Automatic functioning [5-9]. | • Initial investment expenses [38-42, 61].<br>• Decreased pollination [64].<br>• Technological dependency [5, 36, 57-59].<br>• Complete dependency on energy [30, 31, 51, 52]. | • Compatible for almost all herbs [14, 20-22, 43].<br>• Accelerated employment opportunities for amateur farmers [75-77].<br>• Cultivation of various types of crops simultaneously [14, 23]. | • Affected by technical breakdowns and power cuts [27, 28, 109].<br>• Dependency on grid power and power backups [11-13]. |

In [104], an efficient aquaponic system for Tilapia farming was developed using multiple sensors and microcontrollers to monitor essential parameters such as water level, total dissolved solids (TDS), temperature, and pH. The system resulted in a 6% increase in fish growth, along with consistent water circulation and minimal power consumption. Various statistical methods were applied to validate the system's performance, confirming significant fish production improvements. In [105], a study evaluated aquaponic farming for basil and coriander, assessing growth rates and nutrient composition. The results showed significant improvements in plant height, root length, leaf count, and biomass production compared to traditional soil-based methods. The additional nutrients derived from fish waste enhanced plant antioxidant activity, while water-use efficiency exceeded 90%, making aquaponics highly suitable for arid regions.

In [106], an automated system for detecting ammonia toxicity was proposed, integrating pH sensors and computer vision for real-time ammonia level monitoring. The system was connected to an IoT-cloud architecture for autonomous control and remote management. A low-salinity aquaponic system was developed in [107] for shrimp and shallot farming. IoT sensors monitored water quality, salinity, nitrite, temperature, and dissolved oxygen levels. A genetic algorithm was used to train the model, while Arduino-controlled actuators optimized environmental conditions. The system was tested for a month and successfully maintained optimal shrimp and shallot growth conditions, demonstrating IoT and ML's effectiveness in commercial aquaponic farming. A smart fish-feeding system was introduced in [108], enabling remote control via mobile devices. The system monitored tank parameters and allowed farmers to adjust water temperature and automate feeding schedules through an Arduino-cloud dashboard. Additionally, a water level sensor was included to notify users of any critical changes, ensuring stable operation.

### *B. MERITS AND DEMERITS OF SOILLESS CULTIVATION TECHNIQUES*

Although soilless cultivation provides several benefits over traditional soil-based methods, it also comes with certain limitations. Table 5 presents a comparative overview of the merits and demerits of all three types of soilless cultivation methods: aeroponics, hydroponics, and aquaponics.

The strengths, weaknesses, opportunities, and threats (SWOT) of different soilless cultivation techniques are summarized in Table 6. While these systems ensure high yields, efficient water use, and year-round cultivation, they also require high initial investment and technological dependency.

A detailed comparison of aeroponics, hydroponics, and aquaponics in terms of yield, water and nutrient use, energy consumption, space efficiency, scalability, and environmental impact is presented in Table 7. Refer to [77], [78], [79], [80], [87], [90], [91], [92], [93], [94], [95], [96], [100], [101], [102], [103], [104], [105], [106], [107], and [108] for more details.

Aeroponics, hydroponics, and aquaponics optimize yield, water, and nutrient efficiency, but each method has unique advantages. Hydroponics is versatile and scalable, aquaponics is less efficient but more sustainable, while aeroponics is the most space-efficient. However, aeroponics and aquaponics require higher energy consumption due to their complex setups. Advancements in smart farming technologies have significantly improved monitoring, automation, and resource efficiency in soilless cultivation. The key technologies enhancing these methods are listed in Table 8.

### *C. COMPONENTS USED IN SOILLESS TECHNIQUES*

In traditional soil-based cultivation methods, soil provides structural support for plants while supplying water and essential nutrients required for growth [109], [110], [111], [112]. The soilless cultivation methods have to depend upon external substrates to hold the plants in place and nutrient solution separately to ensure proper yield [23], [113], [114], [115]. These are the basic components required to carry out any soilless cultivation.

**TABLE 7.** Comparative analysis of soilless cultivation methods.

| Aspect | Aeroponics | Hydroponics | Aquaponics |
|---|---|---|---|
| Yield | • High yield per unit area due to maximum oxygenation of roots. | • High yields for herbs and some fruits.<br>• Reduced crop cycles due to direct nutrient supply. | • Yields are low as they are influenced by fish growth and the selected plant. |
| Water use efficiency | • Uses negligible amount of water compared to soil-based farming. | • Uses 90-95% less water than soil-based farming.<br>• Water can be recirculated. | • Reduced water use efficiency compared to hydroponics and aeroponics due to fish tank evaporation and cleaning requirements. |
| Nutrient use efficiency | • Extremely efficient nutrient use due to precise delivery.<br>• Minimal nutrient loss. | • Nutrient concentrations are controlled and reused. | • Nutrients come from fish waste.<br>• Requires careful nutrient balancing to avoid damage to fish or plants. |
| Energy use | • High energy requirements for misting systems and maintaining optimal growth conditions. | • Moderate energy use for sensors and actuators in controlled environments. | • High energy use for maintaining aquaculture and hydroponics.<br>• High energy requirements. |
| Space efficiency | • Most space-efficient system.<br>• Can be vertically scaled. | • Space-efficient but requires more area than aeroponics.<br>• Can be vertically scaled. | • Less space-efficient due to the need for fish tanks and plant-growing areas. |
| Scalability | • Scalable, but requires high initial investment and expertise. | • Highly scalable for various farm sizes. | • Challenging to scale due to its complexity. |
| Environmental footprint | • Low environmental footprint | • Low environmental footprint | • Low environmental impact if managed well. |

**TABLE 8.** Most promising smart technologies in soilless cultivation methods.

| References | Technology | Description | Key impact |
|---|---|---|---|
| [82, 86, 90-91, 94-96] | IoT sensors | Real-time monitoring of agronomic variables. | Optimizing plant growth and resource utilization through real-time data-driven decisions. |
| [82, 95, 100] | AI and ML | Analyzing data to predict issues and forecast yields. | Automatic decision-making reduces human intervention and improves yield productivity. |
| [40] | Automated CEA and remote monitoring | Automating irrigation, lighting, and climate control. Enabling alerts, and data visualization via applications, IoT platforms, and web interfaces. | Ensuring consistency, reducing labor costs, and minimizing yield losses, and minimizing human intervention and real-time farm monitoring. |
| [12, 15, 45, 94] | Vertically stacked arrangements | Vertically stacked layers for optimizing cultivation area. | Maximizes cultivation in limited area, especially in urban farming. |

### 1) VARIOUS CULTIVATION SUBSTRATES FOR SOILLESS CULTIVATION

Cultivation substrates are solid materials that play a vital role in providing multifold enhanced growth conditions compared to soil cultivation [116]. These substrates are used in the cultivation of vegetables and ornamental plants. According to [111] and [112], various organic and inorganic mixtures, such as wetting agents, liming products, and fertilizers, are added to substrates in horticultural cultivation setups. However, for commercial-scale vegetable production, an appropriate substrate is selected based on the specific needs of the plant. Substrates may be broadly classified into two types such as organic and inorganic substrates.

#### *a: ORGANIC SUBSTRATES*

These substrates originate from organic sources, though they may be synthetic, e.g., wool-based substrates, peat, etc. [113]. Some of the most common organic substrates are composts [117], [118], [119], peat [120], [121], and bark [122], [123], [124], [125]. Nurseries, ornamental plant production, and the horticulture industry predominantly use peat as a substrate [112]. It has the added advantages of being inexpensive and possessing excellent properties. Additionally, it has low nutrient content, pH, and water-holding capacity. It is also lightweight and highly porous for air [113], [114], [115].

#### *b: INORGANIC SUBSTRATES*

Those These substrates are created through the industrial processing of natural resources before use. Rockwool is one of the most commonly used inorganic substrates due to its lightweight nature and convenience for fruit and vegetable cultivation in controlled structures [126], [127], [128]. Another widely used inorganic substrate is perlite, which is both inexpensive and effective [129], [130], [131]. It has emerged as an alternative to gravel and sand, which were extensively used in older installations but resulted in poor production due to their low porosity. Apart from these, various other inorganic materials have also been used in soilless cultivation setups [132], [133], [134], [135], [136]. The choice of substrate depends on factors such as its physical, chemical, and biological properties, as well as its compatibility with the crop being grown [137]. Additionally, factors like cost, sustainability, and eco-friendliness are considered during substrate selection. Table 9 summarizes the merits and demerits of organic and inorganic substrates used in soilless cultivation.

TABLE 9. Merits and demerits of various organic and inorganic substrates used in soilless cultivation.

| References | Substrates | Original composition | Merits | Demerits |
| --- | --- | --- | --- | --- |
| [110, 112] | Sand | Natural particles of 0.04-2.2 mm. | Cheap, good drainage ability, 66% porous, 0.13 g/cm$^3$ density. | Devoid of nutrients, no water or air retention ability, relatively more weight (1400-1600 kg/m$^3$) |
| [129-131] | Perlite | Volcanic mineral material is heated and processed. | Relatively lightweight (90-130 kg/m$^3$), pH neutral (6.5-7.5), sterile, 90% porous, 0.11 g/cm3 dense, enhanced aeration. | Nutrient-deprived, expensive, and energy-consuming substrate |
| [127-130] | Coconut coir | Natural, coconut by-product | Lightweight (65-110 kg/m$^3$), good air and water retention ability, 5-6.8 pH, 92 % porous, 0.16 g/cm3 density, good aeration. | High salt content, energy-consuming creation process |
| [136, 138] | Hydrochar and biochar | Derived by pyrolysis or hydrolysis of biomass | The energy-independent production process, biologically stable, low EC of hydrochar. | Expensive production process, high pH values of biochar, dusty texture |
| [127, 128] | Rockwool | Derived from silicates melted at high temperatures | Comparatively lightweight (80-90 kg/m$^3$), easy to use, inert in nature, 94% porous, 0.06 g/cm$^3$ density, good aeration. | Problematic disposal, energy-dependent manufacture |
| [24, 115] | Pumice | Derived from volcanic matter | Lightweight (450-670 kg/m$^3$), porous, inexpensive, eco-friendly | High pH, expensive transportation |
| [57, 117-121] | Compost | Naturally manufactured from plant compost | Potassium and micronutrient-rich, disease suppressant, good moisture-retaining ability | Heavyweight (600-950 kg/m$^3$), excessive salt content, delayed compost time, water logging issue |
| [126, 138] | Peat | Naturally processed residues of plants | Physically stable, air and water retention ability, lightweight (60-200 kg/m$^3$), low nutrient content and low pH | $CO_2$ release and environmental footprint during preparation, acidic nature, limited resources for manufacturing, |
| [112] | Vermiculite | Sieved and heated to high temperatures | Lightweight (80-120 kg/m$^3$), good water and nutrient holding capacity, enhanced aeration | The expensive and energy-consuming process of production shrinks in excess moisture |

### 2) NUTRIENT SOLUTIONS USED IN SOILLESS CULTIVATION

Plants cultivated in soilless mediums require nutrients for healthy and optimal growth. These inorganic and organic nutrient solutions are prepared from water-soluble salts. The NPK solution primarily comprises nitrogen (N), phosphorus (P), potassium (K), and micronutrients such as boron (B), zinc (Zn), copper (Cu), iron (Fe), calcium (Ca), and magnesium (Mg) [27]. In [133], nitrogen, potassium, and phosphorus were used as a nutrient mix for plants grown in different organic and inorganic mediums, and their dry and fresh weights were monitored. In [39], a combination of NPK was used to support root growth, along with a stock solution of Fe, Ca, and Mg for the hydroponic cultivation of lettuce. In [139], substrates for ornamental plants were identified, and cultivation was carried out to monitor their growth responses. In [140], barley was cultivated in soilless culture using DWC hydroponics, which utilizes various nutritional salts for optimal growth. It was found that hydroponic plants grow much healthier than those cultivated using conventional soil-based methods, which require significant amounts of fertilizers and pesticides. Air and water supply macronutrients such as oxygen, carbon, and hydrogen. The hydroponic method of soilless cultivation allows for precise monitoring of plant nutrient requirements and their optimal use to produce nutritious plants. The varying application of nutrients influences plant growth based on their characteristics and quantity of use. Important characteristics of nutrient solutions include pH, total dissolved salts (TDS), electrical conductivity (EC), and temperature [141]. The pH range and EC vary according to the plant's requirements.

ML algorithms and statistical methods have been developed in [142] to monitor nutrient content based on plant needs in an aquaponic cultivation setup. A review of smart agricultural practices describing the macronutrients required in soilless cultivation systems is presented in [143]. Table 10 provides a summarized representation of the names, sources, and deficiencies of essential macronutrients.

### 3) ENVIRONMENTAL IMPACTS OF THE MATERIALS USED IN CONSTRUCTING THE SOILLESS SYSTEMS

The various materials used in constructing soilless cultivation systems have certain environmental impacts. Food grade Polyvinyl Chloride (PVC) and Polypropylene (PP) are generally used in pipes, grow tubes, and reservoirs due to their durable and lightweight nature. They have certain adverse environmental impacts that can also be mitigated using certain strategies. Different types and subtypes of soilless cultivation systems have been analyzed, highlighting their advantages and disadvantages. Similarly, the substrates used in these systems also impact the environment.

**TABLE 10.** Macronutrient sources and their deficiencies.

| References | Macro-nutrients | Macro-nutrient source | Deficiency and its ill effects |
|---|---|---|---|
| [39, 143] | Ca | Ca $(NO_3)_2$-Calcium Nitrate, $CaCl_2$-Calcium Chloride | Calcium deficiency results in BER, which is adversely dependent on potassium concentration |
| [143, 144] | Mg | $MgSO_4$-Epsom salt | Pulpy fruit, curled and scorched leaf edges, leaf vein yellowing |
| [125, 144, 145] | S | $K_2SO_4$-Potassium sulphate, $MgSO_4$-Magnesium sulphate | Pae greening in veins and chlorosis |
| [146] | N | Ca $(NO_3)_2$-Calcium nitrate, $KNO_3$-Potassium Nitrate, $NH_4NO_3$-ammonium nitrate | Increased nitrogen results in cracked stems and de-shaped leaves. Soft rot, blossom end rot, enhanced foliar growth, and depressed bulbs. |
| [145, 147] | P | $KH_2PO_4$-Monopotassium phosphate, $H_3PO_4$-Phosphoric acid, rock phosphate, fertilizers, bone meal, and manure | Reduced stem and root growth, reduced nutrient uptake, low fruit yield, and discolored leaves |
| [144] | K | $KNO_3$-Potassium nitrate, $KH_2PO_4$-Monopotassium phosphate, $K_2SO_4$-Potassium sulphate, KCl-potassium chloride | Increased dry matter in fruits, and increased amount of lycopene in tomatoes. |

These environmental effects can be significantly reduced through recycling and the integration of sustainable practices. Implementing such measures provides a more environmentally friendly and less hazardous alternative to traditional agriculture, as summarized in Table 11.

#### 4) KEY MEASUREMENTS IN SOILLESS CULTIVATION SYSTEMS

Efficient monitoring and control are essential for optimizing soilless farming systems. Several key parameters must be measured and regulated to maintain optimal plant growth and resource utilization. These parameters are discussed as follows.

##### *a: MEASUREMENT OF EC IN THE NUTRIENT SOLUTION FOR SOILLESS CULTIVATION*

The nutrient content in the nutrient solution must be measured accurately to detect any necessary adjustments for optimal plant growth [129], [141]. When the nutrient solution is reused in the system, its nutrient concentration decreases over time. Therefore, frequent monitoring is essential to ensure that nutrients are replenished as needed [115], [116], [117].

EC level in the nutrient solution can be computed as [137]:

$$E_x = fE_{ds} + (1 - f)\,E_i \tag{1}$$

Here, $E_x$ shows target EC value needed in the nutrient solution. $f$ is fraction of the recycled drainage solution. $E_{ds}$ shows EC of the recycled drainage solution. $E_i$ indicates EC of the irrigation water.

##### *b: IRRIGATION DURATION CONTROL IN SOILLESS CULTIVATION*

In soilless cultivation, the amount of nutrient solution available in the root zone is limited. Therefore, frequent nutrient supply is essential to ensure maximum plant productivity. This is achieved using actuators within the system [140]. The mathematical representation of irrigation time control can be defined as [137]:

$$T_i = \frac{60 v_f x_P F}{(1 - f)\,y_r} \tag{2}$$

Here, $T_i$ is a time needed for each event of irrigation. $v_f$ shows a fraction of the entire volume of irrigation water. $x_p$ indicates volume of growing medium for each plant. $F$ indicates a fractional quantity of the water, when consumed by the plants, triggers the next event of irrigation. $f$ indicates fraction of the recycled drainage solution. $y_r$ is output flow rate of drippers.

##### *c: MEASUREMENT OF WATER USE EFFICIENCY IN SOILLESS CULTIVATION*

The agriculture sector largely depends on water, which is one of the continuously depleted natural resources. An effective way to conserve water in soilless cultivation systems is to optimize the irrigation process to improve water use efficiency (WUE). The biomass of crops produced per unit of water consumed during the crops' cultivation is termed as WUE. This requires proper identification of the irrigation needs of plants for their maximum yield. A low-cost, sensor-based irrigation technology can be implemented by efficiently calculating the WUE and resource efficiency, to obtain maximum yield causing minimum environmental hazard. The calculation of WUE can be done using equation 3 as suggested by [148] and [149].

$$WUE = \frac{Bt}{Wt} \tag{3}$$

Here, $WUE$ = water use efficiency. $Bt$ shows a total biomass at harvest. $Wt$ indicates a total amount of water in each container of the soilless setup.

##### *d: FRESH YIELD CALCULATION IN SOILLESS CULTIVATION*

One of the key aspects to consider when enhancing WUE in soilless cultivation is maintaining the fresh yield of the crop. In [148], the calculation of fresh yield for agricultural produce

TABLE 11. Environmental impacts and mitigation strategies of materials used in soilless cultivation.

| Material | | Use in soilless cultivation | Environmental impact | Mitigation strategies |
|---|---|---|---|---|
| Plastics | PVC, PP [94-96] | • Grow tubes, hydroponic trays, reservoirs, and pipes. | • Non-biodegradable.<br>• May release microplastics. | • Use bioplastics or durable recycled plastics.<br>• Implement recycling. |
| Substrates | Organic [129-130] | • Act as an anchor for roots. | • Involve water-intensive processing. | • Use organic and renewable options like coconut coir or hemp fiber. |
| | Inorganic [127, 128] | • Act as an anchor for roots. | • Use high energy for production while also being nonbiodegradable. | |

is suggested as follows.

$$FY = \frac{Bt}{a} \quad (4)$$

Here, $FY$ is a fresh yield or the total crop weight immediately after harvest. $Bt$ is a total biomass at harvest. $a$ is cultivation area of crops in $m^2$.

#### e: CALCULATION OF THE ELECTRICAL ENERGY CONSUMED TO CULTIVATE IN A SOILLESS CULTIVATION SETUP

Any sensor-enabled automatic setup relies heavily on electrical energy. The environmental conditions required for plant growth in soilless cultivation are continuously monitored and maintained using sensors, actuators, and microcontrollers, all of which depend entirely on electricity for their operation. While improving yield and WUE in soilless cultivation, minimizing energy consumption is crucial to ensure optimal system performance. In [149], the total electrical energy consumed (in Watt-hours) is calculated using the following equation.

$$E_t = T * P \quad (5)$$

Here, $Et$ is a total electrical energy consumed (Watt/hr). $T$is a target time (hr). $P$is a total power (Watt).

The total power consumed can be calculated by summing up all the phases.

$$P = \sum P_{ph} \quad (6)$$

Here, $P$ is a total power consumed. $P_{ph}$ is a power consumed in each phase.

Power consumption in each phase can be calculated by multiplying the voltage required in each phase, the current in each phase, and the constant power factor. The formula for calculating power consumption in each phase is given as follows.

$$P_{ph} = V_{ph} * I_{ph} * PF \quad (7)$$

Here, $P_{ph}$ is a power of each phase ($W$). $V_{ph}$ is a voltage of each phase ($V$).$I_{ph}$ indicates a current of each phase ($I$). $PF$ is a constant power factor (0.95).

## III. SMART PRECISION FARMING

Emerging modern technologies such as data analytics, ML, and IoT have greatly improved soilless cultivation methods. These advancements make farming smarter and self-sustaining. By integrating these technologies, pesticide-free production is maximized while both the quality and quantity of agricultural yield are enhanced. This innovative approach is known as smart precision farming. In [150], an overview of state-of-the-art innovations, challenges, and future prospects in precision farming is presented. The study discusses the role of aerial vehicles, ML, and IoT in modern agriculture. It also highlights key challenges faced in implementing these technologies. Additionally, in [17], the significance of AI and IoT in precision farming is explored. The study demonstrates how these technologies help increase profitability in the agricultural sector. Furthermore, the advantages of real-time sensors and smart devices in optimizing farm operations are elaborated upon.

Smart precision farming enables the precise application of fertilizers and nutrients based on crop requirements. This approach ensures high nutrient and resource efficiency. Unlike traditional bulk applications, nutrients are supplied frequently and in controlled amounts, optimizing plant growth. The applications of supervised and unsupervised ML algorithms for predicting agricultural yield are also discussed in [151].

In [18], the capabilities and drawbacks of ICT in conventional farming have been highlighted, including the use of IoT, robotics, AI, ML, sensors, and drones for crop monitoring and yield prediction. In [152], brinjal was cultivated in a drip irrigation hydroponic system under CEA using sensors and microcontrollers. The system recorded environmental parameters, analyzed the data in the cloud, and implemented an automated irrigation system using an irrigation scheduling program. Results showed that IoT-enabled hydroponic plants performed better than those grown in an uncontrolled environment. In [153], it was noted that nutrient use efficiency is lower in conventional farming due to bulk nutrient application.

Another advantage of smart precision farming is its independence from manual labor, as it does not rely on soil or land. Instead, actuators automate farming tasks using Fuzzy logic, ML algorithms, and Deep Neural Networks. In [154], an approach was proposed to assist amateur farmers in achieving high productivity at minimal cost by predicting yield and cultivation expenses using ML and DL techniques.

The transition from conventional soil-based cultivation to smart precision farming has been driven by these advancements. Figure 5 illustrates this evolution, showing the shift from traditional substrate-based cultivation to modern precision farming techniques.

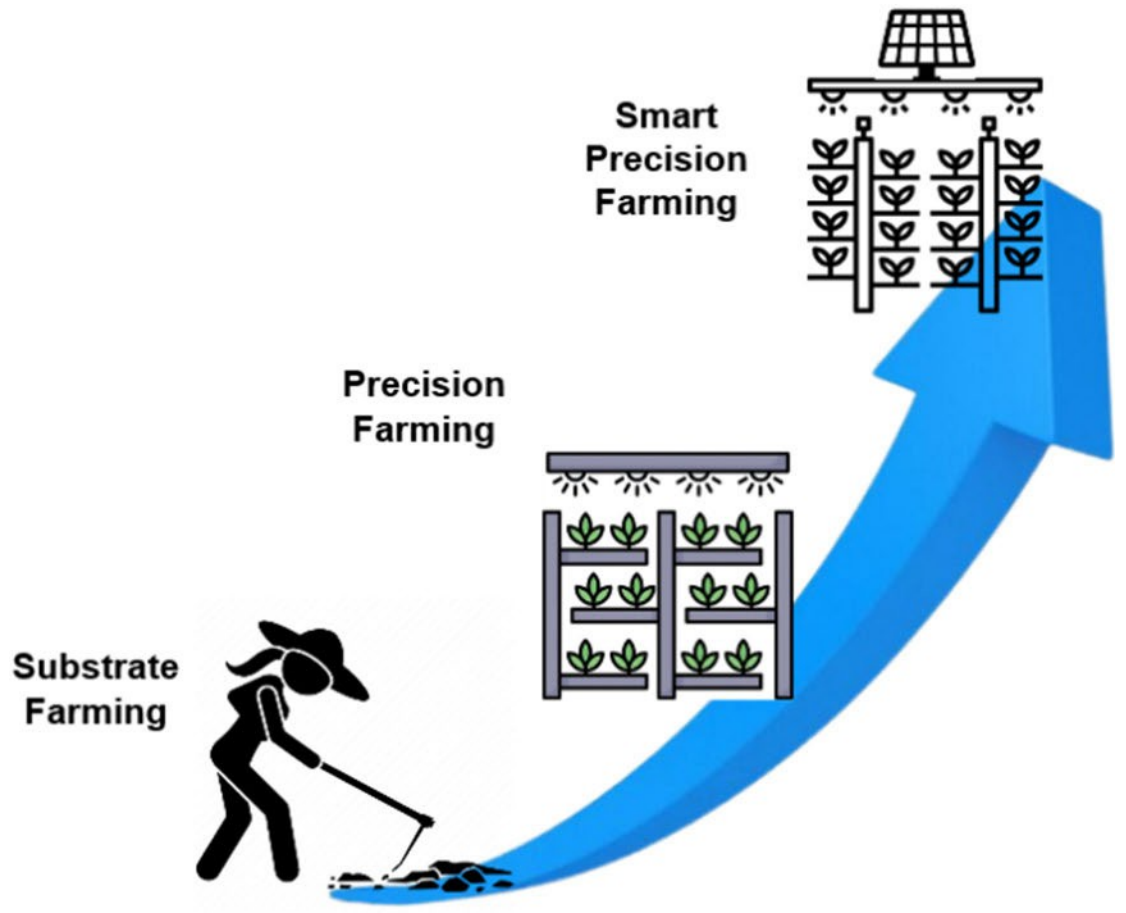

FIGURE 5. Evolution of smart precision farming.

## A. EVOLUTION AND APPLICATIONS OF SMART PRECISION FARMING

The transition from conventional farming to smart precision farming has been largely driven by advancements in automation, AI, and data-driven agriculture. The role of IoT, intelligent sensors, sensor networks, and communication protocols in precision farming is detailed in [155]. Additionally, the use of UAVs for applications such as irrigation, fertilization, pesticide spraying, weed management, plant growth monitoring, crop disease detection, and field-level phenotyping is explored.

A key aspect of smart precision farming is real-time monitoring and data-driven decision-making. Since smart farming systems rely on continuous monitoring, the precise administration of fertilizers, pesticides, nutrients, and water is achieved [156]. A review of 67 articles in [157] discusses the benefits, challenges, and technological advancements of precision farming. One of the major advantages of smart farming is its ability to significantly reduce crop damage caused by diseases. Automated monitoring and control mechanisms enhance crop productivity [158]. Additionally, sensors are deployed to monitor and regulate irrigation, ensuring water is supplied only when needed, preventing wastage and plant stress.

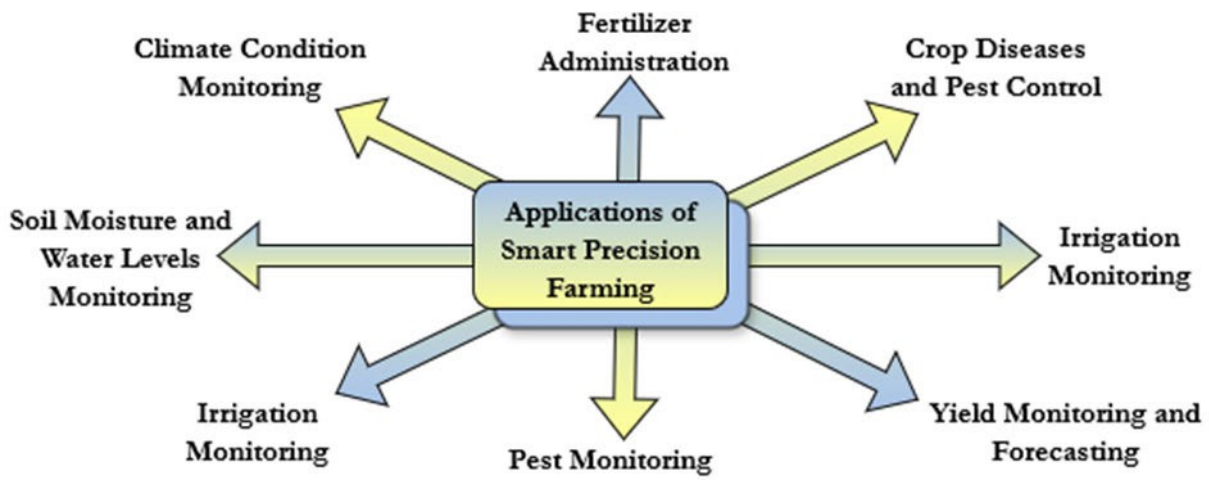

FIGURE 6. Applications of smart precision farming.

Furthermore, AI, ML, and DL have proven to be highly effective when integrated with smart precision farming. These technologies facilitate efficient monitoring and yield prediction based on real-time environmental conditions [17], [18], [148], [154]. Some of the major applications of smart precision farming are depicted in Figure 6. It showcases various advancements such as continuous soil moisture monitoring, climate control, and automated irrigation.

## B. SENSORS USED IN SMART PRECISION FARMING SETUP

In IoT-enabled smart precision farming, sensors play a crucial role in real-time data collection and automated decision-making. These sensors continuously monitor environmental and agronomic parameters, ensuring optimal growing conditions and efficient resource utilization.

A pH sensor measures the acidity or alkalinity of the nutrient solution or soil, which directly impacts nutrient uptake and plant growth. Maintaining a pH range of 5.5–6.5 is essential for optimal growth in hydroponic and soil-based setups [159], [160], [161]. If pH levels deviate from this range, automated systems can adjust them by adding pH-up or pH-down solutions.

A water level sensor ensures proper irrigation by preventing waterlogging or dehydration of plant roots. In hydroponic systems, these sensors maintain a stable nutrient solution level, while in soil-based farming, they help regulate irrigation cycles [162], [163], [164].

Temperature and humidity significantly impact plant health and productivity. DHT11 sensors monitor these factors in real time, preventing heat stress or fungal infections [165], [166], [167]. If temperature or humidity exceeds optimal levels, automated climate control systems can adjust environmental conditions.

In hydroponic setups, EC sensors measure total dissolved solids (TDS) in the nutrient solution. EC levels indicate the concentration of dissolved fertilizers and salts, which affect water absorption and plant health. Continuous monitoring of EC ensures optimal nutrient balance, preventing nutrient toxicity or deficiencies [168], [169], [170].

Hyperspectral sensors detect plant diseases, nutrient deficiencies, and physiological stress before visible symptoms appear. These sensors analyze minute changes in light reflection from plants, helping in early disease detection and precision nutrient management [171], [172], [173].

The different sensors, along with their specifications and applications, are summarized in Table 12.

# IV. CONTROLLED ENVIRONMENT AGRICULTURE

CEA is a farming approach that ensures optimal growing conditions for crops by externally providing essential nutrients and environmental factors. This method minimizes dependency on natural resources and climatic conditions, making it an advanced and efficient alternative to traditional farming. CEA is also referred to as protected farming [174], where agronomic variables are analyzed and modified to meet specific plant requirements. This results in faster growth, higher yields, and maximum profitability while reducing land and water usage. Sensors play a crucial role in monitoring and controlling environmental factors in protected structures such as low polytunnels, insect-free net houses, ventilated high-roof tunnels, playhouses, and greenhouses [175], [176], [177], [178].

TABLE 12. Sensors used in smart precision farming.

| References | Sensor Types | Specifications | Applications |
|---|---|---|---|
| [159-161] | pH sensor | Voltage: 5V<br>Current: 5-10 mA<br>Concentration range: pH 0-14<br>Response time: ≤5sec<br>Power consumption: ≤0.5W<br>Operating temp: 0-60 °C | • Measurement of hydrogen ions in the nutrient solution.<br>• The optimal pH range is 5.5 to 6.5. |
| [162-164] | Water level sensor | Voltage: DC 3-5V<br>Current: <20mA<br>Detection area: 40mm X 16mm<br>Operating temp: 10-30 °C | • Measurement of water level in the nutrient solution in an aquaponic or hydroponic system. |
| [165-167] | Temperature and humidity sensors (DHT11) | Response time: <5s<br>Temp range: 0-50 °C<br>Accuracy temp: ±2<br>Humidity range measured: 20-90%<br>Accuracy hum: ±5 | • Measurement of temperature and humidity in an enclosed setup. |
| [168-170] | Electrical conductivity | Voltage: 3.3V/5V<br>EC range: 0-2000 µs/cm<br>Response time: <10s<br>Operating temp: 5-80 °C<br>Accuracy: 1% | • Measurement of the TDS in the solution used in a hydroponic setup.<br>• EC influences a plant's ability to absorb water. |
| [171-173] | Hyperspectral sensors | Focal length: 14mm<br>Spatial resolution: 0.013 cm<br>Frame per second: 46 fps | • Helps to detect very small changes in the plant physiology. |

CEA has gained popularity in soilless cultivation setups as it allows for precise control of agronomic parameters in smart precision farming systems [174]. This method significantly reduces water consumption, optimizes nutrient levels, and prevents plant malnutrition [175]. The absence of soil eliminates risks associated with pests and rodents [176], further improving crop health and yield. CEA can be incorporated into aeroponics, hydroponics, and aquaponics. Every soilless smart precision farming system includes sensors, actuators, cloud connectivity, end devices, and a communication gateway to ensure accurate monitoring and automated responses [177].

In [178], lettuce was cultivated under CEA, where sensors were deployed to collect real-time agronomic data. These sensors were pre-set within permissible parameter ranges, and the recorded data was transmitted to the cloud through a communication gateway. Any deviation from the acceptable range triggered an alert to the user via end devices, buzzers, or LED indicators. Simultaneously, actuators adjusted conditions to restore optimal levels. Additional hardware components such as LEDs and buzzers can be integrated into the system to notify users about potential issues, ensuring an uninterrupted and well-managed cultivation process.

CEA plays a crucial role in addressing food security concerns. In [179], food adulteration and excessive fertilizer and pesticide use are noted to negatively impact food quality. Implementing CEA in precision farming ensures the delivery of precise agricultural inputs, leading to high-quality, pesticide-free produce throughout the year. Another major advantage of CEA is sustainability. In [180], sustainable plant production through CEA is highlighted as an approach that optimizes resource utilization while minimizing environmental impact, making it an eco-friendly farming method.

Smart precision farming methods integrated with CEA promote water conservation, reduce fertilizer consumption, and minimize environmental impact [181]. These methods shorten the vegetation period while enhancing both qualitative and quantitative crop production compared to soil-based farming. Figure 7 illustrates the classification of various CEA methods, including soilless cultivation techniques with and without growing media.

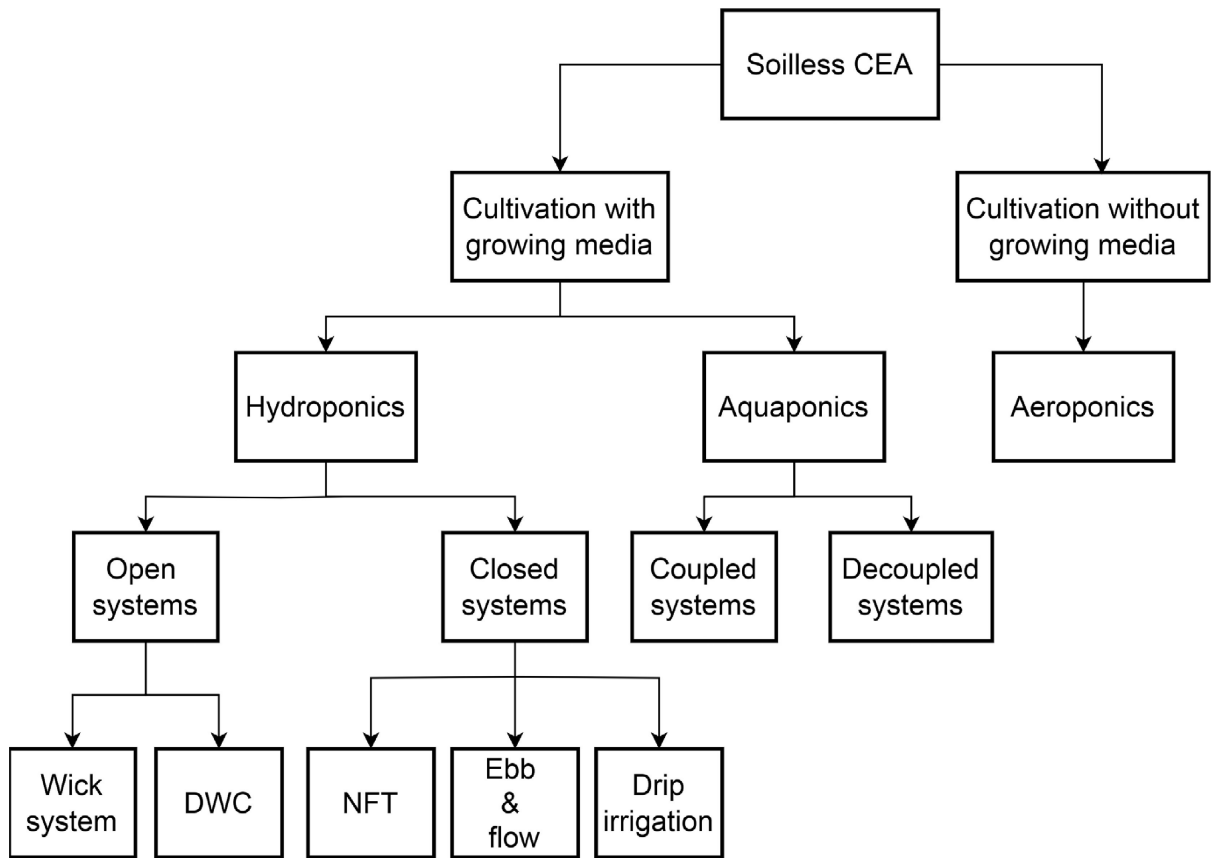

FIGURE 7. Classification of CEA.

**TABLE 13.** Advantages of CEA in smart precision VF systems.

| References | Parameters | | Benefits | | |
|---|---|---|---|---|---|
| | | | Environmental | Social | Economic |
| [58] | Reliable Produce | | Reduction of environmental hazards. | Improved health benefits. | Enhanced economic returns. |
| [58, 125] | Reduced Overheads | Transportation | Lesser pollution. | Faster delivery. | Reduced transportation cost. |
| | | Labor | - | Enhanced employability. | Employability leads to economic benefits |
| | | Water | Water conservation. | - | - |
| [126-128] | Reduced Travel Duration of Food Produce | | Better environmental conservation chances. | Improved food quality. | Reduced cost of transportation. |
| [129] | Reduced use of Chemicals (Pesticides, Fertilizers, etc.) | | Conservation of soil quality. | Improved health benefits. | Enhanced quantitative yield improves the economy. |
| [175-177] | Increased Cultivatable Area | | Enhanced production causes reduced pollution. | Ease of farming in more quality. | Improves economic benefits. |
| [177-181, 195] | Maximized Yield | | Helps in reducing pollution. | Qualitative and quantitative yield. | Improved economy. |
| [180-185, 195] | Faster, Year-Round Produce | | Sustains environmental conditions. | Helpful for habitats. | Promotes economic growth. |
| [66, 72, 73] | Wide Range of Compatible Crops | | Sustainability is ensured. | Ease of social life. | The lesser cost incurred. |
| [187, 188] | Smart and Automatic Concept | | - | Helps urban farmers with less agricultural knowledge. | Enhances yield and thus the economy. |
| [36, 59] | Recycling Organic Waste | | Reduced wastage. | Cleanliness is ensured. | Reduces the cost of organic manure. |
| [126-128] | Minimized Agricultural Losses Due to natural calamities | | Sustains environment. | Helpful for farmers and communities. | Minimized economic losses. |
| [189] | Use of renewable energy | | Eco friendly. | Sustains urban life. | Less grid power is needed. |
| [189-191] | Ease of implementation for amateur farmers | | Environment friendly. | Encourages urban farmers. | Less training is needed for traditional farming. |

CEA can be implemented in both soil-based and soilless farming systems [180]. In soilless farming, cultivation is divided into two types: without a growing medium (aeroponics) [181] and with a growing medium (hydroponics and aquaponics) [71], [73], [172], [182], [183], [184]. Hydroponic systems are further categorized into open and closed systems [185], [186]. Open hydroponic systems do not recycle nutrient-mixed water, as seen in Deep Water Culture (DWC) and Wick Irrigation Systems [187], [188], [189], [190]. In contrast, closed hydroponic systems collect and reuse nutrient-mixed water, minimizing waste. Examples of closed hydroponic systems include Nutrient Film Technique (NFT), Drip Irrigation, and Ebb & Flow Methods [187], [188], [189], [190], [191], [192], [193]. In these setups, the only water lost is due to evaporation and plant transpiration. However, nutrient levels must be continuously monitored to maintain optimal growing conditions.

Aquaponic systems are categorized into coupled and decoupled systems [194]. In a coupled aquaponic system, water continuously circulates between fish tanks and plant beds. Meanwhile, a decoupled aquaponic system features separate water circulation loops for aquaculture and hydroponics, reducing risks and improving system efficiency. Another form of soilless cultivation integrated with CEA is aeroponics, where no growing medium is used. Instead, nutrient solutions are directly sprayed onto plant roots according to their requirements [195]. This technique optimizes nutrient absorption and significantly reduces water consumption.

CEA offers multiple economic, social, and environmental benefits in vertical farming systems [195]. It improves food quality, reduces environmental hazards, and enhances economic returns [58]. By lowering transportation costs and pollution, CEA enables faster delivery of fresh produce while conserving resources [125]. It also reduces the need for chemical fertilizers and pesticides, promoting healthier crops and improving soil sustainability [129]. The technology enables maximized yield with reduced environmental impact, making farming more efficient and profitable [175], [176], [177]. The advantages of CEA in smart precision farming systems are summarized in Table 13.

TABLE 14. Disadvantages of CEA in smart precision VF systems.

| References | Parameters | Harmfulness | | |
|---|---|---|---|---|
| | | Environmental | Social | Economic |
| [66, 67] | Initial setup of CEA enabled VF | The environment is adversely affected during VF setup. | Need for proper technical knowledge and training. | A high initial cost is incurred while setting up a VF system. |
| [180-183, 195] | Expensive and short-lived cladding material | The non-biodegradable cladding material is not environmentally friendly. | Short-lived cladding material needs to be discarded frequently. | The cost incurred in changing the cladding material repeatedly. |
| [190, 191] | Lack of repair tools and maintenance facilities | - | Proper technical knowledge is needed. | Expenses incurred in repair and maintenance. |
| [71-73] | Lack of skilled and technically trained workers | - | The difficulty for farmers is in upgrading their skills. | Cost overhead in outsourcing skilled men. |

Despite its numerous advantages, CEA-enabled vertical farming systems face challenges such as high initial costs, expensive materials, and a lack of skilled labor [66], [71], [72], [73], [182], [183], [184], [185], [186], [187], [188], [189], [190], [191], [192], [193], [194], [195]. The cost of setting up a CEA-based system is significant, and maintaining non-biodegradable cladding materials can add recurring expenses [180], [181], [182], [183], [195]. Furthermore, limited access to repair tools and maintenance facilities can restrict the widespread adoption of CEA, particularly in rural and underdeveloped regions [190], [191]. A shortage of technically trained workers presents an additional barrier, as farmers need proper training to operate and maintain CEA-enabled setups [71], [72], [73]. These challenges are summarized in Table 14.

Although CEA minimizes labor dependency and ensures higher-quality yields, its high energy consumption and reliance on technology remain significant concerns [194]. However, integrating renewable energy sources into CEA can mitigate these issues, making urban farming more sustainable [189].

The opportunities and threats of CEA are depicted in Figure 8, highlighting both its potential and limitations. While initial setup costs and energy consumption are challenges, CEA ensures year-round, high-quality production with reduced labor involvement [184]. The implementation of automated disease management and renewable energy sources further enhances its sustainability, making CEA a promising solution for the future of agriculture [191], [192].

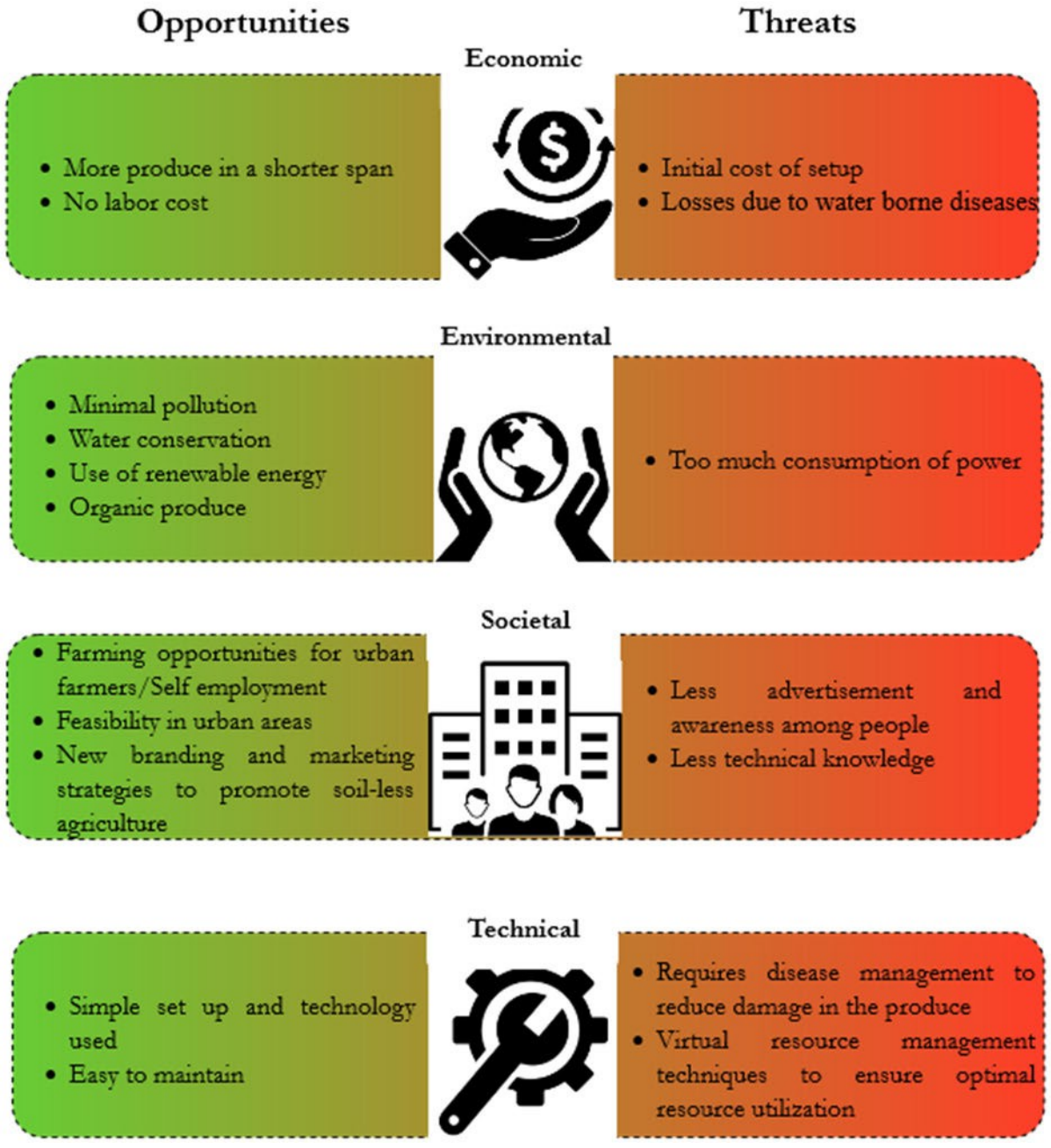

FIGURE 8. Opportunities and threats of CEA.

## V. INTEGRATION AND APPLICATION OF TECHNOLOGIES LIKE IOT, ML, AND DL IN CEA ENABLED SMART PRECISION FARMING

The integration of IoT, AI, ML, and DL has significantly advanced smart precision farming, improving monitoring, automation, and decision-making. These technologies enhance CEA-enabled soilless farming methods, such as hydroponics, aeroponics, and aquaponics [17], [18]. By utilizing these technologies, farming systems can optimize resource use, reduce cultivation time, and increase crop yields.

ML plays a crucial role in smart precision farming by analyzing input data, identifying patterns, and making data-driven decisions. It enables farmers to gather real-time information, predict outcomes, and optimize performance [17]. By combining theories, algorithms, and data analytics, ML helps to reduce production costs while improving crop yield and quality. AI and ML also help in analyzing nonlinear relationships between dependent and independent variables. This improves accuracy and performance compared to conventional regression models [142], [143]. Additionally, larger training datasets result in better accuracy for prediction and classification tasks, which is useful for crop yield estimation and profitability analysis [143], [151].

One major application of AI and ML in precision farming is disease classification in crops. Detecting diseased or infected leaves with high accuracy enables efficient disease management and reduces risks. In [196], the authors applied a classification model to categorize diseased leaves, improving accuracy in plant health monitoring. This reduces agricultural losses and enhances farming efficiency. In [153], a four-tier system using fog computing, edge computing, cloud computing, and sensors was implemented to remotely monitor CO2, moisture, humidity, light, and temperature in greenhouses. The system used IoT-based monitoring with the MQTT protocol to transmit and analyze sensor data.

In [154], IoT and ML were used to optimize the cultivation process by predicting the number of crops needed, allowing for better pre-planning. Recent advancements in DL-based models have further improved precision agriculture. A DL-based AgriLeafNet Model, integrating Few-Shot Learning (FSL) and NASNetMobile, has been used for crop classification and agricultural sustainability [197]. Additionally, in [198], a comprehensive review of ML and DL techniques in precision farming highlighted their potential for improving efficiency and productivity.

The use of prediction algorithms is crucial for forecasting crop yields and preventing food shortages. Various ML algorithms have been studied to recommend optimal crop selection based on environmental conditions. In [199], the authors reviewed recent research on AI, ML, and DL applications in agriculture. The study highlighted challenges related to data processing, scalability, and decision-making. While AI-based decision-making improves resource efficiency and reduces environmental impact, certain issues still need to be addressed.

The role of IoT in precision farming has also been widely explored. In [200], a detailed study on IoT architecture examined various challenges, applications, and solutions. The study emphasized IoT's ability to enable real-time monitoring, automation, and predictive analysis, making farming systems more responsive to environmental changes. Further research in [201] examined DL applications in agriculture, such as pest detection, yield prediction, disease classification, and soil and weed condition analysis. The study also covered IoT architecture, UAV-based sensor systems, and data processing techniques, including data acquisition, annotation, and augmentation. In [202], AI2Farm was proposed as an AI-based decision support system designed to predict and mitigate agricultural risks while enhancing productivity.

The integration of IoT, ML, and DL in CEA-enabled smart precision farming has transformed agriculture. These technologies facilitate automated, data-driven decision-making, mitigating uncertainties in agricultural production. By improving crop monitoring, disease detection, and resource management, AI-powered precision farming contributes to sustainable agriculture. It ensures year-round, faster, and higher-quality crop yields while minimizing waste, environmental impact, and production costs.

## VI. BIBLIOMETRIC ANALYSIS

Bibliometric analysis is a quantitative method used to evaluate the impact of academic publications within a specific research domain. It employs statistical techniques and graphical representations to facilitate data analysis and comprehension. This method provides key insights into academic literature dissemination, helping researchers, publishers, institutions, and policymakers make informed decisions [203].

Researchers can analyze trends and patterns to identify keywords, relevant authors, leading sources, and emerging institutions. Additionally, it helps in assessing international collaborations and making funding decisions, thereby guiding future research directions. Recently, bibliometric analysis has been widely applied across various research fields.

In [204], a statistical frequency analysis was conducted using VOSviewer on 2,334 articles (2003-2021) in the agricultural domain. The study examined research areas, publication trends, journal productivity, and key institutions. In [205], a bibliometric analysis was conducted on 27,725 articles from the Web of Science (WoS) database, focusing on organic farming and voluntary certifications. The most prominent research trends identified included plant factories, CEA, IoT, Life Cycle Assessment (LCA), and vertical farming.

In [206], bibliometric analysis using CiteSpace software was performed on 592 research articles (2004-2022) related to digital agriculture. Similarly, in [207], a study analyzed 2,348 research articles (1999-2022) on ML and pest control, retrieved from the WoS database. In [208], 2,033 research articles (2015-2023) from the WoS database were analyzed using CiteSpace V6.2.R4, focusing on blockchain applications in agriculture.

In [209], a systematic literature review (SLR) was conducted on rice leaf disease detection, integrating a deep learning (DL) model for accurate disease identification. In [210], a bibliometric and thematic analysis was performed on urban and peri-urban farming (2002-2022). A total of 1,257 articles were retrieved from the WoS database, covering topics such as local food systems, urban agriculture, food security, and sustainability, analyzed using VOSviewer and Bibliometrix.

The integration of SLR, thematic analysis, and bibliometric analysis allows researchers to triangulate observations and gain a clearer understanding of research trends. Thematic analysis provides qualitative insights, linking empirical data with research themes. In contrast, bibliometric analysis offers a quantitative perspective, helping identify major research trends and topics [211], [212].

A combined analysis of bibliometric studies, SLR, and thematic analysis has been conducted to explore soilless and soil-based cultivation methods. SLR is widely recognized for conducting systematic literature reviews across various research domains. Table 15 presents a summary of bibliometric analyses, including the software used, databases considered, number of articles retrieved, and study periods.

## A. MATERIALS AND METHODS

This study conducts a bibliometric analysis on research related to soilless cultivation, CEA, and smart precision farming.

The first step involves database selection. The Scopus database was chosen due to its advanced search and analysis features, covering multidisciplinary, peer-reviewed publications, book chapters, and articles. It is a globally recognized research database that provides essential bibliometric parameters, including the h-index, citation counts, and collaboration networks. Scopus also ensures metadata completeness through rigorous validation procedures, making it highly reliable for bibliometric analysis.

Next, data filtering and extraction were performed. A search query was generated using the keywords ''aeroponics,'' ''aquaponics,'' ''hydroponics,'' ''soil cultivation,'' and ''controlled environment agriculture.'' The search string used was:

TITLE-ABS-KEY ((''aeroponics'') OR (''aquaponics'') OR (''hydroponics'') OR (''soil cultivation'') OR (''controlled environment agriculture''))

Filtering research articles by specific keywords provides a comprehensive overview of the research landscape. It also ensures the inclusion of studies covering key aspects such as socio-economic impact and environmental sustainability.

To refine the dataset, inclusion and exclusion criteria were applied. The dataset was limited to articles published between 2013 and 2023. Initially, 8,214 articles met the search criteria. After restricting the language to English only, the final dataset was reduced to 7,764 articles.

Figure 9 illustrates the selection steps and stages used to refine the dataset for analysis.

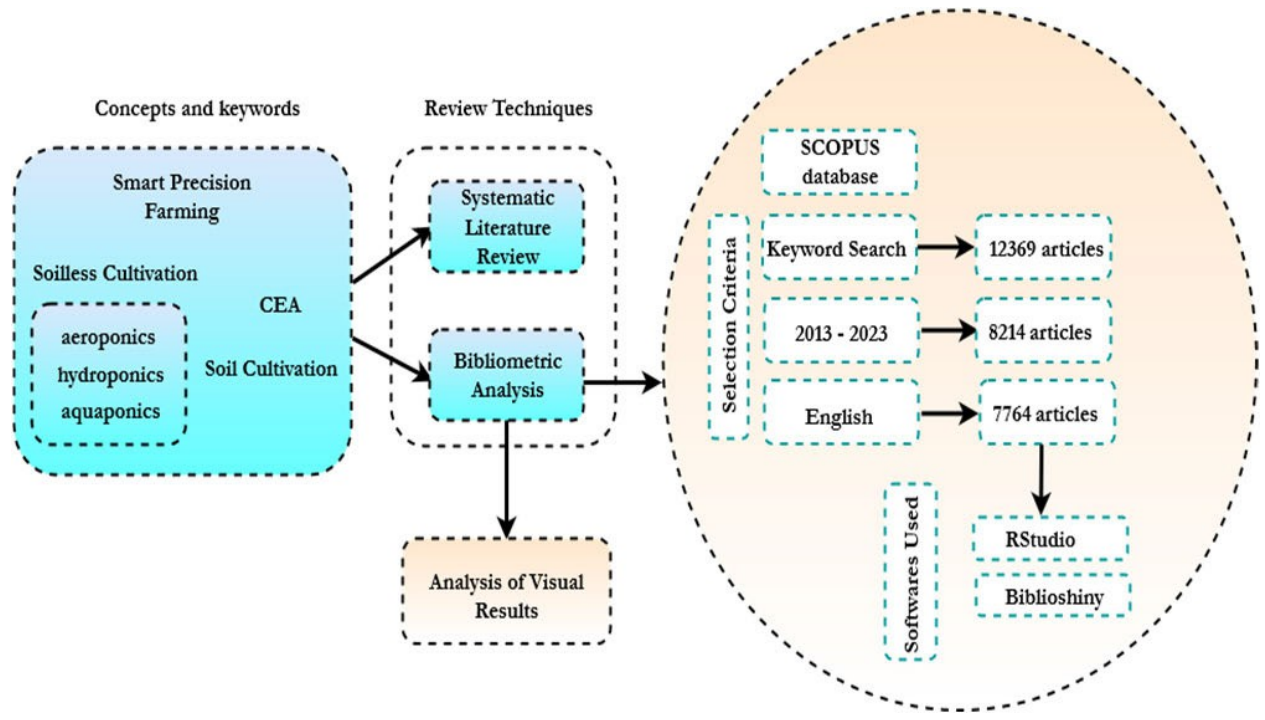

FIGURE 9. Steps for bibliometric analysis.

A CSV file containing details of 7,764 articles was downloaded from the Scopus database and used as input for bibliometric analysis. The analysis was conducted using RStudio (version 4.2.1), with Biblioshiny, a widely used tool for bibliometric studies and research quality assessment.

Biblioshiny facilitates the analysis of large datasets, making it useful for evaluating research output in academia, industry, and research organizations. It enables the examination of publication counts, trends, author contributions, country-wise research distribution, and institutional collaborations.

Table 16 presents key information extracted from the Scopus dataset, summarizing the analyzed data.

## B. SYNTHESIS, ANALYSIS, AND FINDINGS

The dataset retrieved from Scopus was analyzed using the Biblioshiny tool in RStudio, generating visual and network analyses. Key graphical representations include annual scientific production, co-citation networks, most relevant sources, co-occurrence networks, and word clouds.

The annual scientific publication trends highlight the growth of research in this field. A steeper increase in publications indicates rising research interest and relevance. Co-citation analysis examines references that appear together in the same article, identifying relationships between research works. A higher co-citation intensity suggests stronger connections and thematic similarities among studies.

Table 17 presents the distribution of research articles across various categories. The maximum percentage of research publications comprises research articles, conference papers, reviews, book chapters, and others. ''Others'' include notes, editorials, errata, letters, short surveys, retracted papers, and data papers indexed in SCOPUS.

## C. ANNUAL INCREASE IN SCIENTIFIC

Initially, 7,764 articles were retrieved from the SCOPUS database for bibliometric analysis. Figure 10 illustrates a steady increase in annual research publications, rising from 371 articles in 2013 to 1,064 in 2022. A notable surge occurred between 2019 (748 articles) and 2021 (1,053 articles), reflecting growing interest in soilless and soil-based cultivation.

The publication trend, shown in Figure 10, follows an exponential growth pattern, emphasizing the field's increasing relevance. Notably, 14% of all articles were published individually in 2021, 2022, and 2023, while the lowest output (∼5%) occurred in 2013 and 2014. This research directly supports the SDGs by addressing food security, zero hunger, climate resilience, and social equity, contributing to a more sustainable agricultural future.

## D. CITATION ANALYSIS OF SCIENTIFIC PUBLICATIONS

The most cited documents are the research articles that have accumulated the maximum number of citations. Figure 11 is a bubble chart representing the global citations in the x-axis and the list of documents in the y-axis. It shows that the article authored by Singh Sekhon B in the year 2014 has secured the maximum citations of 636. Next in decreasing order of the number of citations was authored by Lin KH in the year 2012. The article received 547 citations.

TABLE 15. Summary of bibliometric analyses on CEA-enabled smart agriculture.

| References / Year | Period | Count of Articles | Database | Software |
|---|---|---|---|---|
| [204]/ 2022 | 2003- 2021 | 2334 | Scopus | VOSviewer |
| [205]/ 2023 | 1983- 2022 | 27,725 | WoS | VOSviewer V 1.6.1 |
| [213]/ 2023 | 2002- 2022 | 1,257 | WoS | VOSviewer and Bibliometrix |
| [206]/ 2023 | 2004- 2022 | 392 | WoS | CiteSpace bibliometric software |
| [207]/ 2024 | 1999- 2022 | 2348 | WoS | VOSviewer |
| [208]/ 2024 | 2013- 2023 | 2033 | WoS | CiteSpace V6.2. R4 |
| [211]/ 2022 | 2014- 2021 | 2011 | Scopus | VOSviewer software V1.6.16 |
| [212]/ 2021 | 2008- 2019 | 1236 | WoS | VOSviewer |

TABLE 16. Database overview for bibliometric analysis.

| Category | Count / Percentage |
|---|---|
| Timespan | 2013:2023 |
| Sources | 1932 |
| Documents | 7764 |
| Annual Growth Rate | 10.85 % |
| Authors | 25937 |
| Authors of single-authored docs | 234 |
| International Co-Authorship | 22.41 % |
| Co-Authors per doc | 5.2 |
| Author's Keywords (DE) | 16243 |
| References | 307624 |
| Document Average Age | 4.85 |
| Average Citations per Doc | 15.65 |

TABLE 17. Category-wise count of articles published in the 'Scopus' database.

| Document type | Count | Percentage |
|---|---|---|
| Article | 5,995 | 77.21 % |
| Conference paper | 1,178 | 15.17 % |
| Review | 291 | 3.74 % |
| Book chapter | 249 | 3.21 % |
| Others | 51 | 0.65 % |

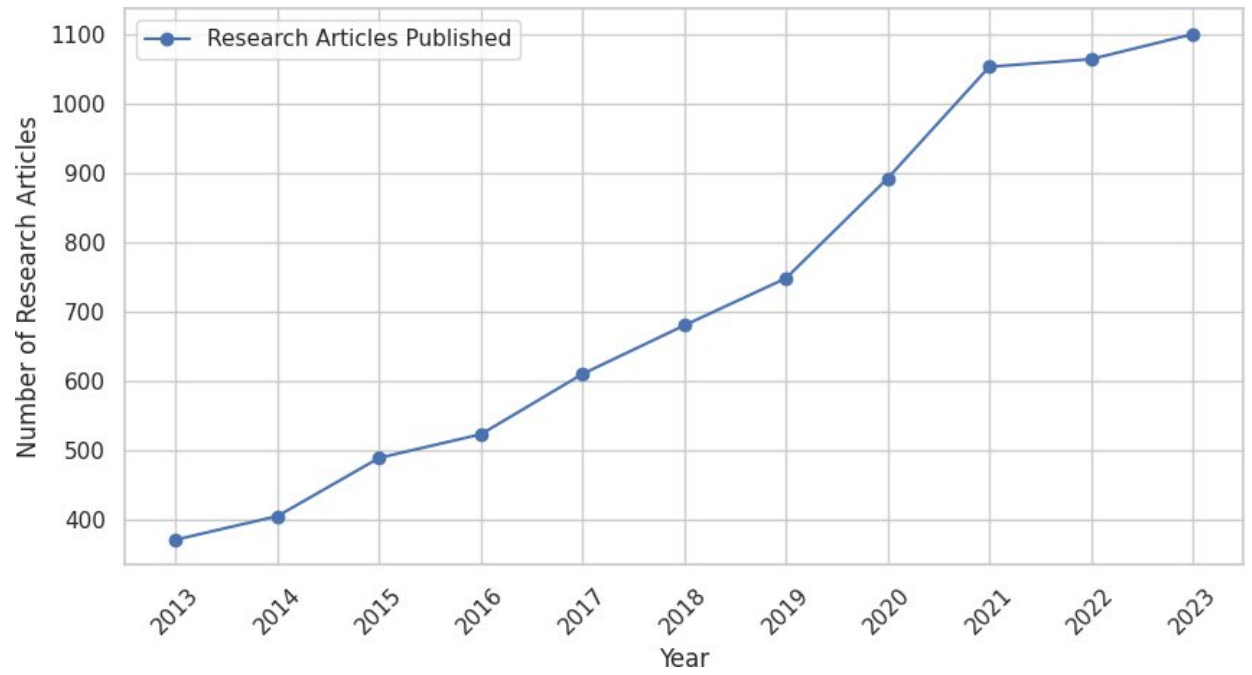

FIGURE 10. Distribution of scientific publications produced per year.

## E. THREE-FIELD PLOT

Figure 12 presents a three-field plot illustrating correlations among authors (AU), keywords (ID), and author countries (AU_CO) in CEA-enabled smart agriculture. Grey links indicate associations, with rectangle sizes representing the number of linked research articles. A dense network signifies strong connections in this research domain.

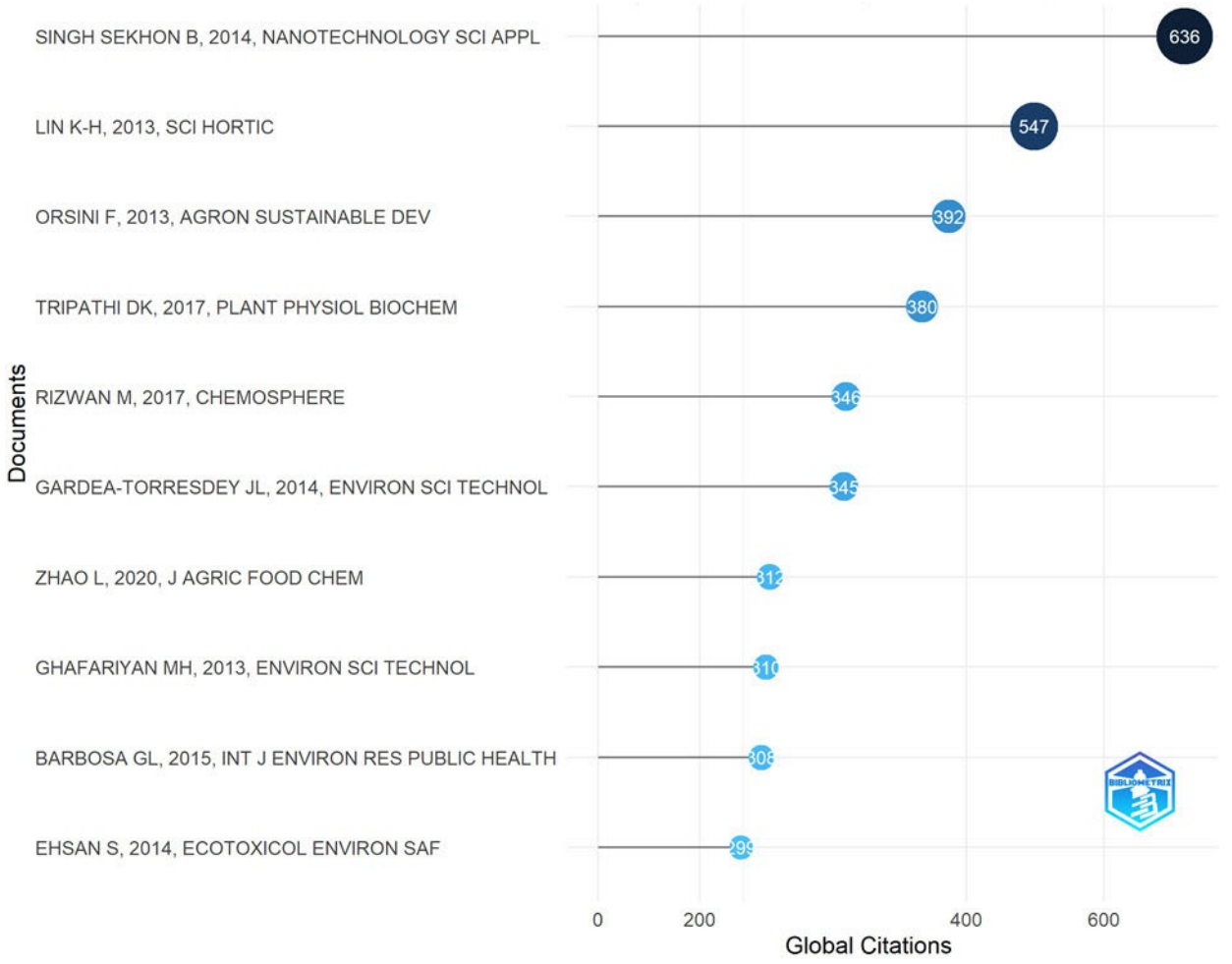

FIGURE 11. Distribution of citations of scientific publications.

Wang Y is the most relevant author, frequently using the keyword *cadmium* and citing *Bradford M.M.* the most. Zhang also prominently uses *cadmium*. The plot helps identify key authors, commonly used keywords, and citation patterns.

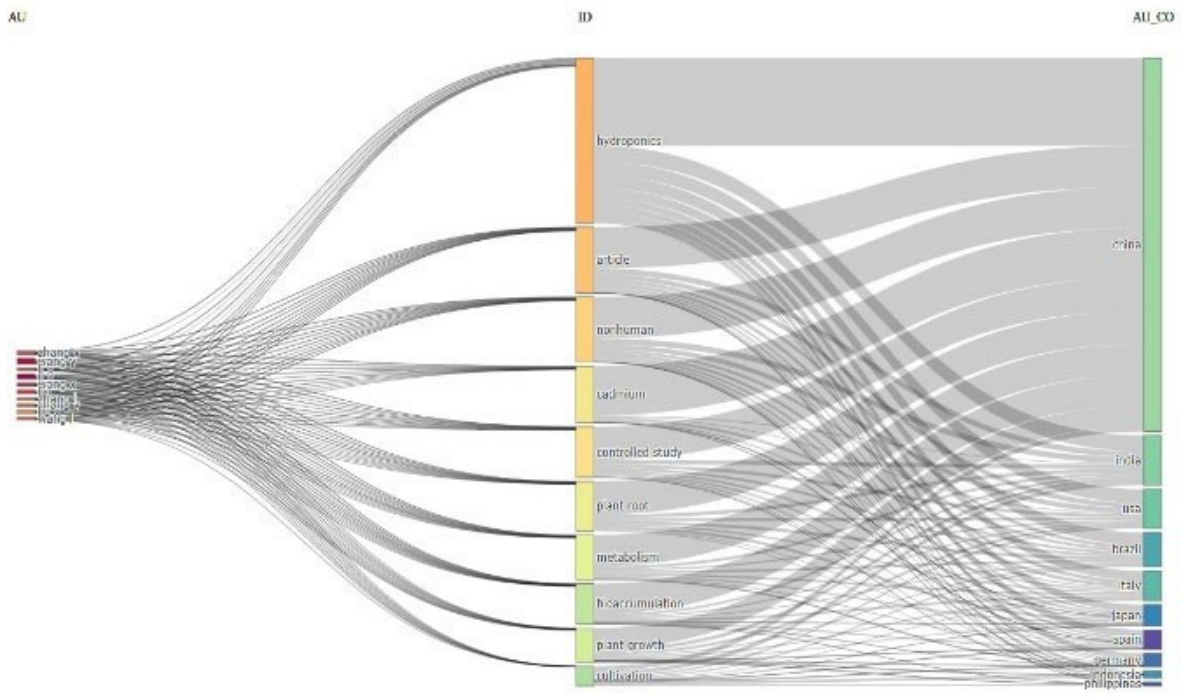

FIGURE 12. Sankey diagram for top 10 authors, keywords, and author countries in a three-field plot.

## F. PERFORMANCE ANALYSIS

This section analyzes research sources, key authors, keywords, and institutions. This section examines the most productive research sources, their publishing trends, and their impact. Key factors include total citations, article count, h-index, and m-index.

### 1) MOST PRODUCTIVE SOURCES, THEIR DYNAMICS, AND MOST RELEVANT SOURCES

Figure 13 shows the top 10 sources with the highest number of publications in soilless cultivation, soil-based cultivation, and CEA. Among them, Acta Horticulturae is the most productive, publishing 209 articles by 2023. Environmental Science and Pollution Research follows with 199 articles, and Ecotoxicology and Environmental Safety ranks third with 170 articles. PLoS ONE and Horticulturae have published fewer articles, with 99 and 98 publications, respectively.

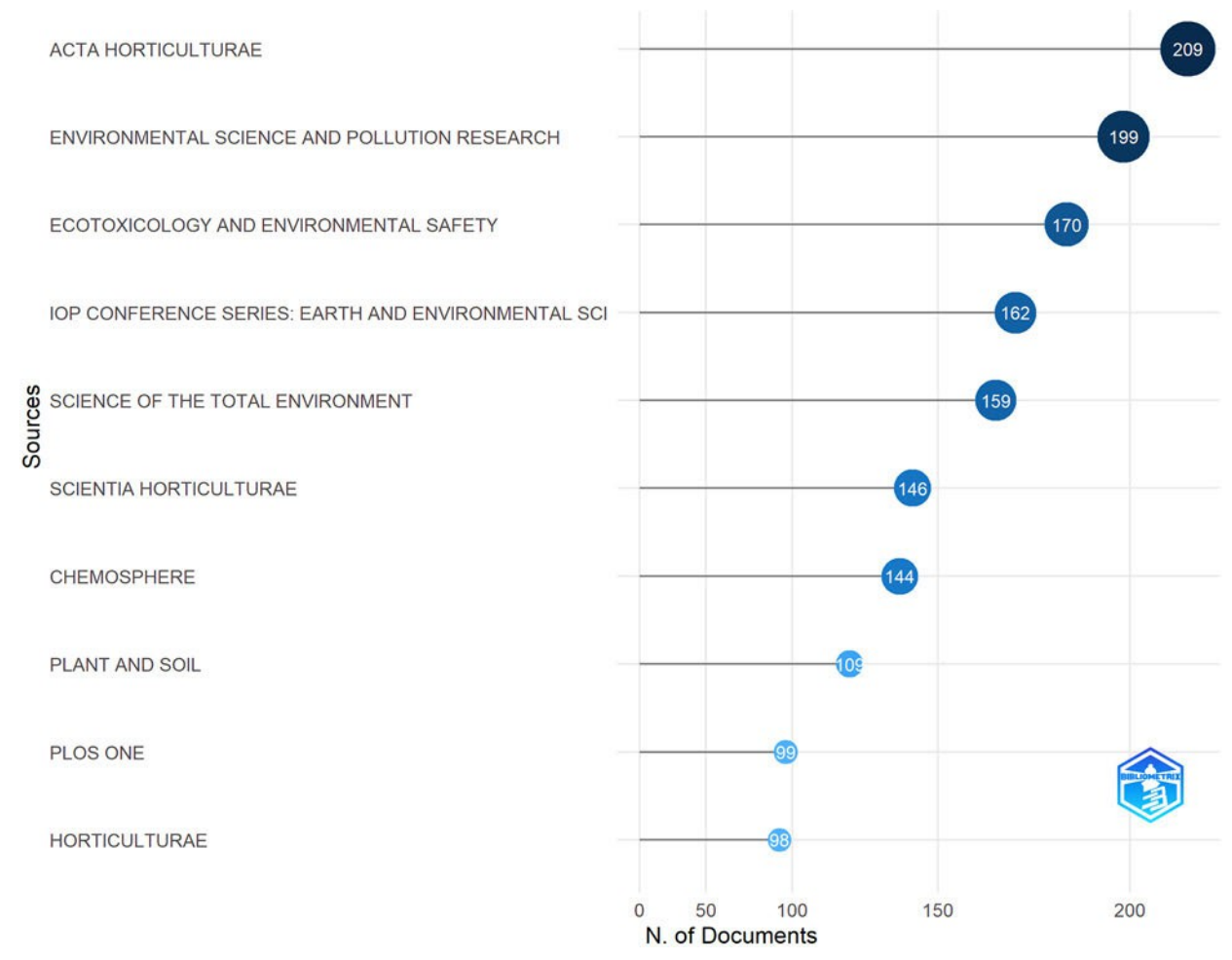

FIGURE 13. Most productive sources.

Figure 14 illustrates the publishing trends of the top five most productive research sources from 2013 to 2023. It tracks the annual publication count for each source, highlighting their growth and consistency over time. This helps identify key trends. Acta Horticulturae has shown steady growth, from 15 articles in 2013 to 209 in 2023. Environmental Science and Pollution Research started with 9 articles in 2013 and grew to 199 by 2023. Ecotoxicology and Environmental Safety began with 5 articles in 2013 and reached 170 in 2023. IOP Conference Series: Earth and Environmental Sciences showed notable growth, starting at 0 in 2013 and reaching 162 in 2023. These trends help researchers identify key sources and predict emerging influential journals.

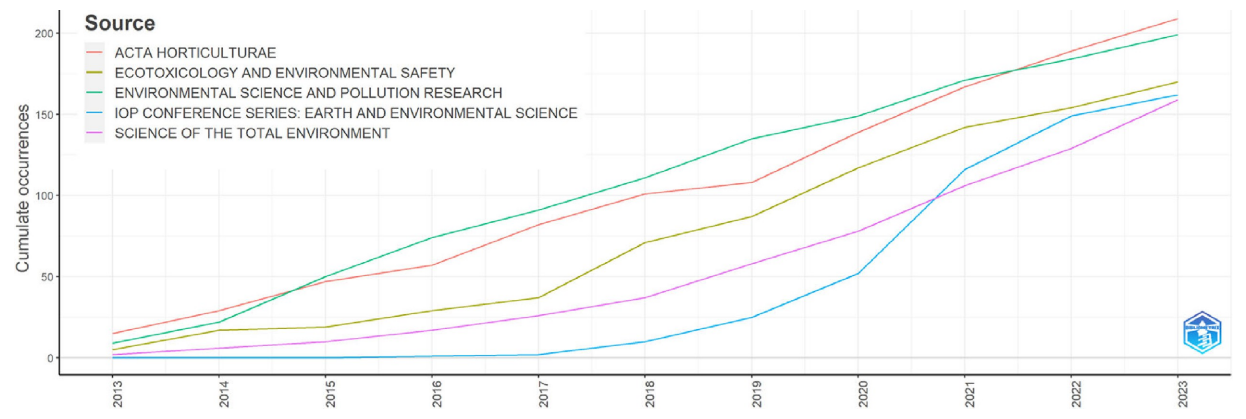

FIGURE 14. Sources production over time.

The top ten sources have been further analyzed based on total citations (TC), h-index, and number of publications (NP), as shown in Table 18. All the sources began publishing in 2013. Among them, Ecotoxicology and Environmental Safety has received the highest number of citations, while PLoS ONE holds the highest h-index.

TABLE 18. Category-wise analysis of sources.

| Sources | TC | h-index | NP |
|---|---|---|---|
| Ecotoxicology and environmental safety | 5575 | 161 | 170 |
| Environmental science and pollution research | 5361 | 154 | 199 |
| Environmental science and technology | 4281 | 131 | 77 |
| Chemosphere | 4110 | 288 | 144 |
| Journal of hazardous materials | 5291 | 329 | 98 |
| Science of the total environment | 4420 | 317 | 159 |
| Scientia horticulturae | 4285 | 134 | 140 |
| Environmental pollution | 2828 | 275 | 85 |
| Plant and soil | 2920 | 212 | 109 |
| PLoS ONE | 5174 | 404 | 99 |

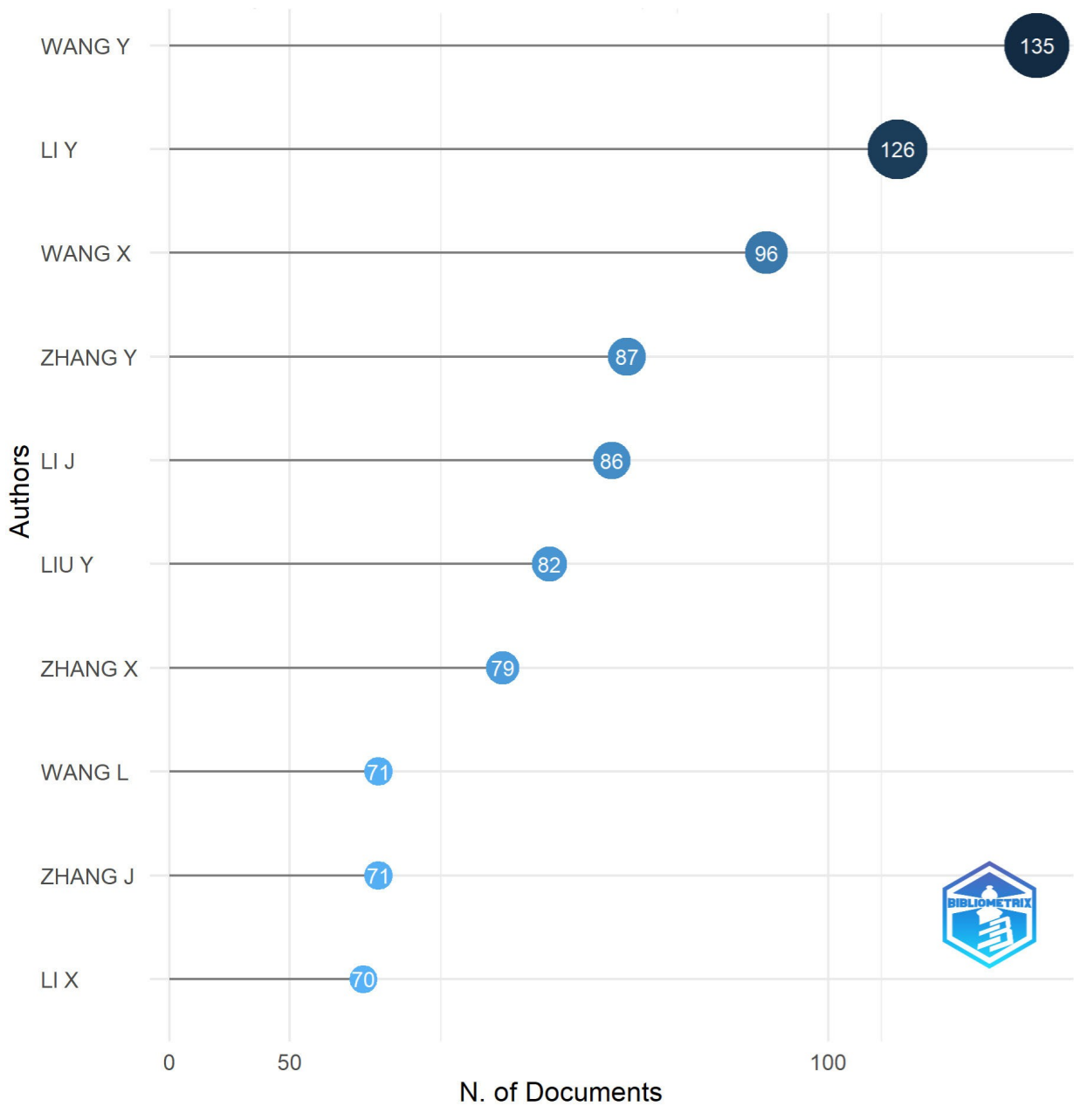

FIGURE 15. Most productive authors.

### 2) MOST PRODUCTIVE AUTHORS

Authors play a crucial role in advancing research within a specific field. Bibliometric analysis of research authors is divided into two key sections, i.e., the most relevant authors in the field and their research productivity over time. In the domain of soilless agriculture and CEA, researchers who have consistently contributed were identified, and their publication trends were analyzed. Figure 15 highlights the top ten most productive authors, with Wang Y emerging as the most prolific, having published 135 articles by 2023. Li Y follows closely with 126 publications, while Wang X ranks third with 96 research articles. Other notable contributors include Zhang X (87 articles), Li J (86 articles), Liu Y (82 articles), Zhang X (79 articles), Wang I (71 articles), Zhang J (71 articles), and Li X (70 articles).

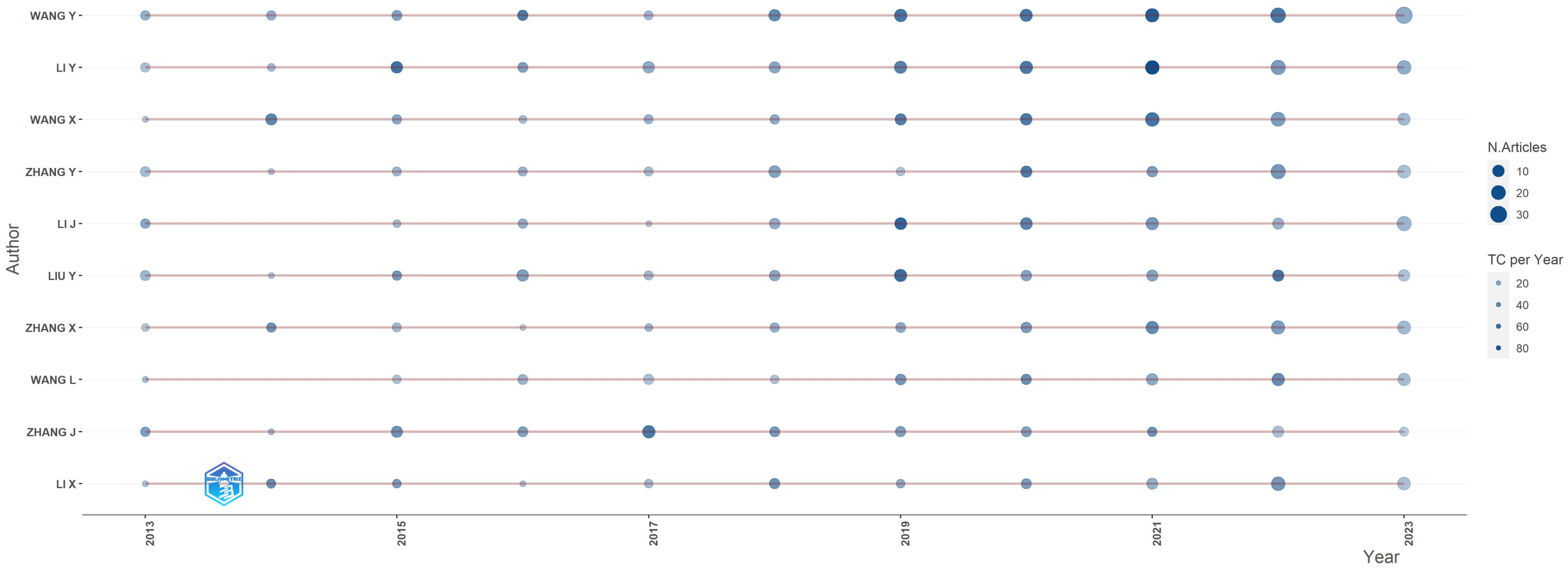

FIGURE 16. Authors' production over time.

### 3) AUTHORS' PRODUCTION OVER TIME

This section analyzes the consistency of research and review article production in modern and smart agriculture over time. Understanding these trends helps emerging researchers identify key contributors in the field and recognize those with sustained productivity. The top 10 authors identified earlier were evaluated for their research output. Wang Y led with 33 publications in 2022, followed by Li Y and Zhang Y with 21 articles each in the same year. Wang X produced 20 articles in 2022, while Li L peaked at 21 publications in 2023. Other notable contributors include Liu Y (12 articles in 2019), Zhang X (17 in 2022), Wang L (13 in 2022), Zhang J (13 in 2017), and Li X (18 in 2022).

While many authors have shown a steady increase in research output, some, such as Liu Y, Zhang J, and Li X, have seen a decline in publication trends over time. Figure 16 illustrates the publication trends of the top 10 most relevant authors in this domain.

Table 19 provides a detailed analysis of these authors based on their h-index, number of publications (NP), and total citations (TC). Ali S has the highest total citations (3262), while Wang Y leads in total publications (135). The highest h-indices (29) belong to Li Y, Wang Y, and Zhang J, indicating their strong influence in the field.

### 4) AUTHOR COLLABORATION NETWORK

Analyzing author collaborations helps identify emerging research trends and key contributors in the field. Figure 17 presents the collaboration network of the 30 most eminent authors in soilless agriculture and CEA. The network consists of seven major clusters: red, blue, green, purple, pink, orange, and brown. Among them, the red, blue, and green clusters are the most significant, containing top researchers such as Wang Y, Wang I, Li X, Zhang X, Zhang Y, and Li J.

TABLE 19. Analysis of the top 10 most relevant authors.

| Authors | H_index | NP | TC |
|---|---|---|---|
| Li Y | 29 | 120 | 2404 |
| Wang Y | 29 | 135 | 2080 |
| Zhang J | 29 | 71 | 2212 |
| Ali S | 25 | 41 | 3262 |
| Liu Y | 25 | 82 | 1788 |
| Wang X | 24 | 96 | 2046 |
| Li J | 25 | 86 | 1519 |
| Wang Q | 25 | 60 | 1220 |
| Chen Y | 21 | 48 | 1481 |
| Li X | 21 | 70 | 1621 |

In the network, authors are represented as nodes, and their collaborations are shown as connecting edges. The thicker the edge, the stronger the collaboration between those authors. The densest collaboration is observed within the red cluster, specifically between Li C and Li Y. The blue cluster also shows strong collaboration, particularly between Wang Y and Wang Q.

## G. COUNTRIES' NETWORK ANALYSIS

This section examines the global research landscape in soilless and soil cultivation using CEA. It focuses on three aspects: corresponding authors' countries, scientific production by country, and international collaboration networks.

### 1) CORRESPONDING AUTHORS' COUNTRIES

Leading researchers from one country often collaborate with those from other nations who share similar research interests. Analyzing these collaborations helps understand the global distribution of research and international cooperation patterns. Figure 18 presents the top 10 countries in terms of research output, categorized into Single Country Production (SCP) and Multiple Country Production (MCP).

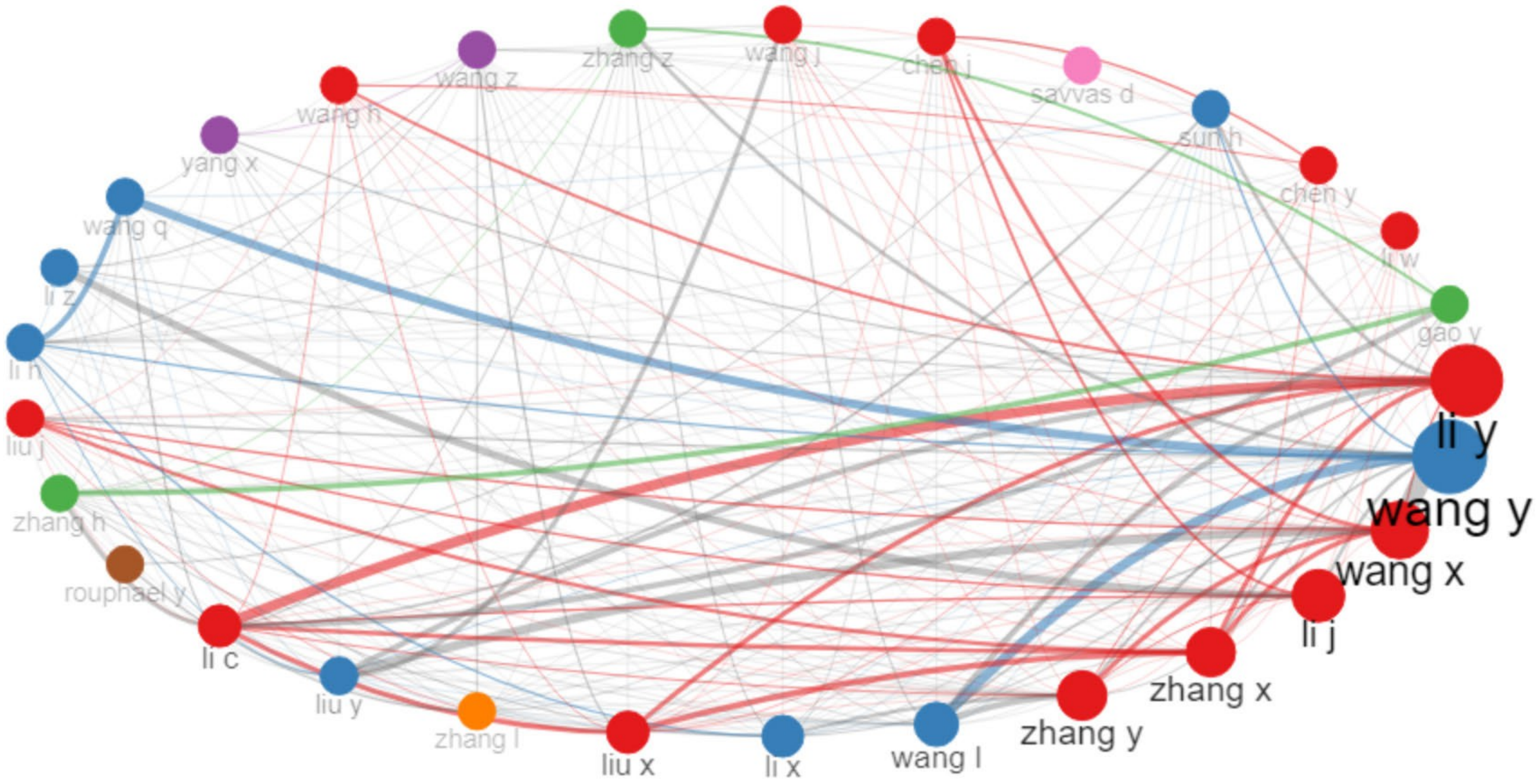

FIGURE 17. Author collaboration network.

China is the most influential country in this research domain, producing the highest number of articles. Other major contributors include India, the USA, Brazil, Italy, Indonesia, Germany, Japan, and Spain. However, many of these countries have a lower MCP than SCP, indicating that most research is conducted domestically with limited international collaboration. Expanding international partnerships could enhance research impact and knowledge exchange.

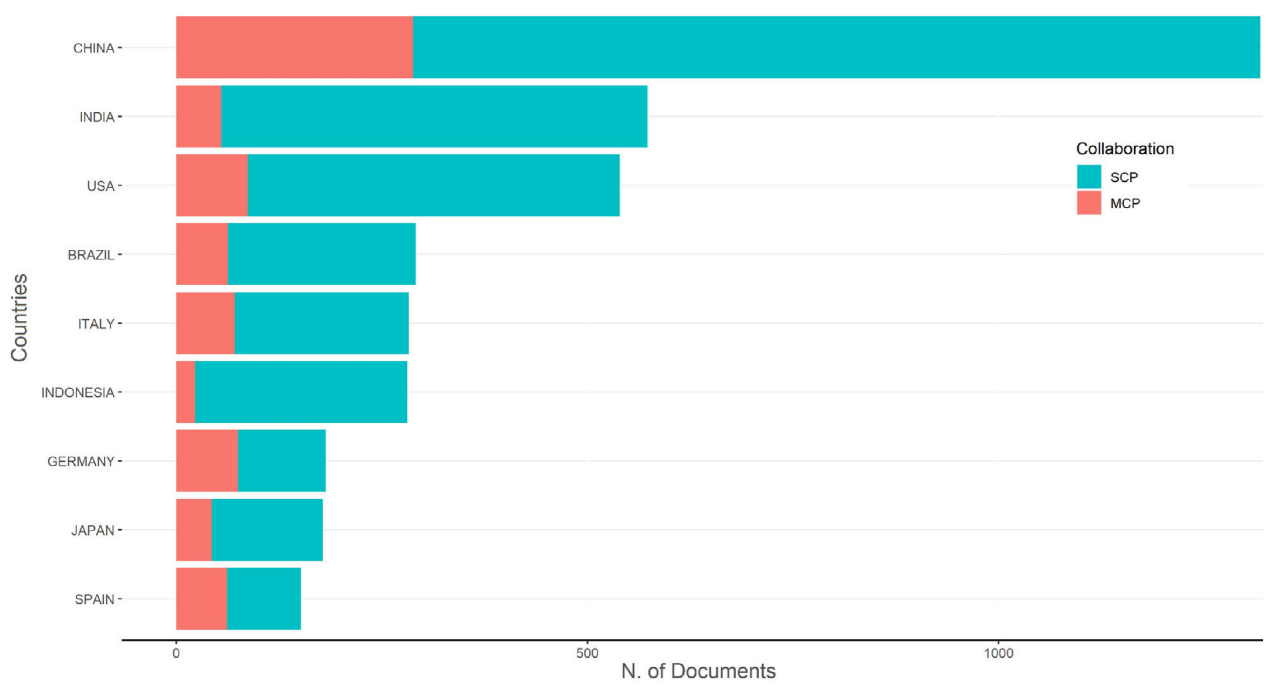

FIGURE 18. Corresponding authors' countries.

### 2) COUNTRY SCIENTIFIC PRODUCTION

This section examines global research trends in soilless and soil cultivation using CEA, highlighting each country's contribution. Figure 19 represents worldwide research output, where dark blue indicates the highest number of publications,

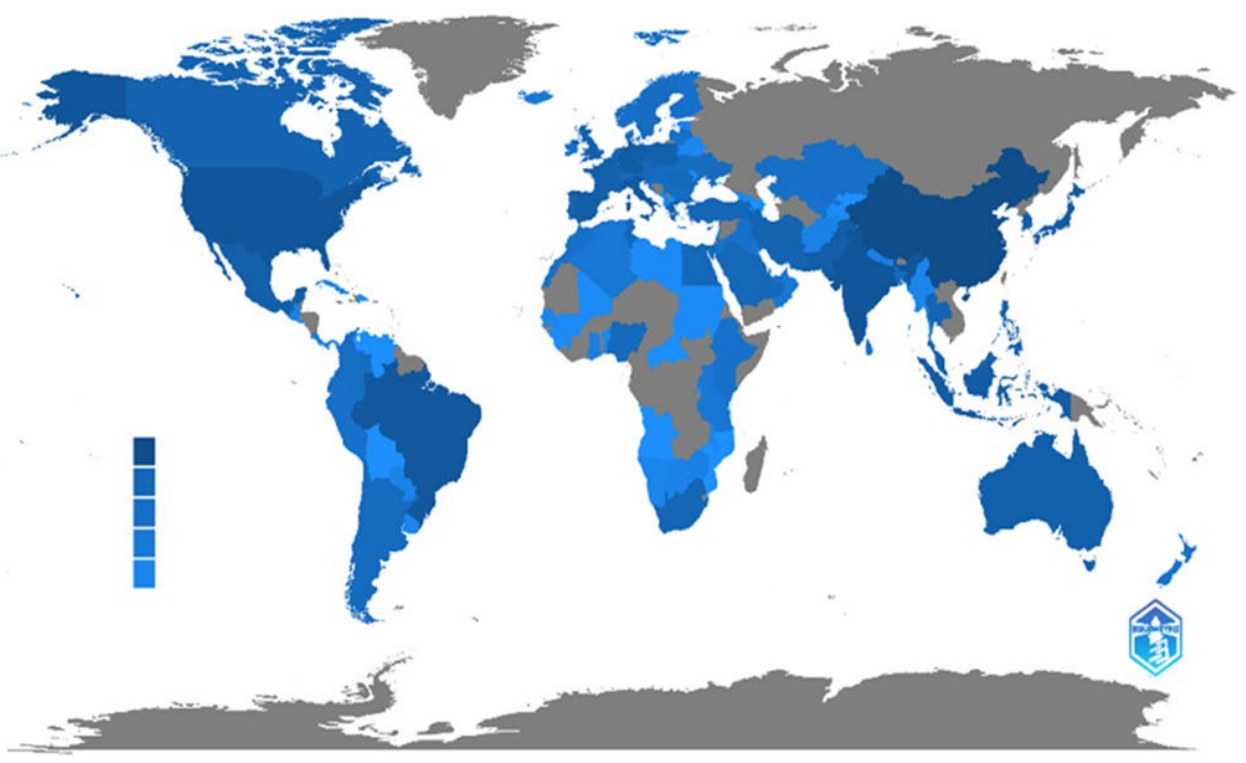

FIGURE 19. Country scientific production.

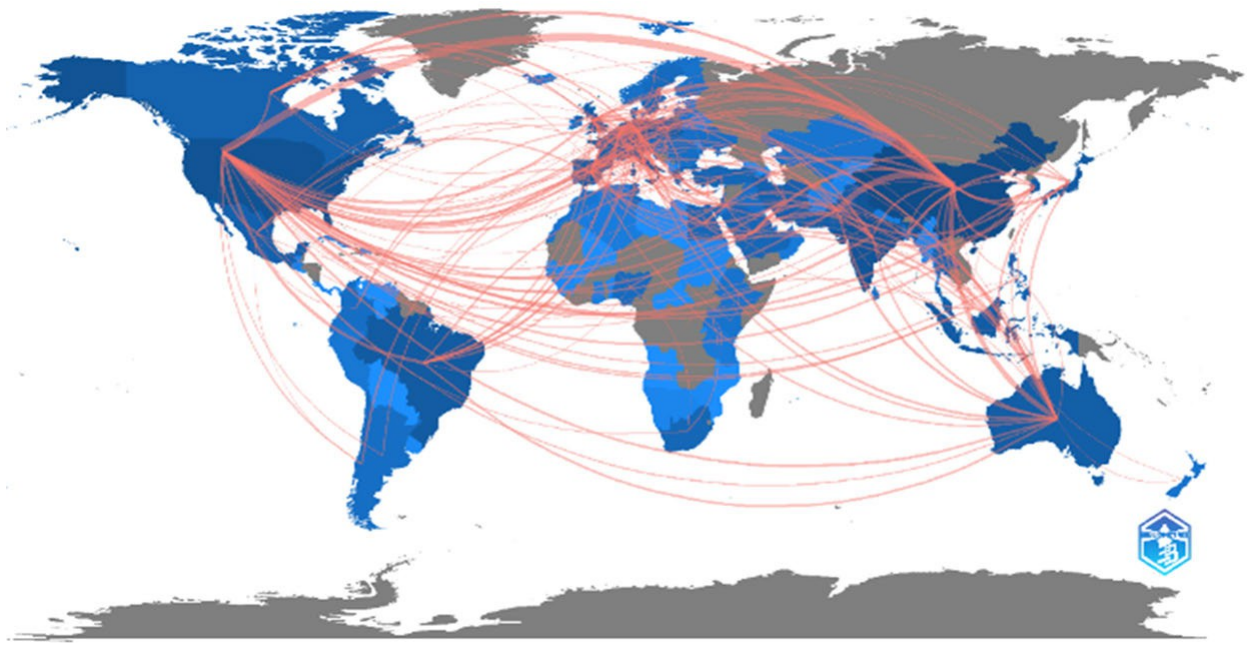

FIGURE 20. Country collaboration map.

lighter blue represents lower contributions, and grey denotes no research output. China leads with 10,651 publications, followed by the USA (2,947) and India (2,829). Other key

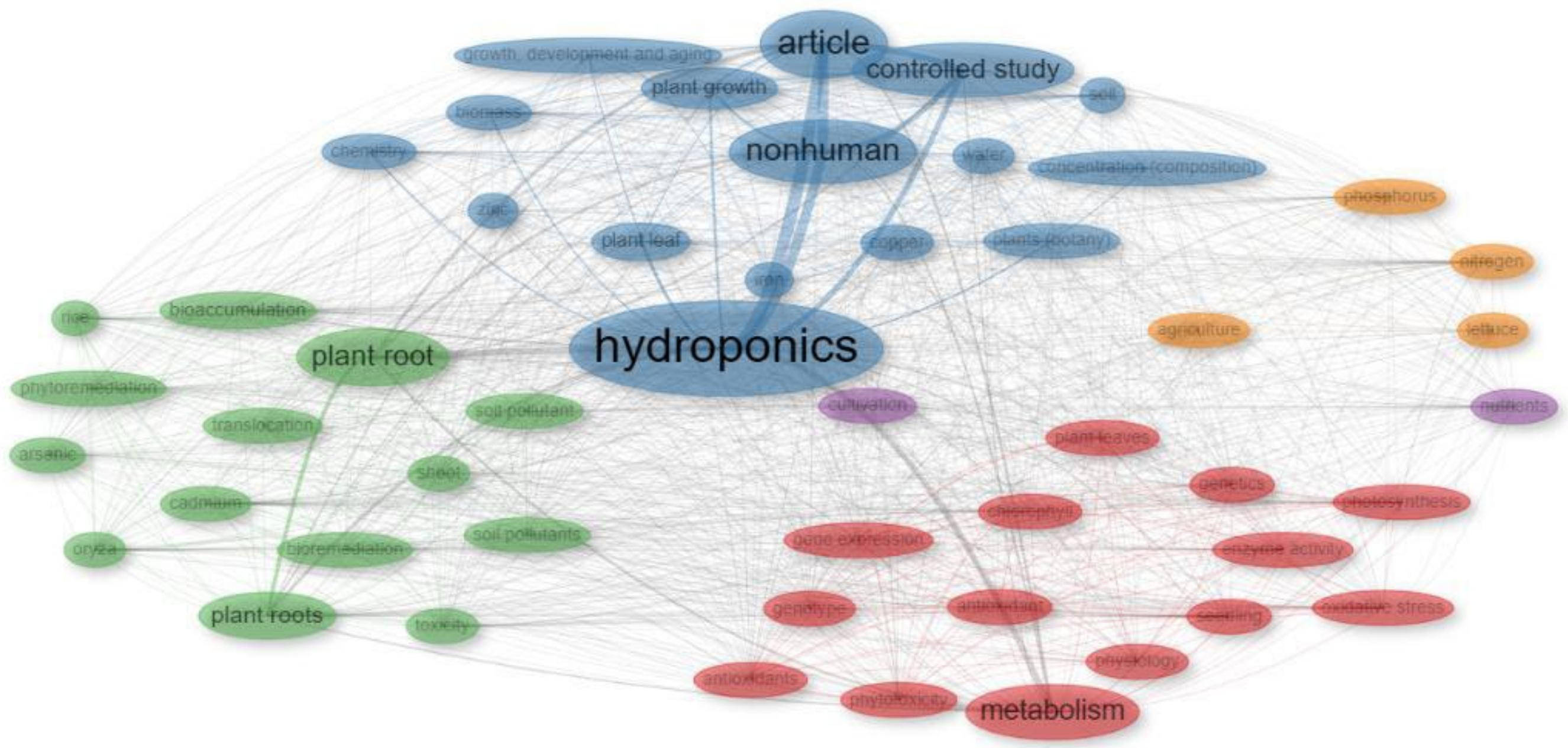

FIGURE 21. Network of co-occurrence of authors' keywords.

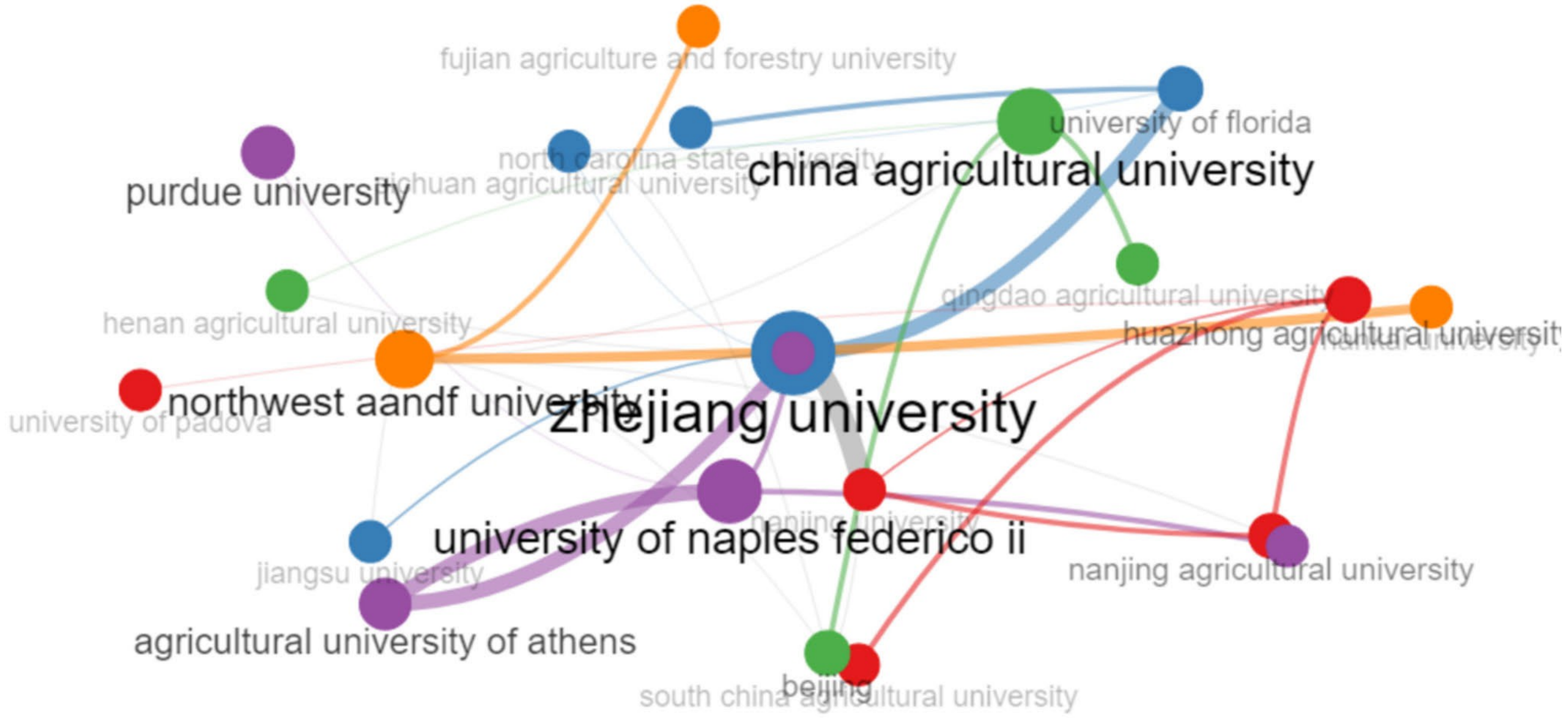

FIGURE 22. Institution collaboration network.

contributors include Brazil, Italy, Japan, Germany, Indonesia, Spain, and Pakistan.

Table 20 ranks the top 10 countries based on research output, emphasizing their role in advancing scientific knowledge and innovation in this field.

### 3) COUNTRY COLLABORATION MAP

This analysis examines international collaborations in soilless and soil cultivation research. Figure 20 maps these partnerships, where thicker links indicate stronger collaboration between countries. The most frequent collaboration occurs between Australia and Austria, followed by Australia's partnerships with Denmark and Canada, highlighting strong research ties. Argentina emerges as the most collaborative country, forming partnerships with Cuba, Austria, Bangladesh, Costa Rica, Croatia, Denmark, Egypt, Finland, France, and Hungary. Table 21 provides a detailed breakdown of these collaborations, showcasing global research connectivity in this domain.

**TABLE 20.** Top 10 countries and the number of articles they produced.

| Countries | Articles |
| --- | --- |
| China | 10651 |
| Usa | 2947 |
| India | 2829 |
| Brazil | 1901 |
| Italy | 1787 |
| Japan | 1290 |
| Germany | 1215 |
| Indonesia | 1107 |
| Spain | 891 |
| Pakistan | 840 |

**TABLE 21.** Frequencies of collaborations among countries.

| Countries | | Frequencies |
| --- | --- | --- |
| Argentina | Cuba | 3 |
| Argentina | Austria | 3 |
| Argentina | Bangladesh | 3 |
| Argentina | Costa Rica | 2 |
| Argentina | Croatia | 3 |
| Argentina | Denmark | 3 |
| Argentina | Egypt | 5 |
| Argentina | Finland | 3 |
| Argentina | France | 4 |
| Argentina | Hungary | 1 |

## *H. INFLUENTIAL ANALYSIS*

The keywords used in research articles are crucial for identifying trends and exploring related fields. Their co-occurrence helps analyze the scope and connections between different research areas. This is particularly valuable for researchers working on both established and emerging topics.

### 1) KEYWORD CO-OCCURRENCE NETWORK

Figure 21 visually represents the co-occurrence network of frequently used keywords, grouping them into four clusters. The blue cluster is the largest and most significant, with ''hydroponics'' as the dominant keyword. The red cluster follows, where ''metabolism'' appears most frequently, while the green cluster ranks third, with ''plant root'' as the key term. The strongest keyword association is within the blue cluster, where ''hydroponics'' is closely linked to related terms. Additionally, ''hydroponics'' frequently co-occurs with ''metabolism'' from the red cluster and ''plant root'' from the green cluster. In contrast, the yellow and purple clusters are relatively less significant.

### 2) INSTITUTIONS COLLABORATION NETWORK

Global institutional collaborations play a crucial role in advancing research by connecting researchers with supporting institutions and potential partners worldwide. Figure 22 presents the institution collaboration network, revealing five clusters in descending order of size: blue, purple, green, red, and orange. Zhejiang University leads the blue cluster, making it the most influential institution, alongside others such as the University of Naples, China Agricultural University, and Northwest A&F University. Additionally, Figure 23 illustrates the thematic evolution of research topics over the past decade (2013–2023), divided into three time slices: 2013–2016, 2017–2020, and 2021–2023. The rectangular shapes in each slice represent the significance of different research concepts during that period, while grey links indicate associations between research fields over time. Notably, ''hydroponics'' remains a densely linked keyword throughout all periods, frequently appearing alongside terms like ''soil cultivation'' and ''climate change.''

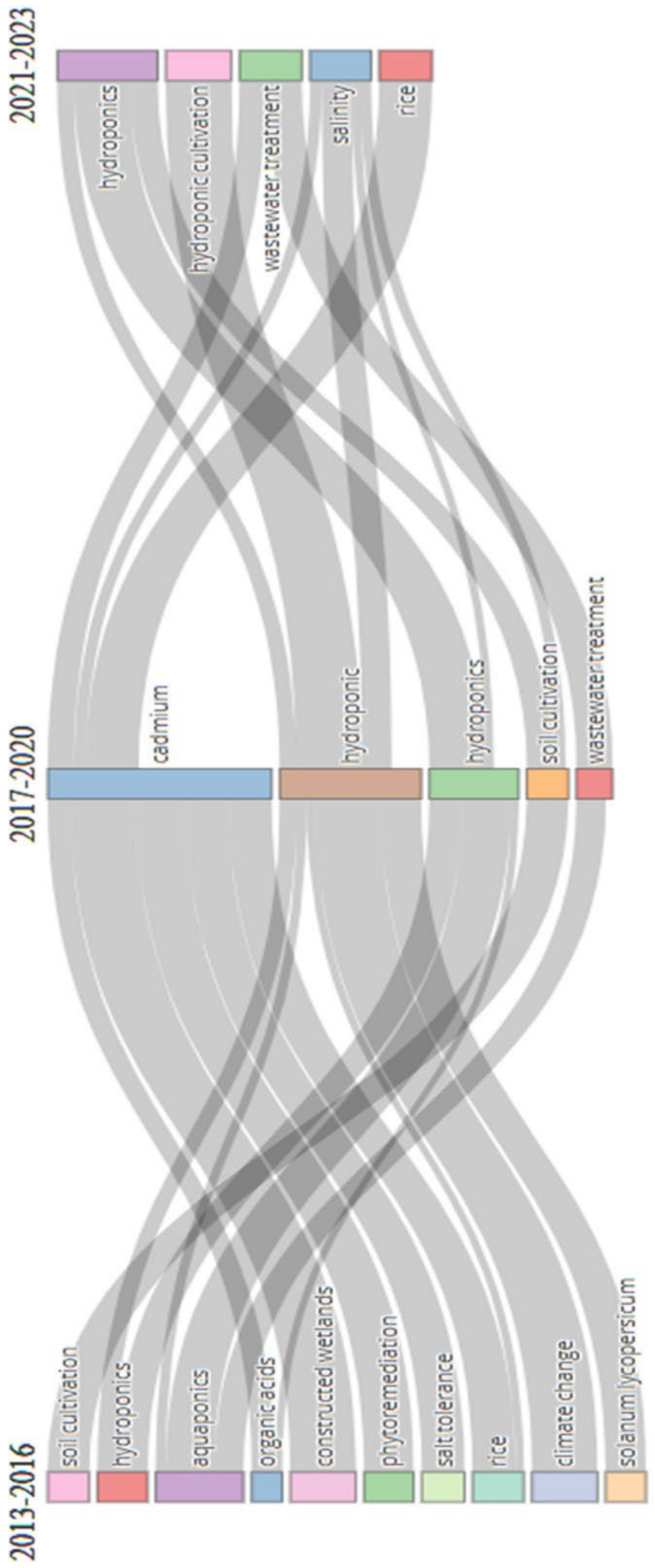

**FIGURE 23.** Thematic evolution.

### 3) WORD CLOUD

A word cloud is a visual representation of the most frequently used author keywords in a research domain. It provides insight into the key terms shaping the field. Figure 24 illustrates a word cloud based on the 50 most relevant keywords in smart soilless agriculture using CEA. Among these, ‘‘hydroponics’’ appears as the most frequently used keyword, followed by ‘‘aquaponics,’’ ‘‘controlled environment agriculture,’’ ‘‘IoT,’’ and ‘‘lettuce.’’ Table 22 presents the frequency of occurrence for the top five keywords, highlighting their significance in this research domain.

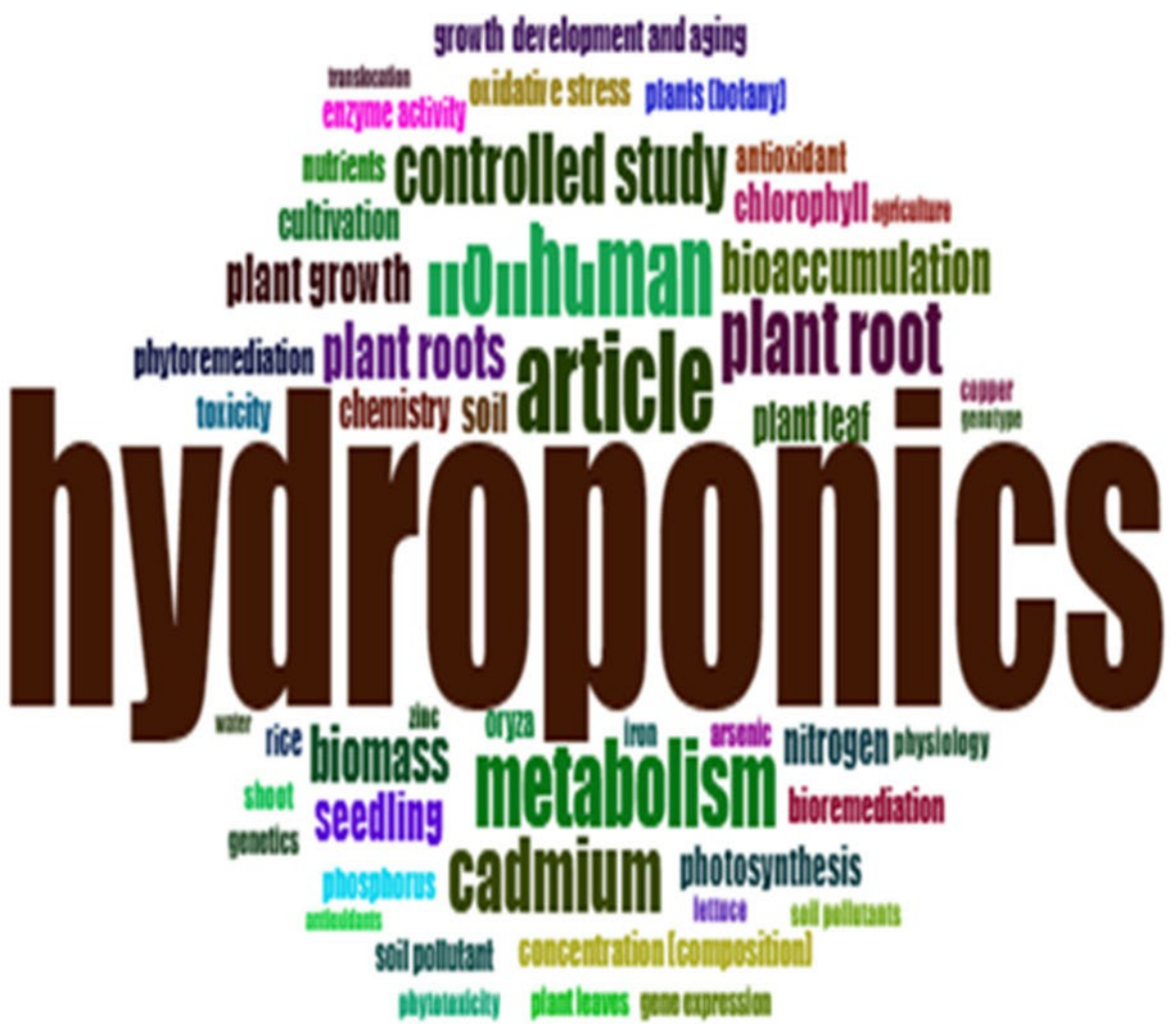

FIGURE 24. Word cloud.

TABLE 22. Frequencies of top 5 authors’ keywords occurring in the word cloud.

| Countries | Frequencies |
|---|---|
| Hydroponics | 1402 |
| Aquaponics | 602 |
| CEA | 221 |
| IoT | 210 |
| Lettuce | 208 |

## VII. DISCUSSIONS

Soilless cultivation methods offer numerous advantages over traditional soil-based farming. When integrated with smart technologies, they enhance various aspects of precision farming, including yield optimization, fertilizer and pest management, moisture regulation, climate adaptation, and irrigation efficiency [14], [15], [16]. Among the three primary soilless farming techniques, such as hydroponics, aeroponics, and aquaponics, hydroponics has been found to be the most efficient. It offers cost-effectiveness, ease of setup and maintenance, plant compatibility, and substrate versatility [80], [87]. With ongoing advancements in CEA-enabled smart farming, research in this field has significantly expanded over the past decade.

A bibliometric analysis was conducted using 7,764 articles retrieved from the Scopus database, filtered based on relevant keywords and limiting factors. The results indicate that 5,964 publications (67.2%) fall under Agricultural and Biological Sciences, with 3,850 articles specifically in this discipline. The most cited research, published in 2014 with DOI 10.2147/NSA.S39406, received 636 citations. The journal Acta Horticulturae was identified as the most productive source, publishing 209 articles by 2023. It also demonstrated a steady growth trend in research contributions. Additionally, Ecotoxicology and Environmental Safety accumulated an impressive 5,575 citations, while PLoS ONE achieved the highest h-index of 404.

In terms of authorship, Wang Y was recognized as the most prolific researcher, with 135 publications in this domain by 2023. The most active international collaboration was between Argentina and Cuba, while Zhejiang University emerged as the leading institution conducting research in CEA-enabled smart precision farming. At the national level, China was identified as the most productive country. It contributed 10,651 research articles from 2013-2023, with the highest SCPs and MCPs. This study integrates systematic literature analysis and bibliometric research to track the evolution of smart farming techniques and assess their impact on CEA-based precision agriculture.

The rapidly growing global population and declining cultivable land pose serious threats to food security. Climate change and the depletion of natural resources further worsen these challenges. In response, IoT-enabled smart precision farming in soilless agriculture has emerged as a sustainable solution. Techniques such as hydroponics, aeroponics, and aquaponics optimize plant growth by automating environmental control, improving resource efficiency, and reducing human intervention. This research explores the opportunities and challenges of IoT-driven soilless farming, emphasizing its role in urban agriculture, vertical farming, and global food security.

The integration of real-time monitoring, automation, and AI-driven decision-making has revolutionized modern agriculture. These advancements enable consistent, high-yield crop production while optimizing resource utilization and minimizing waste. However, challenges such as high initial investment costs, energy consumption, and technological dependency remain significant barriers. The need for specialized expertise also limits the large-scale adoption of these technologies.

## VIII. CONCLUSION AND FUTURE DIRECTIONS

The rapid growth of the global population and the decline in cultivable land pose serious threats to food security. These challenges are further worsened by climate change and shrinking availability of natural resources. In response, IoT-enabled smart precision farming in soilless agriculture has emerged as a sustainable solution. Techniques such as hydroponics, aeroponics, and aquaponics optimize plant growth by automating environmental control, improving resource efficiency, and reducing human intervention. This paper has explored the opportunities and challenges of IoT-driven soilless farming. It highlights its role in urban agriculture, vertical farming, and global food security.

The integration of real-time monitoring, automation, and AI-driven decision-making has transformed modern agriculture. These advancements enable consistent and high-yield crop production throughout the year while optimizing resource utilization and minimizing waste. However, high initial investment costs, energy consumption, and technological dependency remain major obstacles. The need for specialized expertise also limits large-scale adoption. Through an extensive bibliometric analysis, this research has identified global trends, key contributors, and emerging research gaps. The study highlights leading nations and institutions at the forefront of IoT-driven soilless farming. It also underscores the importance of interdisciplinary research, policy support, and technological advancements. These factors will be essential for scaling smart farming solutions globally.

To maximize the impact and scalability of IoT-based soilless agriculture, future research should focus on the following directions. In the near future, we aim to explore low-power IoT sensors to reduce energy consumption. We also plan to integrate renewable energy sources such as solar and wind power to make farming systems more sustainable. Additionally, we intend to explore nutrient recycling systems to minimize waste and improve resource conservation. Furthermore, we seek to implement AI-powered crop monitoring systems for disease detection, yield forecasting, and precision irrigation. Lastly, we aim to utilize edge and fog computing to enable faster data processing and reduce reliance on cloud infrastructure.

## REFERENCES

[1] C. Liu, L. Chen, Z. Li, and D. Wu, ''The impact of digital financial inclusion and urbanization on agricultural mechanization: Evidence from counties of China,'' *PLoS ONE*, vol. 18, no. 11, Nov. 2023, Art. no. e0293910.

[2] A.-L. Balogun, N. Adebisi, I. R. Abubakar, U. L. Dano, and A. Tella, ''Digitalization for transformative urbanization, climate change adaptation, and sustainable farming in Africa: Trend, opportunities, and challenges,'' *J. Integrative Environ. Sci.*, vol. 19, no. 1, pp. 17–37, Dec. 2022.

[3] A. Raihan, D. A. Muhtasim, S. Farhana, M. A. U. Hasan, M. I. Pavel, O. Faruk, M. Rahman, and A. Mahmood, ''Nexus between economic growth, energy use, urbanization, agricultural productivity, and carbon dioxide emissions: New insights from Bangladesh,'' *Energy Nexus*, vol. 8, Dec. 2022, Art. no. 100144.

[4] *United Nations Population | United Nations*. Accessed: Apr. 22, 2024. [Online]. Available: https://www.un.org/en/global-issues/population

[5] L. Rota-Aguilera and M. José, ''A new socio-ecological integrated analysis for agricultural land use planning in metropolitan areas: Applications for tropical (Cali, Colombia) and Mediterranean (Barcelona, Spain) biocultural landscapes,'' Universitat Autònoma de Barcelona, Tech. Rep., 2022.

[6] A. Balkrishna, D. Srivastava, M. Chauhan, G. Sharma, A. Kumar, and V. Arya, ''Insights of agri-food in India: Present trends, challenges and proposed solutions,'' *Indian Ecol. Soc.*, vol. 49, pp. 987–993, Jun. 2022.

[7] C. Seijger and P. Hellegers, ''How do societies reform their agricultural water management towards new priorities for water, agriculture, and the environment?'' *Agricult. Water Manage.*, vol. 277, Mar. 2023, Art. no. 108104.

[8] J. F. V. Muñoz, J. Aznar Sánchez, B. L. Felices, and G. Balacco, ''Adopting sustainable water management practices in agriculture based on stakeholder preferences,'' Departamento de Economía y Empresa, Artículos de revista Dpto. Economía y Empresa, Tech. Rep., 2022. [Online]. Available: http://hdl.handle.net/10835/14897

[9] G. Hurduzeu, R. L. Paˇnzaru, D. M. Medelete, A. Ciobanu, and C. Enea, ''The development of sustainable agriculture in EU countries and the potential achievement of sustainable development goals specific targets (SDG 2),'' *Sustainability*, vol. 14, no. 23, p. 15798, Nov. 2022.

[10] C. M. Viana, D. Freire, P. Abrantes, J. Rocha, and P. Pereira, ''Agricultural land systems importance for supporting food security and sustainable development goals: A systematic review,'' *Sci. Total Environ.*, vol. 806, Feb. 2022, Art. no. 150718.

[11] D. Joshi, A. Nainabasti, R. Bhandari, P. Awasthi, D. Banjade, S. Malla, and B. Subedi, ''A review on soilless cultivation: The hope of urban agriculture,'' *Arch. Agricult. Environ. Sci.*, vol. 7, no. 3, pp. 473–481, Sep. 2022.

[12] S. Oh and C. Lu, ''Vertical farming–smart urban agriculture for enhancing resilience and sustainability in food security,'' *J. Horticultural Sci. Biotechnol.*, vol. 98, no. 2, pp. 133–140, Mar. 2023.

[13] N. Monisha, V. R. Maraiah, M. Jafari, and K. B. Raju, ''An analysis of different smart agricultural system using IoT,'' *Int. J. Hum. Comput. Intell.*, vol. 2, pp. 169–175, Jun. 2023.

[14] D. Tality, A. J. Serevo, A. L. Alipio, and M. A. Rosete, ''Cost-benefit analysis of soilless cultivation system in Tagaytay City, Philippines,'' *Int. J. Social Manage. Stud.*, vol. 3, pp. 140–156, Feb. 2022.

[15] L. Kumar, P. Ahlawat, P. Rajput, R. I. Navsare, and P. K. Singh, ''Internet of Things (IoT) for smart precision farming and agricultural systems productivity: A review,'' *IJEAST*, vol. 5, pp. 141–146, Jan. 2021.

[16] H. Bach and W. Mauser, ''Sustainable agriculture and smart farming,'' in *Earth Observation Open Science and Innovation* (SSI Scientific Report Series). Bern, Switzerland: International Space Science Institute (ISSI), 2018, pp. 261–269.

[17] E. F. I. Raj, M. Appadurai, and K. Athiappan, ''Precision farming in modern agriculture,'' in *Smart Agriculture Automation Using Advanced Technologies: Data Analytics and Machine Learning, Cloud Architecture, Automation and IoT*. Berlin, Germany: Springer, 2022, pp. 61–87.

[18] T. Ayoub Shaikh, T. Rasool, and F. Rasheed Lone, ''Towards leveraging the role of machine learning and artificial intelligence in precision agriculture and smart farming,'' *Comput. Electron. Agricult.*, vol. 198, Jul. 2022, Art. no. 107119.

[19] R. Shofiyati, M. I. Habibie, D. Cahyana, and Z. Susanti, ''Precision farming to achieve sustainable and climate smart agriculture,'' in *Technological Approaches for Climate Smart Agriculture*. Cham, Switzerland: Springer, 2024, pp. 229–248.

[20] S. Khan, S. Satypal, K. Yuvraj, S. Sonia, and S. Bijender, ''Soilless cultivation-hydroponic techniques: A review,'' *Pharma Innov. J.*, vol. 12, no. 6, pp. 6900–6904, 2023.

[21] N. Tzortzakis, S. Nicola, D. Savvas, and W. Voogt, ''Soilless cultivation through an intensive crop production scheme. Management strategies, challenges and future directions,'' *Frontiers Plant Sci.*, vol. 11, p. 363, Apr. 2020.

[22] M. S. Gumisiriza, P. A. Ndakidemi, Z. Nampijja, and E. R. Mbega, ''Soilless urban gardening as a post COVID-19 food security salvage technology: A study on the physiognomic response of lettuce to hydroponics in Uganda,'' *Sci. Afr.*, vol. 20, Jul. 2023, Art. no. e01643.

[23] A. Fussy and J. Papenbrock, ''An overview of soil and soilless cultivation techniques—Chances, challenges and the neglected question of sustainability,'' *Plants*, vol. 11, no. 9, p. 1153, Apr. 2022.

[24] D. Kalaivanan, G. Selvakumar, and A. C. Rathinakumari, ''Soilless cultivation to secure the vegetable demand of urban and peri-urban population,'' in *Soilless Culture*. London, U.K.: IntechOpen, 2022.

[25] P. Sambo, C. Nicoletto, A. Giro, Y. Pii, F. Valentinuzzi, T. Mimmo, P. Lugli, G. Orzes, F. Mazzetto, S. Astolfi, R. Terzano, and S. Cesco, ''Hydroponic solutions for soilless production systems: Issues and opportunities in a smart agriculture perspective,'' *Frontiers Plant Sci.*, vol. 10, p. 923, Jul. 2019.

[26] S. G. Verdoliva, D. Gwyn-Jones, A. Detheridge, and P. Robson, ''Controlled comparisons between soil and hydroponic systems reveal increased water use efficiency and higher lycopene and β-carotene contents in hydroponically grown tomatoes,'' *Scientia Horticulturae*, vol. 279, Mar. 2021, Art. no. 109896.

[27] M. Lykogianni, E. Bempelou, I. Karavidas, C. Anagnostopoulos, K. A. Aliferis, and D. Savvas, ''Impact of sodium hypochlorite applied as nutrient solution disinfectant on growth, nutritional status, yield, and consumer safety of tomato (*Solanum lycopersicum* L.) fruit produced in a soilless cultivation,'' *Horticulturae*, vol. 9, no. 3, p. 352, Mar. 2023.

[28] Z. Siddiqui, D. Hagare, Z.-H. Chen, V. Jayasena, A. A. Shahrivar, O. Panatta, W. Liang, and N. Boyle, ''Growing lettuce and cucumber in a hydroponic system using food waste derived organic liquid fertiliser,'' *Environ. Sustainability*, vol. 5, no. 3, pp. 325–334, Jun. 2022.

[29] H. C. Sousa, G. G. D. Sousa, P. B. C. Cambissa, C. I. N. Lessa, G. F. Goes, F. D. B. D. Silva, F. D. S. Abreu, and T. V. D. A. Viana, ''Gas exchange and growth of zucchini crop subjected to salt and water stress,'' *Revista Brasileira de Engenharia Agrícola e Ambiental*, vol. 26, no. 11, pp. 815–822, Nov. 2022.

[30] L. L. Meena, A. K. Verma, K. K. Krishnani, D. Reang, M. H. Chandrakant, and V. C. John, ''Effects of foliar application of macronutrients (K, P) and micronutrient (Fe) on the growth of okra (*Abelmoschus esculentus* (L.) Moench) and pangasius (*Pangasianodon hypophthalmus*) in a recirculating aquaponic system,'' *South Afr. J. Botany*, vol. 160, pp. 384–393, Sep. 2023.

[31] G. Sajiv, A. Anburani, and D. Venkatakrishnan, ''Study on the growth of eggplant (*Solanum melongena* L.) under hydroponics with modified Hoagland solution,'' *J. Curr. Res. Food Sci.*, vol. 4, pp. 4–6, Jan. 2023.

[32] K. Kusnierek, P. Heltoft, P. J. Møllerhagen, and T. Woznicki, ''Hydroponic potato production in wood fiber for food security,'' *Npj Sci. Food*, vol. 7, no. 1, p. 24, Jun. 2023.

[33] G. Pennisi, D. Gasperi, S. Mancarella, L. V. Antisari, G. Vianello, F. Orsini, and G. Gianquinto, ''Soilless cultivation in urban gardens for reduced potentially toxic elements (PTEs) contamination risk,'' in *Proc. 6th Int. Conf. Landscape Urban Horticulture*, vol. 1189, 2016, pp. 377–382.

[34] H. Can, M. Paksoy, and C. S. Uulu, ''The effects of different solid growing media on onion fresh leaf yields in soilless culture,'' *Proc. Book*, vol. 148, pp. 148–152, Oct. 2020.

[35] M. G. D. Silva, L. F. D. Costa, T. M. Soares, and H. R. Gheyi, ''Growth and yield of cauliflower with brackish waters under hydroponic conditions,'' *Revista Brasileira de Engenharia Agrícola e Ambiental*, vol. 27, no. 9, pp. 663–672, Sep. 2023.

[36] H. Hartley, M. M. Slabbert, and M. Maboko, ''Comparative performance of hydroponically grown broccoli and cauliflower cultivars in different growth seasons,'' in *Proc. 31st Int. Horticultural Congr. (IHC), Int. Symp. Innov. Technol. Production*, vol. 1377, 2022, pp. 671–678.

[37] Y. Liu, Y. Zhu, B. Mu, L. Zong, X. Wang, and A. Wang, ''One-step green construction of granular composite hydrogels for ammonia nitrogen recovery from wastewater for crop growth promotion,'' *Environ. Technol. Innov.*, vol. 33, Feb. 2024, Art. no. 103465.

[38] E. Okudur and Y. Tuzel, ''Effect of EC levels of nutrient solution on glasswort (*Salicornia perennis* Mill.) production in floating system,'' *Horticulturae*, vol. 9, no. 5, p. 555, May 2023.

[39] M. Dutta, D. Gupta, S. Sahu, S. Limkar, P. Singh, A. Mishra, M. Kumar, and R. Mutlu, ''Evaluation of growth responses of lettuce and energy efficiency of the substrate and smart hydroponics cropping system,'' *Sensors*, vol. 23, no. 4, p. 1875, Feb. 2023.

[40] L. Y. Lynn, N. A. M. Amin, M. A. Arif, R. Ibrahim, H. Dzinun, and N. I. M. Ismail, ''Small-scale aquaponics and hydroponics systems: Pak choy and spinach growth rate comparison,'' *Multidisciplinary Appl. Res. Innov.*, vol. 3, no. 1, pp. 22–28, 2022.

[41] M. G. D. Silva, T. M. Soares, H. R. Gheyi, C. C. D. Santos, and M. G. B. D. Oliveira, ''Hydroponic cultivation of coriander intercropped with rocket subjected to saline and thermal stresses in the root-zone,'' *Revista Ceres*, vol. 69, no. 2, pp. 148–157, Apr. 2022.

[42] D. Memiş, G. Tunçelli, M. Tinkir, and M. H. Erk, ''Investigation of different lighting (LED, HPS and FLO) in aquaponics systems for joint production of different plants (lettuce, parsley and cress) and koi carp,'' *Aquatic Res.*, vol. 6, no. 1, pp. 43–51, 2023.

[43] R. Shanmugabhavatharani, A. Sankari, and R. K. Kaleeswari, ''Comparative performance of mint in different hydroponics systems,'' *Madras Agricult. J.*, vol. 109, p. 1, Jan. 2022.

[44] Y. E. Fimbres-Acedo, S. Traversari, S. Cacini, G. Costamagna, M. Ginepro, and D. Massa, ''Testing the effect of high pH and low nutrient concentration on four leafy vegetables in hydroponics,'' *Agronomy*, vol. 13, no. 1, p. 41, Dec. 2022.

[45] H. R. Atherton and P. Li, ''Hydroponic cultivation of medicinal plants—Plant organs and hydroponic systems: Techniques and trends,'' *Horticulturae*, vol. 9, no. 3, p. 349, Mar. 2023.

[46] P. K. Maurya, R. Y. Karde, A. S. Bayskar, N. Charitha, and S. Rai, ''Innovative technologics such as vertical farming and hydroponics to grow crops in controlled environments,'' *Adv. Farming Technol.*, vol. 84, pp. 84–107, Jan. 2023.

[47] C. O. Ossai, S. C. Akpeji, I. S. Alama, and F. N. Emuh, ''Hydroponics production of cucumber as soil farming alternative in Nigeria,'' *Amer. Int. J. Agricult. Stud.*, vol. 6, pp. 8–11, Jun. 2022.

[48] A. K. Jurayev, U. A. Jurayev, B. N. Atamurodov, K. S. Sobirov, and M. M. Najmiddinov, ''Growing tomatoes hydroponically in greenhouses,'' *Sci. Innov.*, vol. 1, no. 2, pp. 87–90, 2022.

[49] F. Behtash, H. S. Hajizadeh, and B. Tarighi, ''Modulation of nutritional and biochemical status of hydroponically grown *Cucurbita pepo* L. by calcium nitrate under saline conditions,'' *Horticultural Sci.*, vol. 50, no. 2, pp. 127–141, Jun. 2023.

[50] G. Hoza, M. Dinu, A. Becherescu, and A. Ionica, ''The impact of foliar fertilizers on the production of field-grown zucchini,'' *Sci. Papers Ser. B, Horticulture*, vol. 66, Jan. 2022.

[51] P. F. N. Sousa, M. V. Dantas, G. S. Lima, L. A. A. Soares, H. R. Gheyi, L. A. Silva, K. P. Lopes, and P. D. Fernandes, ''Hydroponic cultivation of okra using saline nutrition solutions under application of salicylic acid,'' *Revista Caatinga*, vol. 36, no. 4, pp. 916–928, 2023.

[52] N. A. Wahocho, M. F. Jamali, N.-U.-N. Memon, W. Ahmad, K. H. Talpur, P. A. Shar, A. Talpur, S. A. Otho, and F. A. Jamali, ''Hydro-priming durations improve the germination and vegetative growth of okra (abelmoschus esculentus var. Sabz pari),'' *Pakistan J. Biotechnol.*, vol. 20, no. 2, pp. 293–300, Nov. 2023.

[53] S. Karak, A. Kundu, and U. Thapa, ''Growth and yield of potato (*Solanum tuberosum* L.) as influenced by biostimulant under soilless culture system,'' Dept. Agronomy, Fac. Agricult., BCKV, West Bengal, India, Tech. Rep., pp. 2317–2320, 2023, vol. 12, no. 3.

[54] P. Pathania, V. Katoch, A. Sandal, and N. Sharma, ''Role of substrate media in growth and development of selected microgreens,'' *Biol. Forum-Int. J.*, vol. 14, pp. 1357–1361, Jan. 2022.

[55] B. Singh, ''New systems of vegetable production: Protected cultivation, hydroponics, aeroponics, vertical, organic, microgreens,'' in *Vegetables for Nutrition and Entrepreneurship*. Berlin, Germany: Springer, pp. 31–56.

[56] D. Saxena and G. Stotzky, ''Bt toxin is not taken up from soil or hydroponic culture by corn, carrot, radish, or turnip,'' *Plant Soil*, vol. 239, pp. 165–172, Feb. 2002.

[57] A. Akter, M. F. Hossain, J.-A. Mahmud, N. Sultana, M. M. Hasan, and T. Begum, ''Cauliflower cultivation on rooftop garden using various composts: Morphological and economic analysis,'' *J. Agroforestry Environ.*, vol. 16, no. 1, pp. 145–159, Jul. 2023.

[58] A. V. Singh, V. B. Rajwade, S. E. Topno, and A. Kerketta, ''Effect of different levels of liquid nitrogen, phosphorus and potassium on growth, yield and quality of broccoli (*Brassica oleraceae* var. Italica) in hydroponic,'' *Int. J. Environ. Climate Change*, vol. 13, no. 9, pp. 1921–1927, Jul. 2023.

[59] S. Kathi, H. Laza, S. Singh, L. Thompson, W. Li, and C. Simpson, ''Vitamin C biofortification of broccoli microgreens and resulting effects on nutrient composition,'' *Frontiers Plant Sci.*, vol. 14, Mar. 2023, Art. no. 1145992.

[60] A. Sharma, H. Pandey, V. S. Devadas, B. D. Kartha, and A. Vashishth, ''Phytoremediation, stress tolerance and bio fortification in crops through soilless culture,'' *Crop Design*, vol. 2, no. 1, Jun. 2023, Art. no. 100027.

[61] T. Hardiyanto, R. Riswana, and A. Arneta, ''Feasibility analysis and break even point celery farming (Apium graveolens L.) hydroponic deep flow technique (DFT) system (Case study on a farmer in Kayuambon Village, Lembang District, West Bandung Regency),'' *Int. J. Quant. Res. Model.*, vol. 4, no. 2, pp. 65–70, Jun. 2023.

[62] W. Liu, M. Muzolf-Panek, and T. Kleiber, ''Effect of varied nitrogen sources and type of cultivation on the yield and physicochemical parameters of flowering Chinese cabbage (*Brassica campestris* L. ssp. chinensis var. utilis Tsen et Lee),'' *Appl. Sci.*, vol. 13, no. 9, p. 5691, May 2023.

[63] A. L. Dela Cruz, C. P. Sinco, M. L. S. Cantor, M. T. Viña, J. R. G. Bolaños, and R. J. B. Bordios, ''Effects of varying nutrient solution to brassica rapa 'Bokchoy' grow under hydroponic system,'' *Amer. J. Environ. Climate*, vol. 1, no. 2, pp. 31–37, Aug. 2022.

[64] A. Mourantian, M. Aslanidou, E. Mente, N. Katsoulas, and E. Levizou, ''Basil functional and growth responses when cultivated via different aquaponic and hydroponics systems,'' *PeerJ*, vol. 11, Jul. 2023, Art. no. e15664.

[65] F. B. de Oliveiraa, ''A typology review for vertical plant farms: Classifications, configurations, business models and economic analyses,'' *Unpublished Manuscript*, vol. 2, p. 2, Oct. 2022.

[66] S. Caputo, ''History, techniques and technologies of soil-less cultivation,'' *Small Scale Soil-Less Urban Agricult. Eur.*, vol. 1, pp. 45–86, Jun. 2022.

[67] J. Garzón, L. Montes, J. Garzón, and G. Lampropoulos, ''Systematic review of technology in aeroponics: Introducing the technology adoption and integration in sustainable agriculture model,'' *Agronomy*, vol. 13, no. 10, p. 2517, Sep. 2023.

[68] M. Suresh, R. Singh, and A. Gehlot, ''Automated aeroponics system using IoT for smart agriculture,'' in *Internet of Things for Agriculture 4.0: Impact and Challenges*. Boca Raton, FL, USA: CRC Press, 2022, pp. 207–239.

[69] G. O. Korter, F. O. Ehiagwina, S. K. Kolawole, A. S. Awe, T. E. Akinola, and P. O. Lawal, ''Development of aeroponics stands using locally sourced materials,'' *Iconic Res. Eng. J.*, vol. 7, pp. 19–23, Jan. 2023.

[70] R. S. Velazquez-Gonzalez, A. L. Garcia-Garcia, E. Ventura-Zapata, J. D. O. Barceinas-Sanchez, and J. C. Sosa-Savedra, ''A review on hydroponics and the technologies associated for medium- and small-scale operations,'' *Agriculture*, vol. 12, no. 5, p. 646, Apr. 2022.

[71] K. Monisha, H. K. Selvi, P. Sivanandhini, A. S. Nachammai, C. T. Anuradha, S. R. Devi, A. K. Sri, N. R. Neya, M. Vaitheeswari, and G. S. Hikku, ''Hydroponics agriculture as a modern agriculture technique,'' *J. Achievements Mater. Manuf. Eng.*, vol. 116, no. 1, pp. 25–35, Jan. 2023.

[72] S. R. Sathyanarayana, W. V. Gangadhar, M. G. Badrinath, R. M. Ravindra, and A. U. Shriramrao, ''Hydroponics: An intensified agriculture practice to improve food production,'' *Rev. Agricult. Sci.*, vol. 10, pp. 101–114, Jan. 2022.

[73] D. I. Pomoni, M. K. Koukou, M. G. Vrachopoulos, and L. Vasiliadis, ''A review of hydroponics and conventional agriculture based on energy and water consumption, environmental impact, and land use,'' *Energies*, vol. 16, p. 1690, Jan. 2023.

[74] E. G. Ikrang, P. O. Ehiomogue, and U. I. Udoumoh, ''Hydroponics in precision agriculture—A review,'' *Ann. Fac. Eng. Hunedoara*, vol. 20, pp. 143–148, May 2022.

[75] G. F. M. Baganz, R. Junge, M. C. Portella, S. Goddek, K. J. Keesman, D. Baganz, G. Staaks, C. Shaw, F. Lohrberg, and W. Kloas, ''The aquaponic principle—It is all about coupling,'' *Rev. Aquaculture*, vol. 14, no. 1, pp. 252–264, Jan. 2022.

[76] L. H. David, S. M. Pinho, F. Agostinho, J. I. Costa, M. C. Portella, K. J. Keesman, and F. Garcia, ''Sustainability of urban aquaponics farms: An emergy point of view,'' *J. Cleaner Prod.*, vol. 331, Jan. 2022, Art. no. 129896.

[77] Z. Broćić, J. Oljača, D. Pantelić, J. Rudić, and I. Momčilović, ''Potato aeroponics: Effects of cultivar and plant origin on minituber production,'' *Horticulturae*, vol. 8, no. 10, p. 915, Oct. 2022.

[78] J. B. Silva Filho, P. C. R. Fontes, J. F. S. Ferreira, P. R. Cecon, and E. Crutchfield, ''Optimal nutrient solution and dose for the yield of nuclear seed potatoes under aeroponics,'' *Agronomy*, vol. 12, no. 11, p. 2820, Nov. 2022.

[79] M. Paponov, J. Ziegler, and I. A. Paponov, ''Light exposure of roots in aeroponics enhances the accumulation of phytochemicals in aboveground parts of the medicinal plants artemisia annua and hypericum perforatum,'' *Frontiers Plant Sci.*, vol. 14, Jan. 2023, Art. no. 1079656.

[80] B. Ghosh, S. Roy, N. Ahmed, and D. De, ''Dew aeroponics: Dew-enabled smart aeroponics system in agriculture 4.0,'' in *Dew Computing: The Sustainable IoT Perspectives*. Berlin, Germany: Springer, 2023, pp. 261–287.

[81] N. Sadek, N. Kamal, and D. Shehata, ''Internet of Things based smart automated indoor hydroponics and aeroponics greenhouse in Egypt,'' *Ain Shams Eng. J.*, vol. 15, no. 2, Feb. 2024, Art. no. 102341.

[82] B. Fasciolo, A. Awouda, G. Bruno, and F. Lombardi, ''A smart aeroponic system for sustainable indoor farming,'' *Proc. CIRP*, vol. 116, pp. 636–641, Jan. 2023.

[83] H. A. Méndez-Guzmán, J. A. Padilla-Medina, C. Martínez-Nolasco, J. J. Martinez-Nolasco, A. I. Barranco-Gutiérrez, L. M. Contreras-Medina, and M. Leon-Rodriguez, ''IoT-based monitoring system applied to aeroponics greenhouse,'' *Sensors*, vol. 22, no. 15, p. 5646, Jul. 2022.

[84] S. Setiowati, R. N. Wardhani, E. B. A. Siregar, R. Saputra, and R. A. Sabrina, ''Fertigation control system on smart aeroponics using Sugeno's fuzzy logic method,'' in *Proc. 8th Int. Conf. Sci. Technol. (ICST)*, Sep. 2022, pp. 1–6.

[85] G. Rajendiran and J. Rethnaraj, ''Smart Aeroponic farming system: Using IoT with LCGM-boost regression model for monitoring and predicting lettuce crop yield,'' *Int. J. Intell. Eng. Syst.*, vol. 16, pp. 251–262, Sep. 2023.

[86] V. Kumara, T. A. Mohanaprakash, S. Fairooz, K. Jamal, T. Babu, and B. Sampath, ''Experimental study on a reliable smart hydroponics system,'' in *Human Agro-Energy Optimization for Bus. and Industry*. Hershey, PA, USA: IGI Global, 2023, pp. 27–45.

[87] J. D. Stevens, D. Murray, D. Diepeveen, and D. Toohey, ''Development and testing of an IoT spectroscopic nutrient monitoring system for use in micro indoor smart hydroponics,'' *Horticulturae*, vol. 9, no. 2, p. 185, Feb. 2023.

[88] Y. L. K. Macayana, I. C. Fernandez, K. Ung, I. Austria, M. T. De Leon, C. V. Densing, J. J. Eslit, P. Magpantay, C. Miras, D. Ong, C. Santos, M. Talampas, N. M. Tiglao, and M. Rosales, ''Internet of Things-based indoor smart hydroponics farm monitoring system,'' *Int. J. Electr. Comput. Eng. (IJECE)*, vol. 13, no. 2, p. 2326, Apr. 2023.

[89] K. E. Lakshmiprabha and C. Govindaraju, ''Hydroponic-based smart irrigation system using Internet of Things,'' *Int. J. Commun. Syst.*, vol. 36, no. 12, p. e4071, Aug. 2023.

[90] V. Mamatha and J. C. Kavitha, ''Remotely monitored web based smart hydroponics system for crop yield prediction using IoT,'' in *Proc. IEEE 8th Int. Conf. Converg. Technol. (ICT)*, Apr. 2023, pp. 1–6.

[91] M. Venkatraman and R. Surendran, ''Design and implementation of smart hydroponics farming for growing lettuce plantation under nutrient film technology,'' in *Proc. 2nd Int. Conf. Appl. Artif. Intell. Comput. (ICAAIC)*, May 2023, pp. 1514–1521.

[92] V. Mamatha and J. C. Kavitha, ''Machine learning based crop growth management in greenhouse environment using hydroponics farming techniques,'' *Meas., Sensors*, vol. 25, Feb. 2023, Art. no. 100665.

[93] G. V. Danush Ranganath, R. Hari Sri Rameasvar, and A. Karthikeyan, ''Smart hydroponics system for soilless farming based on Internet of Things,'' in *Smart Technologies in Data Science and Communication: Proceedings of SMART-DSC*. Berlin, Germany: Springer, 2023, pp. 271–280.

[94] M. K. Shukla, A. Kanwar, S. Raul, A. V. Sai, and B. Verma, ''IoT based monitoring of environment in a smart hydroponic system,'' in *Proc. IEEE Devices Integr. Circuit (DevIC)*, Apr. 2023, pp. 257–260.

[95] L. S. Kondaka, R. Iyer, S. Jaiswal, and A. Ali, ''A smart hydroponic farming system using machine learning,'' in *Proc. Int. Conf. Intell. Innov. Technol. Comput. Elect. Electron. (IITCEE)*, 2023, pp. 357–362.

[96] J. Mehare and A. Gaikwad, ''Secured framework for smart farming in hydroponics with intelligent and precise management for IoT with blockchain technology,'' *J. Data Acquisition Process.*, vol. 38, no. 2, p. 127, 2023.

[97] B. Basumatary, A. K. Verma, and M. K. Verma, ''Global research trends on aquaponics: A systematic review based on computational mapping,'' *Aquaculture Int.*, vol. 31, no. 2, pp. 1115–1141, Apr. 2023.

[98] V. T. Okomoda, S. A. Oladimeji, S. G. Solomon, S. O. Olufeagba, S. I. Ogah, and M. Ikhwanuddin, ''Aquaponics production system: A review of historical perspective, opportunities, and challenges of its adoption,'' *Food Sci. Nutrition*, vol. 11, no. 3, pp. 1157–1165, Mar. 2023.

[99] B. Nemade and D. Shah, ''An IoT-based efficient water quality prediction system for aquaponics farming,'' in *Computational Intelligence*. Berlin, Germany: Springer, 2023, pp. 311–323.

[100] A. H. Eneh, C. N. Udanor, N. I. Ossai, S. O. Aneke, P. O. Ugwoke, A. A. Obayi, C. H. Ugwuishiwu, and G. E. Okereke, ''Towards an improved Internet of Things sensors data quality for a smart aquaponics system yield prediction,'' *MethodsX*, vol. 11, Dec. 2023, Art. no. 102436.

[101] T. B. Pramono, N. I. Qothrunnada, F. Asadi, T. W. Cenggoro, and B. Pardamean, ''Water quality monitoring system for aquaponic technology using the Internet of Things (IoT),'' *Commun. Math. Biol. Neurosci.*, vol. 2023, p. 120, Jun. 2023.

[102] A. Siddique, J. Sun, K. J. Hou, M. I. Vai, S. H. Pun, and M. A. Iqbal, ''SpikoPoniC: A low-cost spiking neuromorphic computer for smart aquaponics,'' *Agriculture*, vol. 13, no. 11, p. 2057, Oct. 2023.

[103] W. P. H. Ubayasena, N. S. C. Hemantha, and W. Wijayakoon, ''Smart aquaponics system,'' *Int. Res. J. Innov. Eng. Technol.*, vol. 7, p. 693, Jan. 2023.

[104] C. J. E. Egnalig, O. M. Jamero, A. P. D. Tampong, R. R. Bacarro, F. A. M. Dumaguit, and L. A. R. Cañete, ''Smart aquaponics system for oreochromis niloticus production,'' *Adv. Res.*, vol. 24, no. 5, pp. 100–123, Jun. 2023.

[105] D. Nishanth, M. H. S. Alshamsi, A. M. K. A. Alkaabi, A. H. A. AlKaabi, S. H. K. Alnuaimi, C. S. Nair, and A. Jaleel, ''Aquaponics as a climate-smart technology for sustainable food production: A comparison with conventional production system in united Arab Emirates,'' *J. World Aquaculture Soc.*, vol. 55, no. 2, Apr. 2024, Art. no. e13049.

[106] P. G. Arepalli and K. J. Naik, ''Water contamination analysis in IoT enabled aquaculture using deep learning based AODEGRU,'' *Ecol. Informat.*, vol. 79, Mar. 2024, Art. no. 102405.

[107] M. Ula, M. Muliani, and H. A. K. Aidilof, ''Optimization of water quality in shrimp-shallot aquaponic systems: A machine learning-integrated IoT approach,'' *Int. J. Intell. Syst. Appl. Eng.*, vol. 12, pp. 480–491, Jan. 2024.

[108] T. Divya, R. Ramyadevi, and V. V. Chamundeeswari, ''Smart fish monitoring system using IoT,'' in *Artificial Intelligence, Blockchain, Computing and Security*, vol. 2. Boca Raton, FL, USA: CRC Press, 2024, pp. 270–273.

[109] E. Fathidarehnijeh, M. Nadeem, M. Cheema, R. Thomas, M. Krishnapillai, and L. Galagedara, ''Current perspective on nutrient solution management strategies to improve the nutrient and water use efficiency in hydroponic systems,'' *Can. J. Plant Sci.*, vol. 104, no. 2, pp. 88–102, Apr. 2024.

[110] F. P. Karagöz, A. Dursun, and M. Karaşal, ''A review: Use of soilless culture techniques in ornamental plants,'' *Ornamental Horticulture*, vol. 28, no. 2, pp. 172–180, Apr. 2022.

[111] C. Bihari, S. Ahamad, M. Kumar, A. Kumar, A. D. Kamboj, S. Singh, V. Srivastava, and P. Gautam, ''Innovative soilless culture techniques for horticultural crops: A comprehensive review,'' *Int. J. Environ. Climate Change*, vol. 13, no. 10, pp. 4071–4084, Sep. 2023.

[112] D. Prisa and S. Caro, ''Alternative substrates in the cultivation of ornamental and vegetable plants,'' *GSC Biol. Pharmaceutical Sci.*, vol. 24, no. 1, pp. 209–220, Jul. 2023.

[113] A. Verma, ''Soilless cultivation techniques in protected structures,'' in *Protected Cultivation*. Milton, U.K.: Academic Press, 2024, pp. 91–112.

[114] M. I. Majdalawi, A. A. Ghanayem, A. A. Alassaf, S. Schlüter, M. A. Tabieh, A. Z. Salman, M. W. Akash, and R. C. Pedroso, ''Economic efficient use of soilless techniques to maximize benefits for farmers,'' *AIMS Agricult. Food*, vol. 8, no. 4, pp. 1038–1051, 2023.

[115] P. B. Kadiog˘ lu, ''Soil regulators and soilless agricultural techniques,'' in *A Sustainable Increase in Food Security Requires Agricultural Biodiversity*, vol. 71. Turkey: Eastern Anatolia Agricultural Research Institute Directorate, 2023.

[116] D. Savvas, G. Gianquinto, Y. Tuzel, and N. Gruda, ''Soilless culture,'' in *Good Agricultural Practices for Greenhouse Vegetable Crops*, vol. 303. Rome, Italy: Food and Agriculture Organization of the United Nations (FAO-UN), 2013.

[117] F. F. Montesano, A. Parente, F. Grassi, and P. Santamaria, ''Posidonia-based compost as a growing medium for the soilless cultivation of tomato,'' *Acta Horticulturae*, vol. 1034, pp. 277–282, May 2014.

[118] P. Mazuela, M. Urrestarazu, and E. Bastias, ''Vegetable waste compost used as substrate in soilless culture,'' in *Crop Production Technologies*. Chile, 2012, pp. 179–198.

[119] B. Vandecasteele, J. Debode, K. Willekens, and T. Van Delm, ''Recycling of P and K in circular horticulture through compost application in sustainable growing media for fertigated strawberry cultivation,'' *Eur. J. Agronomy*, vol. 96, pp. 131–145, May 2018.

[120] N. Kitir, E. Yildirim, Ü. Sahin, M. Turan, M. Ekinci, S. Ors, R. Kul, H. Ünlü, and H Ünlü, *Peat Use in Horticulture*, B. Topcuoglu and M. Turan, Eds., London, U.K.: IntechOpen, 2018, pp. 75–90.

[121] Y. Amha, H. Bohne, G. Schmilewski, P. Picken, and O. Reinikainen, ''Microbial activity of ten horticultural peats un-der different incubation conditions,'' in *Proc. Int. Symp. Growing Media Composting*, vol. 891, 2009, pp. 33–39.

[122] C. Depardieu, V. Prémont, C. Boily, and J. Caron, ''Sawdust and bark-based substrates for soilless strawberry production: Irrigation and electrical conductivity management,'' *PLoS ONE*, vol. 11, no. 4, Apr. 2016, Art. no. e0154104.

[123] L. Sabatino, ''Increasing sustainability of growing media constituents and stand-alone substrates in soilless culture systems—An editorial,'' *Agronomy*, vol. 10, no. 9, p. 1384, Sep. 2020.

[124] C. C. Wolcott, ''Characterizing the physical and hydraulic properties of pine bark soilless substrates,'' Virginia Polytech. Inst. State Univ., Tech. Rep., 2023.

[125] M. J. Maher and D. Thomson, ''Growth and manganese content of tomato (*Lycopersicon esculentum*) seedlings grown in sitka spruce (*Picea sitchensis* (Bong.) Carr.) bark substrate,'' *Scientia Horticulturae*, vol. 48, nos. 3–4, pp. 223–231, Nov. 1991.

[126] A. Nerlich and D. Dannehl, ''Soilless cultivation: Dynamically changing chemical properties and physical conditions of organic substrates influence the plant phenotype of lettuce,'' *Frontiers Plant Sci.*, vol. 11, Jan. 2021, Art. no. 601455.

[127] C. B. An and J. H. Shin, ''Comparion of rockwool, reused rockwool and coir medium on tomato (*Solanum lycopersicum*) growth, fruit quality and productivity in greenhouse soilless culture,'' *J. Bio-Environment Control*, vol. 30, no. 3, pp. 175–182, Jul. 2021.

[128] R. M. Giurgiu, A. Dumitraş, G. Morar, P. Scheewe, and F. G. Schröder, ''A study on the biological control of fusarium oxysporum using trichoderma spp., on soil and rockwool substrates in controlled environment,'' *Notulae Botanicae Horti Agrobotanici Cluj-Napoca*, vol. 46, no. 1, pp. 260–269, Jan. 2018.

[129] M. Sharaf-Eldin, F. El-Aidy, N. Hasssan, A. Masoud, and E. El-Khateeb, ''Comparison between soil and soilless cultivation of autumn tomato production under Spanish net-house conditions,'' *Egyptian J. Soil Sci.*, vol. 63, no. 4, Dec. 2023.

[130] D. M. Papadimitriou, I. N. Daliakopoulos, I. Louloudakis, T. I. Savvidis, I. Sabathianakis, D. Savvas, and T. Manios, ''Impact of container geometry and hydraulic properties of coir dust, perlite, and their blends used as growing media, on growth, photosynthesis, and yield of golden thistle (S. hispanicus L.),'' *Scientia Horticulturae*, vol. 323, Jan. 2024, Art. no. 112425.

[131] A. Hassan, E. Abou El-Salehein, M. El Hamady, and M. Sobh, ''Effect of different substrate media and irrigation on flowering and production of strawberry (Fragaria spp),'' *J. Productiv. Develop.*, vol. 26, no. 4, pp. 1053–1069, Oct. 2021.

[132] J. M. Kovačič, T. Ciringer, J. Ambrožič-Dolinşek, and S. Kovačič, ''Use of emulsion-templated, highly porous polyelectrolytes for in vitro germination of chickpea embryos: A new substrate for soilless cultivation,'' *Biomacromolecules*, vol. 23, no. 8, pp. 3452–3457, Aug. 2022.

[133] R. Zabaleta, E. Sánchez, P. Fabani, G. Mazza, and R. Rodriguez, ''Almond shell biochar: Characterization and application in soilless cultivation of eruca sativa,'' *Biomass Convers. Biorefinery*, vol. 14, no. 15, pp. 18183–18200, Aug. 2024.

[134] K. Frączek, K. Bulski, and T. Zaleski, ''The effect of silicon and calcium additives on the growth of selected groups of microorganisms in substrate used in soilless cultivation of strawberries,'' *Acta Scientiarum Polonorum Hortorum Cultus*, vol. 21, no. 4, pp. 53–66, Aug. 2022.

[135] A. Gelsomino, B. Petrovičová, and M. R. Panuccio, ''Exhausted fire-extinguishing powders: A potential source of mineral nutrients for reuse and valorisation in compost enrichment for soilless cultivation,'' *Sci. Total Environ.*, vol. 906, Jan. 2024, Art. no. 167633.

[136] R. Abd Ghani, S. Omar, M. Jolánkai, Á. Tarnawa, Z. Kende, N. Khalid, C. Gyuricza, and M. K. Kassai, ''Soilless culture applications for early development of soybean crop (*Glycine max* l. Merr),'' *Agriculture*, vol. 13, no. 9, p. 1713, Aug. 2023.

[137] D. Savvas and N. Gruda, ''Application of soilless culture technologies in the modern greenhouse industry—A review,'' *Eur. J. Horticultural Sci.*, vol. 83, no. 5, pp. 280–293, Nov. 2018.

[138] D. Zhang, Q. Peng, R. Yang, W. Lin, H. Wang, W. Zhou, Z. Qi, and L. Ouyang, ''Slight carbonization as a new approach to obtain peat alternative,'' *Ind. Crops Products*, vol. 202, Oct. 2023, Art. no. 117041.

[139] H. Kiani, N. Khakipour, S. K. Jari, and S. S. Sar, ''Introduction of a suitable cultivation substrate for the optimal growth of the ornamental plant *Crassula capitella*,'' Dept. Horticultural Sci., Sci. Res. Branch, Islamic Azad Univ., Tehran, Iran, Tech. Rep. 20230210503, 2023.

[140] M. Dutta, D. Gupta, Y. Javed, K. Mohiuddin, S. Juneja, Z. I. Khan, and A. Nauman, ''Monitoring root and shoot characteristics for the sustainable growth of barley using an IoT-enabled hydroponic system and AquaCrop simulator,'' *Sustainability*, vol. 15, no. 5, p. 4396, Mar. 2023.

[141] M. H. Saleem, J. Afzal, M. Rizwan, Z.-U.-H. Shah, N. Depar, and K. Usman, ''Chromium toxicity in plants: Consequences on growth, chromosomal behavior andmineral nutrient status,'' *Turkish J. Agricult. Forestry*, vol. 46, no. 3, pp. 371–389, Jan. 2022.

[142] S. B. Dhal, M. Bagavathiannan, U. Braga-Neto, and S. Kalafatis, ''Nutrient optimization for plant growth in aquaponic irrigation using machine learning for small training datasets,'' *Artif. Intell. Agricult.*, vol. 6, pp. 68–76, Jan. 2022.

[143] S. Ragaveena, A. Shirly Edward, and U. Surendran, ''Smart controlled environment agriculture methods: A holistic review,'' *Rev. Environ. Sci. Bio/Technol.*, vol. 20, no. 4, pp. 887–913, Dec. 2021.

[144] X. Guan, X. Wang, B. Liu, C. Wu, C. Liu, D. Liu, C. Zou, and X. Chen, ''Magnesium supply regulate leaf nutrition and plant growth of soilless cultured cherry tomato-interaction with potassium,'' China Agricult. Univ., Southwest Univ., Tech. Rep., 2020, doi: 10.21203/rs.3.rs-15977/v1.

[145] M. Tavan, B. Wee, S. Fuentes, A. Pang, G. Brodie, C. G. Viejo, and D. Gupta, ''Biofortification of kale microgreens with selenate-selenium using two delivery methods: Selenium-rich soilless medium and foliar application,'' *Scientia Horticulturae*, vol. 323, Jan. 2024, Art. no. 112522.

[146] T. Barbagli, J. van Ruijven, W. Voogt, and A. van Winkel, ''Soilless USDA-organic cultivation of tomato with 'natural nitrogen': A comparison study between 'Natural nitrogen' and organic nitrogen,'' Wageningen Univ. Res., Tech. Rep. 596369, 2022.

[147] D. Sausen, I. R. Carvalho, F. A. Neis, S. C. Uliana, R. B. Mambrin, R. Schwalbert, M. D. S. Tavares, and F. T. Nisoloso, ''Time of potato plant harvest in soilless cultivation for screening phosphorus use efficient genotypes,'' *Commun. Plant Sci.*, vol. 12, no. 2022, pp. 1–6, 2022.

[148] M. Tavan, B. Wee, G. Brodie, S. Fuentes, A. Pang, and D. Gupta, ''Optimizing sensor-based irrigation management in a soilless vertical farm for growing microgreens,'' *Frontiers Sustain. Food Syst.*, vol. 4, Jan. 2021, Art. no. 622720.

[149] A. Sagheer, M. Mohammed, K. Riad, and M. Alhajhoj, ''A cloud-based IoT platform for precision control of soilless greenhouse cultivation,'' *Sensors*, vol. 21, no. 1, p. 223, Dec. 2020.

[150] E. M. B. M. Karunathilake, A. T. Le, S. Heo, Y. S. Chung, and S. Mansoor, ''The path to smart farming: Innovations and opportunities in precision agriculture,'' *Agriculture*, vol. 13, no. 8, p. 1593, Aug. 2023.

[151] D. K. Kwaghtyo and C. I. Eke, ''Smart farming prediction models for precision agriculture: A comprehensive survey,'' *Artif. Intell. Rev.*, vol. 56, no. 6, pp. 5729–5772, Jun. 2023.

[152] C. D. Singh, K. V. Rao, M. Kumar, and Y. A. Rajwade, ''Development of a smart IoT-based drip irrigation system for precision farming,'' *Irrigation Drainage*, vol. 72, no. 1, pp. 21–37, 2023.

[153] A. Rokade, M. Singh, P. K. Malik, R. Singh, and T. Alsuwian, ''Intelligent data analytics framework for precision farming using IoT and regressor machine learning algorithms,'' *Appl. Sci.*, vol. 12, no. 19, p. 9992, Oct. 2022.

[154] S. K. S. Durai and M. D. Shamili, ''Smart farming using machine learning and deep learning techniques,'' *Decis. Analytics J.*, vol. 3, Jun. 2022, Art. no. 100041.

[155] A. D. Boursianis, M. S. Papadopoulou, P. Diamantoulakis, A. Liopa-Tsakalidi, P. Barouchas, G. Salahas, G. Karagiannidis, S. Wan, and S. K. Goudos, ''Internet of Things (IoT) and agricultural unmanned aerial vehicles (UAVs) in smart farming: A comprehensive review,'' *Internet Things*, vol. 18, May 2022, Art. no. 100187.

[156] M. Dhanaraju, P. Chenniappan, K. Ramalingam, S. Pazhanivelan, and R. Kaliaperumal, ''Smart farming: Internet of Things (IoT)-based sustainable agriculture,'' *Agriculture*, vol. 12, no. 10, p. 1745, Oct. 2022.

[157] I. Bhakta, S. Phadikar, and K. Majumder, ''State-of-the-art technologies in precision agriculture: A systematic review,'' *J. Sci. Food Agricult.*, vol. 99, no. 11, pp. 4878–4888, Aug. 2019.

[158] E. S. Mohamed, A. A. Belal, S. K. Abd-Elmabod, M. A. El-Shirbeny, A. Gad, and M. B. Zahran, ''Smart farming for improving agricultural management,'' *Egyptian J. Remote Sens. Space Sci.*, vol. 24, pp. 971–981, Dec. 2021.

[159] N. Yaqin, A. F. Zuhri, and T. Hariono, ''Automatic control of hydroponic plant ph levels using sensor sku sen 0161,'' *NEWTON, Netw. Inf. Technol.*, vol. 1, no. 3, pp. 125–131, Feb. 2022.

[160] M. Safira, M. W. Lim, and W. S. Chua, ''Design of control system for water quality monitoring system for hydroponics application,'' *IOP Conf. Ser., Mater. Sci. Eng.*, vol. 1257, Oct. 2022, Art. no. 012027.

[161] A. Sidibe, R. Ndeda, E. Murimi, and U. N. Mutwiwa, ''Implementation of a microcontroller-based neural network for prediction of pH and EC levels in hydroponics,'' *J. Sustain. Res. Eng.*, vol. 7, pp. 106–117, Aug. 2023.

[162] D. Simanjuntak and L. Hakim, ''Design and control system of temperature and water level in hydroponic plants,'' *J. Phys., Conf.*, vol. 2193, no. 1, Feb. 2022, Art. no. 012018.

[163] A. Abu Sneineh and A. A. A. Shabaneh, ''Design of a smart hydroponics monitoring system using an ESP32 microcontroller and the Internet of Things,'' *MethodsX*, vol. 11, Dec. 2023, Art. no. 102401.

[164] A. Z. M. Noor, M. S. Dzulkafle, N. M. Bohri, M. H. M. Khaireez, S. S. Jupri, and M. H. F. M. Fauadi, ''Development of smart agriculture (smart hydroponic) to monitor soil humidity level,'' in *IT Applications for Sustainable Living*. Berlin, Germany: Springer, 2023, pp. 25–34.

[165] Y. V. Bhargava, P. K. Chittoor, and C. Bharatiraja, ''IoT powered smart hydroponic system for autonomous irrigation and crop monitoring,'' *ECS Trans.*, vol. 107, no. 1, pp. 11937–11944, Apr. 2022.

[166] M. Niswar, ''Design and implementation of an automated indoor hydroponic farming system based on the Internet of Things,'' *Int. J. Comput. Digit. Syst.*, vol. 15, no. 1, pp. 337–346, Jan. 2024.

[167] F. Fadillah, R. Buaton, and S. Ramadani, ''IoT-based hydroponic plant monitoring system,'' *J. Artif. Intell. Eng. Appl. (JAIEA)*, vol. 3, no. 1, pp. 38–43, Oct. 2023.

[168] S. Arnett-Robinson and R. Tatro, ''Smart hydroponic garden,'' College Eng. Comput. Sci., Tech. Rep. EEE 193B/CPE 191, 2022.

[169] S. Kousar, S. Shahid, A. Ullah, A. G. Tabsum, and M. Tubassam, ''Monitoring and control system for precision agriculture using wireless sensor network,'' *Asian J. Res. Comput. Sci.*, vol. 16, no. 4, pp. 95–103, Oct. 2023.

[170] P. Srivani, C. R. Y. Devi, S. H. Manjula, and K. R. Venugopal, ''Monitoring of environmental parameters using Internet of Things and analysis of correlation between the parameters in a DWC hydroponic technique,'' *Int. J. Adv. Intell. Paradigms*, vol. 26, no. 1, pp. 60–72, 2023.

[171] A. Ahaddi and Z. Al-Husseini, ''The use of hyperspectral sensors for quality assessment: A quantitative study of moisture content in indoor vertical farming,'' Mälardalen Univ., School Innov., Des. Eng., Tech. Rep., 2023.

[172] V. Venkadesh, V. Kamat, S. Bhansali, and K. Jayachandran, ''Advanced multi-functional sensors for in-situ soil parameters for sustainable agriculture,'' *Electrochemical Soc. Interface*, vol. 32, no. 4, pp. 55–60, Dec. 2023.

[173] R. A. Dinesh, J. Shanmugam, and K. Biswas, ''Integration of technology and nanoscience in precision agriculture and farming,'' in *Contemporary Developments in Agricultural Cyber-Physical Systems*. Hershey, PA, USA: IGI Global, 2023, pp. 149–171.

[174] A. Dsouza, L. Newman, T. Graham, and E. D. G. Fraser, ''Exploring the landscape of controlled environment agriculture research: A systematic scoping review of trends and topics,'' *Agricult. Syst.*, vol. 209, Jun. 2023, Art. no. 103673.

[175] J. Luo, B. Li, and C. Leung, ''A survey of computer vision technologies in urban and controlled-environment agriculture,'' *ACM Comput. Surv.*, vol. 56, no. 5, pp. 1–39, May 2024.

[176] C. I. Gan, R. Soukoutou, and D. M. Conroy, ''Sustainability framing of controlled environment agriculture and consumer perceptions: A review,'' *Sustainability*, vol. 15, no. 1, p. 304, Dec. 2022.

[177] M. H. Talbot, D. Monfet, T. Lalonde, and D. Haillot, ''Impact of modelling thermal phenomena in a high-density controlled environment agriculture (CEA-HD) space,'' *Building Simul.*, vol. 17, pp. 1357–1364, Sep. 2022.

[178] L. Casey, B. Freeman, K. Francis, G. Brychkova, P. McKeown, C. Spillane, A. Bezrukov, M. Zaworotko, and D. Styles, ''Comparative environmental footprints of lettuce supplied by hydroponic controlled-environment agriculture and field-based supply chains,'' *J. Cleaner Prod.*, vol. 369, Oct. 2022, Art. no. 133214.

[179] A. N. Hamilton, K. E. Gibson, M. A. Amalaradjou, C. W. Callahan, P. D. Millner, S. Ilic, M. L. Lewis Ivey, and A. M. Shaw, ''Cultivating food safety together: Insights about the future of produce safety in the US controlled environment agriculture sector,'' *J. Food Protection*, vol. 86, no. 12, Dec. 2023, Art. no. 100190.

[180] M. Ezzeddine, ''Towards a sustainable lifecycle in controlled environment agriculture (CEA),'' Doctoral dissertation, Fac. Graduate School, Cornell Univ., USA, 2023.

[181] N. Tongphanpharn, C.-Y. Guan, W.-S. Chen, C.-C. Chang, and C.-P. Yu, ''Evaluation of long-term performance of plant microbial fuel cells using agricultural plants under the controlled environment,'' *Clean Technol. Environ. Policy*, vol. 25, pp. 633–644, Nov. 2021.

[182] S. Marvin, L. Rickards, and J. Rutherford, ''The urbanisation of controlled environment agriculture: Why does it matter for urban studies?'' *Urban Stud.*, vol. 61, no. 8, pp. 1430–1450, Jun. 2024.

[183] M. McClure et al., ''An outbreak investigation of Salmonella Typhimurium illnesses in the United States linked to packaged leafy greens produced at a controlled environment agriculture indoor hydroponic operation—2021,'' *J. Food Protection*, vol. 86, no. 5, May 2023, Art. no. 100079.

[184] P. Kumar, B. Sampath, S. Kumar, B. H. Babu, and N. Ahalya, ''Hydroponics, aeroponics, and aquaponics technologies in modern agricultural cultivation,'' in *Trends, Paradigms, and Advances in Mechatronics Engineering*. Hershey, PA, USA: IGI Global, 2023, pp. 223–241.

[185] J. Kim, H.-J. Kim, M.-S. Gang, D.-W. Kim, W.-J. Cho, and J. K. Jang, ''Closed hydroponic nutrient solution management using multiple water sources,'' *J. Biosyst. Eng.*, vol. 48, no. 2, pp. 215–224, Jun. 2023.

[186] S. S. Saldinger, V. Rodov, D. Kenigsbuch, and A. Bar-Tal, ''Hydroponic agriculture and microbial safety of vegetables: Promises, challenges, and solutions,'' *Horticulturae*, vol. 9, no. 1, p. 51, Jan. 2023.

[187] F. Kabiru, C. Nina, and B. Kosgei, ''Feasibility study of the best CEA system for school feeding in mukuru informal settlement, Nairobi,'' Mukuru Informal Settlement, Nairobi, Tech. Rep., 2023.

[188] M. A. C. Griffith, G. P. Buss, P. A. Carroll, X. Yang, J. L. Griffis, G. Papkov, S. Bauer, K. Jackson, and A. K. Singh, ''The comparative performance of nutrient film technique and deep-water culture hydroponics method using GREENBOX technology,'' in *Proc. ASABE Annu. Int. Meeting*, 2023, p. 1.

[189] A. N. Hamilton, Z. Topalcengiz, and K. E. Gibson, ''Growing safer greens: Exploring food safety practices and challenges in indoor, soilless production through thematic analysis of leafy greens grower interviews,'' *J. Food Protection*, vol. 86, no. 11, Nov. 2023, Art. no. 100163.

[190] Á. Szepesi, ''Alternative production systems ('roof-top,' vertical, hydroponic, and aeroponic farming),'' in *Agroecological Approaches for Sustainable Soil Management*. USA: Wiley, 2023, pp. 261–275.

[191] Q. Arshad, S. Ahmad, H. F. Gabriel, Z. H. Dahri, M. Shahid, U. Ullah, and A. Ullah, ''Conserving water: Cost and productivity analysis of responsive drip and conventional irrigation,'' *Environ. Eng. Manage. J.*, vol. 22, no. 4, pp. 639–649, 2023.

[192] Y. N. Rahmat and J. Neilson, ''The ebb and flow of capital in Indonesian coastal production systems,'' *Singap. J. Tropical Geography*, vol. 44, pp. 300–321, May 2023.

[193] V. Lobanov, ''Nutrient transfer in aquaponic systems-optimizing microbial processes for greater circularity and economic viability,'' Fac. Sci., Dept. Marine Sci., Tech. Rep., 2023. [Online]. Available: https://hdl.handle.net/2077/75607

[194] L. Wang, G. Lian, Z. Harris, M. Horler, Y. Wang, and T. Chen, ''The controlled environment agriculture: A sustainable agrifood production paradigm empowered by systems engineering,'' in *Computer Aided Chemical Engineering*. Amsterdam, The Netherlands: Elsevier, pp. 2167–2172.

[195] A. Min, N. Nguyen, L. Howatt, M. Tavares, and J. Seo, ''Aeroponic systems design: Considerations and challenges,'' *J. Agricult. Eng.*, vol. 54, pp. 19–30, Sep. 2023.

[196] S. B. Kasturi, C. H. Ellaji, D. Ganesh, K. Somasundaram, and B. Sreedhar, ''IoT and machine learning approaches for classification in smart farming,'' *J. Surv. Fisheries Sci.*, vol. 10, pp. 3373–3385, May 2023.

[197] S. Saleem, M. I. Sharif, M. I. Sharif, M. Z. Sajid, and F. Marinello, ''Comparison of deep learning models for multi-crop leaf disease detection with enhanced vegetative feature isolation and definition of a new hybrid architecture,'' *Agronomy*, vol. 14, no. 10, p. 2230, Sep. 2024.

[198] Y. Akkem, S. K. Biswas, and A. Varanasi, ''Smart farming using artificial intelligence: A review,'' *Eng. Appl. Artif. Intell.*, vol. 120, Apr. 2023, Art. no. 105899.

[199] M. T. Linaza, J. Posada, J. Bund, P. Eisert, M. Quartulli, J. Döllner, A. Pagani, I. G. Olaizola, A. Barriguinha, T. Moysiadis, and L. Lucat, ''Data-driven artificial intelligence applications for sustainable precision agriculture,'' *Agronomy*, vol. 11, no. 6, p. 1227, Jun. 2021.

[200] M. Sharif, ''A comprehensive survey on applications, challenges, threats and solutions in IoT environment and architecture challenges,'' in *Threats and Solutions in IoT Environment and Architecture*. Washington, DC, USA: Science Publications, Jan. 2024.

[201] T. Saranya, C. Deisy, S. Sridevi, and K. S. M. Anbananthen, ''A comparative study of deep learning and Internet of Things for precision agriculture,'' *Eng. Appl. Artif. Intell.*, vol. 122, Jun. 2023, Art. no. 106034.

[202] M. J. Williams, M. N. K. Sikder, P. Wang, N. Gorentala, S. Gurrapu, and F. A. Batarseh, ''The application of artificial intelligence assurance in precision farming and agricultural economics,'' in *AI Assurance*. Amsterdam, The Netherlands: Elsevier, 2023, pp. 501–529.

[203] N. Rao, S. Patil, C. Singh, P. Roy, C. Pryor, P. Poonacha, and M. Genes, ''Cultivating sustainable and healthy cities: A systematic literature review of the outcomes of urban and peri-urban agriculture,'' *Sustain. Cities Soc.*, vol. 85, Oct. 2022, Art. no. 104063.

[204] M. E. Latino, M. Menegoli, and A. Corallo, ''Agriculture digitalization: A global examination based on bibliometric analysis,'' *IEEE Trans. Eng. Manag.*, vol. 71, pp. 1330–1345, 2022.

[205] I. S. Brumă, A. R. Jelea, S. Rodino, P. E. Bertea, A. Butu, and M. A. Chițea, ''A bibliometric analysis of organic farming and voluntary certifications,'' *Agriculture*, vol. 13, no. 11, p. 2107, Nov. 2023.

[206] Y. Wang and Y. Xia, ''International research hotspots and trends analysis of digital agriculture based on CiteSpace,'' *Hubei Agricult. Sci.*, vol. 62, no. 12, pp. 178–187, 2023.

[207] J. Wang, Y. Chen, J. Huang, X. Jiang, and K. Wan, ''Leveraging machine learning for advancing insect pest control: A bibliometric analysis,'' *J. Appl. Entomol.*, vol. 2024, pp. 112–125, Jan. 2024.

[208] X. Su, S. Wang, and R. Yu, ''A bibliometric analysis of blockchain development in industrial digital transformation using CiteSpace,'' *Peer-Peer Netw. Appl.*, vol. 17, no. 2, pp. 739–755, Mar. 2024.

[209] A. Kaur, K. Guleria, and N. K. Trivedi, ''Rice leaf disease detection: A review,'' in *Proc. 6th Int. Conf. Signal Process., Comput. Control (ISPCC)*, Oct. 2021, pp. 418–422.

[210] A. Kaur, K. Guleria, and N. K. Trivedi, ''A deep learning based model for rice leaf disease detection,'' in *Proc. 10th Int. Conf. Rel., INFOCOM Technol. Optim.*, Oct. 2022, pp. 1–5.

[211] A. Kumar, S. K. Mangla, and P. Kumar, ''An integrated literature review on sustainable food supply chains: Exploring research themes and future directions,'' *Sci. Total Environ.*, vol. 821, May 2022, Art. no. 153411.

[212] M. W. Barbosa, ''Uncovering research streams on agri-food supply chain management: A bibliometric study,'' *Global Food Secur.*, vol. 28, Mar. 2021, Art. no. 100517.

[213] K. Srinivasan and V. K. Yadav, ''An integrated literature review on urban and peri-urban farming: Exploring research themes and future directions,'' *Sustain. Cities Soc.*, vol. 99, Dec. 2023, Art. no. 104878.